\documentclass[useAMS,usenatbib]{mn2e}
\usepackage{amssymb,graphicx,amsmath}

\newcommand{\cov}[2]{\ensuremath{{\rm Cov}(#1,#2)}}

\newcommand{\e}[1]{\ensuremath{{\rm E}(#1)}}
\newcommand{\ed}[2]{\ensuremath{{\rm E}_{#1}(#2)}}

\newcommand{\var}[1]{\ensuremath{{\rm Var}(#1)}}
\newcommand{\vard}[2]{\ensuremath{{\rm Var}_{#1}(#2)}}

\newcommand{\mycomment}[1]{}

\providecommand{\eqref}[1]{(\ref{#1})}

\def\kms{\hbox{km}\,\hbox{s}^{-1}}

\def\Fbh{F_{\rm bh}}
\def\dstab{\epsilon_{\rm disk}}
\def\acool{\alpha_{\rm cool}}

\def\arh{\alpha_{\rm reheat}}
\def\alphahot{\alpha_{\rm hot}}
\def\vhotdisk{V_{\rm hot, disk}}
\def\vhotburst{V_{\rm hot, burst}}
\def\alphareheat{{\arh}}
\def\alphacool{{\acool}}
\def\epsilonStar{\epsilon_{\star}}
\def\invepsilonStar{\epsilonStar^{-1}}
\def\alphastar{\alpha_{\star}}
\def\taumrg{f_{\rm df}}
\def\VCUT{v_{\rm cut}}
\def\ZCUT{z_{\rm cut}}
\def\fellip{f_{\rm ellip}}
\def\fburst{f_{\rm burst}}
\def\FSMBH{{\Fbh}}
\def\yield{p_{\rm yield}}
\def\stabledisk{f_{\rm stab}}
\def\epsilonSMBHEddington{\epsilon_{\rm Edd}}

\def\talphahot{\tilde{\alpha}_{\rm hot}}
\def\tvhotdisk{\tilde{V}_{\rm hot, disk}}
\def\tvhotburst{\tilde{V}_{\rm hot, burst}}
\def\talphareheat{\tilde{\alpha}_{\rm reheat}}
\def\talphacool{\tilde{\alpha}_{\rm cool}}
\def\tepsilonStar{\tilde{\epsilon}_{\star}}
\def\tinvepsilonStar{\tepsilonStar^{-1}}
\def\talphastar{\tilde{\alpha}_{\star}}
\def\ttaumrg{\tilde{\tau}_{0, \rm mrg}}
\def\tyield{\tilde{p}_{\rm yield}}
\def\tstabledisk{\tilde{f}_{\rm stab}}
\def\tepsilonSMBHEddington{\tilde{\epsilon}_{\rm Edd}}

\def\GALFORM{{\sc GALFORM}}
\def\mx{\mathcal{X}}

\def\bJ{b_{\rm J}}    
\def\vx{{\bf x}}     
\def\vf{{\bf f}}
\def\vy{{\bf y}}
\def\vz{{\bf z}}
\def\vepsilon{{\boldsymbol\epsilon}}
\def\vPhi{{\bf \Phi}}

\def\apj{ApJ}
\def\mnras{MNRAS}

\def\rgb{}   

\begin{document}

\title{The Parameter Space of Galaxy Formation}

\author[Bower et al.]
{
\parbox[t]{\textwidth}{
\vspace{-1.0cm}
R. G. Bower$^{1}$,
I. Vernon$^{2}$,
M. Goldstein$^{2}$,
A. J. Benson$^{3}$,
C. G. Lacey$^{1}$,\\
C. M. Baugh$^{1}$,
S. Cole$^{1}$,
C. S. Frenk$^{1}$
} 
\vspace*{6pt}\\
$^{1}$Institute for Computational Cosmology, Department of Physics, 
University of Durham, South Road, Durham, DH1 3LE, UK.\\ ({\it e-mail: r.g.bower@durham.ac.uk})\\
$^{2}$Department of Mathematics, University of Durham, Durham, UK\\
$^{3}$Mail Code 350-17, California Institute of Technology, Pasadena, CA 91125, U.S.A.\\
\vspace*{-0.5cm}}

\maketitle

\begin{abstract}
Semi-analytic models are a powerful tool for studying the formation of galaxies.
However, these models inevitably involve  a significant number of poorly constrained 
parameters that must be adjusted to provide an acceptable match to the observed 
universe. In this paper, we set out to quantify the degree to which observational
data-sets can constrain the model parameters. By revealing degeneracies in the
parameter space we can hope to better understand the key physical processes
probed by the data.  We use novel mathematical techniques to explore
the  parameter space of the GALFORM semi-analytic model. We base our investigation on the 
Bower et al.\ 2006 version of GALFORM, adopting the same methodology
of selecting model parameters based on an acceptable match to the local $b_J$ and $K$ luminosity
functions. {\rgb Since the GALFORM model is inherently approximate, we explicitly include
a model discrepancy term when deciding if a match is acceptable or not.}
The model contains 16 parameters that are poorly constrained by our prior understanding
of the galaxy formation processes and that can plausibly be adjusted between reasonable limits.
We investigate this parameter space using the Model Emulator technique, constructing
a Bayesian approximation to the GALFORM model that can be rapidly evaluated at any point in 
parameter space. The emulator returns both an expectation for the GALFORM model and an uncertainty
which allows us to eliminate regions of parameter space in which it is implausible that
a GALFORM run would match the luminosity function data. By combining successive waves of 
emulation, we show that only 0.26\% of the initial volume is of interest for further exploration.
However, within this region we show that the Bower et al.\ 2006 model is only one choice from an extended 
sub-space of model parameters that can provide equally acceptable fits to the luminosity function data. 
We explore the geometry of this region and begin to explore the physical connections between 
parameters that are exposed by this analysis. We also consider the impact of adding additional 
observational data to further constrain the parameter space. We see that the known tensions existing
in the Bower et al.\ 2006 model lead to a further reduction in the successful parameter space.

\end{abstract}

\section{Introduction}

Semi-analytic galaxy formation models are a successful tool for exploring the 
physical processes responsible for galaxy formation. In essence this  technique
aims to understand the formation of galaxies by breaking the problem down into a discrete
set of (typically non-linear) differential equations describing each physical
process. 
For example, the amount of gas able to cool from the halo depends
non-linearly on the halo mass and its gas content.
These discrete processes are 
then coupled through a set of interactions. For example, the cold gas mass grows as
a result of gas accretion and cooling and decreases as a result of star formation and 
gas ejection. In simple cases, the network of equations can be integrated analytically
to make quantitative predictions for the properties of the galaxy population. In 
more complex cases, the set of equations must be solved numerically, but the computational
task is still minor compared to integrating fundamental physical laws on a particle 
by particle (or cell-by-cell) basis, as required by a fully numerical approach 
(for a few state of the art examples of the fully numerical approach see 
Crain et al.\ 2009; Schaye et al.\ 2009; Gnedin et al.\ 2009).
However, although the description of each individual component of the semi-analytic model may
appear simple, the complex interplay between the components means that the 
outcome of a model is notoriously hard to predict.

Nevertheless, such models have been very successful
in defining our current picture of how galaxies form. Initial models, such as White \& Frenk
(1991), Lacey 
\& Silk (1991), Kauffmann, White, 
\& Guiderdoni (1993) and  Cole et al. (1994) showed how the formation of
galaxies resulted from a competition between gas cooling 
and accretion, and the ejection of gas from galaxies in supernova-driven winds. This type
of feedback explained the observed paucity of faint galaxies compared
to the high abundance of low mass  cold dark 
matter haloes. By incorporating these effects into a realistic model for the growth
of dark matter haloes and  galaxies, these models were able to
make a quantitative connection between the assumptions about gas cooling, star formation, 
feedback, merging and other physical ingredients, and the observed properties of galaxies. Over 
the past two decades, the sophistication of these models has increased, allowing them
to make predictions for many more observational properties such as galaxy sizes, colours,
infrared luminosities and correlation functions (eg., Kauffmann et al.\ 1999; 
Somerville \& Primack 1999; Cole et al.\ 2000; Granato et al. 2000, 2004; Baugh et al.\ 2005;
Menci et al.\ 2005, 2006; Cattaneo et al.\ 2006; Kang, Jing, \& Silk 2006;  Monaco 2007). 
At the same time,
the improvement in our knowledge of the cosmological parameters has tied down some of the
major uncertainties in the input physical description (eg., Dunkley et al.\ 2009). 
As a result, the comparative power of the models has increased. 

A particular issue that has been revealed is the need for 
additional physics to match the sharp break at the bright end of the
galaxy luminosity function.
A number of additional physical processes have been proposed (c.f. Benson et al 2003a) but
the currently favored explanation centres on an additional feedback channel motivated
by observations of the interactions between radio galaxies and the surrounding
IGM in clusters. Although implementations differ, the aim of this ``radio-mode'' feedback is to
suppress cooling in the most massive haloes leading to the sharp break in the luminosity 
function (Bower et al. 2006 [Bow06]; Croton et al. 2006; Cattaneo et
al. 2006; Somerville et al. 2008). 
An important result of implementing this type of feedback in the models is that
they then predict that much of the star formation in the largest galaxies will be completed
relatively early in the history of the universe, in many cases above redshift 2. This 
has largely eased the conflict between observations of a large population of passive
galaxies at high redshift and the tendency for Cold Dark Matter (CDM) models to form 
the largest dark matter structures only recently (Bow06). 

Despite these successes, the semi-analytic technique has been criticised for
a perceived lack of predictive power. Each component of the model must
simply encapsulate the physical process that it describes. However, since the processes
are poorly understood, this almost inevitably involves parameterising the process in
such a way that our limited knowledge or understanding can be included by allowing parameters
to vary between plausible limits. By comparing the model to a limited set
of observational results, the model can be calibrated and then, with the values
of the parameters fixed, the model can be tested against additional observational constraints.
While the traditional approach, such as that used in Cole et al.\ (2000)
or Bow06, is {\rgb iterate on an intital guess to find a single set of parameters
that adequately match the calibration data}, this is clearly a Bayesian problem in
which we should seek to use the observational data to successively constrain the
parameter space of acceptable models.

In this paper we set out to make a systematic exploration of the parameter space
of the Bow06 version of the GALFORM model. This contains 16
parameters which can reasonably be adjusted over a plausible range. 
We note earlier GALFORM models have considered an even larger parameter
space: for example, the Baugh et al.\ (2005) model uses a different parameterization
for the disk star formation timescale, includes a mode of
superwind feedback (eg., Benson et al. 2003a), allows for a different
IMF in starbursts from that in disk star formation, and uses different
descriptions of gas cooling and gas reheating (cf., Bow06). These
differences are not considered here --- our purpose is to compare the
parameter set identified by Bow06 with the full parameter space available in
that model. We explore the effect of introducing additional physical
processes in Benson \& Bower 2009.

A variety of strategies for calibrating the model parameters have been
adopted in published semi-analytical models. The majority of models
have used observational data on selected galaxy properties at $z=0$ to
choose a ``best fit'' set of parameters, and have then made predictions for
higher redshifts, but some models have also supplemented the $z=0$
constraints with observational data on high-redshift galaxies when
choosing the ``best fit'' model parameters. Different authors have made different
choices as to what is the best set of $z=0$ properties to use in
setting the model parameters. For example, Kauffmann et al. (1993),
Kauffmann et al. (1999), Somerville \& Primack (1999) and De~Lucia et
al. (2004) used the normalization of the Tully-Fisher relation and the
gas masses of Milky Way-like galaxies as their primary observational
constraints. On the other hand, Cole et al. (1994), Cole et al. (2000),
Nagashima et al. (2001), Kang et al. (2005), Baugh et al. (2005) and
Bow06 all used the galaxy optical and near-IR luminosity functions as
their primary constraints. In addition, most
models have used additional $z=0$ properties beyond their ``primary''
constraint in choosing best-fit parameters. For example, Cole et
al.(2000) used the gas fractions and sizes of galaxy disks, together
with the ratio of early to late morphological types and the stellar
metallicities of elliptical galaxies, in addition to the $B$- and
$K$-band luminosity functions. 
In contrast to this, Benson et al.\ 2003a and Bow06 chose to focus on 
obtaining a good match to the $z=0$ $B$- and $K$-band luminosity 
functions {\rgb (but it is important to note that the starting point for
iteration was taken from the Cole et al. 2000 model)}. Several subsequent
papers have explored the performance of the Bow06 model with respect to
additional data sets (eg., Bower et al.\ 2008; Font et al.\ 2008; Gonzalez-Perez et al.\ 2008; 
Seek Kim et al.\ 2009; Gonzalez et al.\ 2009).

These different strategies for calibrating the model parameters have
different advantages and drawbacks. The Bow06 approach has the advantage
of simplicity, and that the $z=0$ luminosity functions are
measured accurately and (largely) free from observational
selection effects. Furthermore, the model outputs do not require a
highly complex layer of additional processing to cast them into the 
observed quantities {\rgb (of course population synthesis models are still required)}. 
A disadvantage is
that the present-day optical and near-IR luminosity functions are
relatively insensitive to some model parameters, such as those controlling the
star formation timescale (e.g. Cole et al. 2000).
For this reason, it is helpful to
introduce additional observational constraints to which these other
model parameters are more sensitive. For example, Cole et al. 2000 found that
the gas mass vs luminosity relation for disk galaxies provides very
good constraints on the model parameters for star formation. 
Potential drawbacks of introducing extra observational constraints
beyond the $z=0$ luminosity functions are that they may be less
accurately determined observationally, and that a subjective
decision is required to assign relative weights to the different
observational constraints. In addition, if all the available
data-sets are used to constrain the model, no observations
will be immediately available to independently test its validity. This is a deep
philosophical issue that we will not tackle here, but clearly we should seek
a strategy in which the physical role of each constraint is clear.

Thus, given the wide variety of observational data that could be used to
constrain semi-analytical models, each with their own random and
systematic errors, what is needed is some more objective procedures for
evaluating what is the range of model parameters consistent with a
particular combination of observational constraints, and what is the
effect on this range of adding or removing a particular observational
constraint. In this way, we hope to end up with an objective measure
of how robust different predictions from the model are, including how
sensitive they are to including different model ingredients and
different observational constraints.

This paper is a first step in this program. We introduce a new method
of exploring the model parameter space to identify those regions that
produce acceptable matches to the observational data. For simplicity,
in this paper we follow the approach of Bow06 and use only the $b_J$
and $K$-band $z=0$ luminosity functions to directly constrain the acceptable
regions of parameter space. Once we have identified these regions, we briefly examine the performance of
the model with respect to additional $z=0$ data sets, but these are not used to define the
initial search criterion. Furthermore, we use the same version of
GALFORM as in Bow06,
making it simple to compare the unique parameter set presented in Bow06 with the full parameter space
that we identify in our search here. Since the Bow06 model is implemented on the
Millennium N-body simulation (Springel et al.\ 2005), we  adopt the
same fixed cosmological 
parameter set. In principle, the methods we present here could be extended to allow the
cosmological background model to vary.

Our investigation aims to address some key questions: How large is the range of 
parameter space that produces acceptable fits? Is the parameter set selected in Bow06 in
some sense typical or optimal? 
It is unlikely that there is a single ``best value''. Given the relatively large
number of model parameters, there will be a range of parameter values giving acceptable fits.
Moreover, we should be careful to define what we mean by an ``acceptable'' fit. Since
the GALFORM model is only an approximation to reality, we would not 
expect the model to exactly reproduce all the observational data, even if the model's 
parameters were set to their ``best'' values. The Bayesian approach we adopt requires us to formalize
this uncertainty by introducing a ``model discrepancy'' term ($\epsilon_{md}$) into our comparison
with the data. This has the effect of ensuring that we do not reject a region of parameter space
if the comparison with the data is sufficiently good that future improvements
in the model (which reduce the degree of approximation) may result in improved agreement with the 
data (in that region).
{\rgb 
This approach is fundamentally different from simply requiring that we find the region of agreement 
within the observational uncertainties --- it recognises that the model is itself approximate. 
Ignoring $\epsilon_{md}$ will lead us to focus on an unjustifiably narrow region of parameter 
space. In this case, reducing the level of approximation in the model would cause new regions of 
acceptable parameter space to appear in areas that were previously deemed implausible. 

Of course, estimation of $\epsilon_{md}$ is uncertain. In principle, one could hope to arrive at a value
by tracking changes to the model as the level of approximation is reduced. This
approach is an active subject in the statistical litterature (eg.\ Goldstein \& Rougier, 2009),
but the methods are not yet suitable for application to GALFORM. Instead, we addressed
the model discrepancy term by constructing a series of test luminosity functions
and asking ourselves whether we would comfortably reject the corresponding region of parameter space
on the basis of the comparison and our previous experience of improvements to the GALFORM code. 
Reassuringly, our estimate of $\epsilon_{md}$ results in the Bow06 being marginally 
acceptable.  Thus the model discrepancy is consistent with our aim of searching parameter
space for parameter sets that perform comparably to (or better than) Bow06.}

Our task is therefore to evaluate the GALFORM model over the input parameter
space,  identifying the portion of this  space for which fits to
the local luminosity functions are acceptable. Unfortunately, the
16-dimensional parameter space (that is introduced below) is extremely large. 
Dividing each axis into (just) 5 values
and exploring all possible parameter combinations would require $10^{11}$ evaluations
of the model (and hence require computing time in excess of $10^8$ cpu-years). 
Even if this were possible, the resulting grid would be of such low resolution that it
would give little indication of the GALFORM parameter space. Clearly a much better 
targeted strategy can be devised.  

In this paper,  we use the ``model emulator'' technique (eg., Craig et al.\ 1997; Kennedy \& O'Hagan 2001; 
Vernon et al.\ 2010) to explore the parameter space. This technique has been specifically developed within the statistics 
community in order to analyze models that possess high-dimensional parameter spaces. It 
involves constructing a stochastic model that emulates the output of the GALFORM model.
The emulator is constructed so as to reproduce the results of known runs and  
statistically interpolate between them taking into account the appropriate correlation
length of the model. At each new point, the model provides an expectation value
for the outcome of a GALFORM evaluation and a variance reflecting the degree of
uncertainty in the emulator output. An evaluation of the emulator is of order $10^7$
times faster than an evaluation of the full model, and the emulator can therefore
be used to eliminate regions of parameter space for which it is implausible that 
an evaluation of GALFORM will result in an
acceptable match to the observational data. By proceeding in waves of emulation,
we successively reduce the volume that must be investigated at each level until
the volume that must be directly evaluated is a tiny fraction (less
than 0.3\%) of the original parameter 
space. The primary advantage of the emulator is its speed, which allows us to
investigate the full parameter space efficiently and restricts time-consuming
evaluations of the GALFORM model to regions of parameter space where the outcome
cannot be predicted with sufficient accuracy by the emulator. Combining the emulator 
method with an efficient strategy for sparsely sampling the parameter space, we can explore
the parameter space of galaxy formation with around a month of CPU time. These 
techniques are gaining widespread acceptance in  the climate research community where full 
evaluations of the computer model are prohibitively expensive.  The parallel with the
galaxy formation problem is powerful and illustrative (Vernon et al.\ 2010).
Our work is also closely related to studies of the galaxy formation parameter
space that are based on exploration with Monte Carlo Markov chain techniques
(Henriques et al.\ 2009, Kampakoglou et al.\ 2008).
These methods currently consider lower dimensionality than we address here,
and it should be noted that, in general, MCMC techniques may face problems when dealing 
with high-dimensional input spaces. {\rgb We should also stress that the dimensionality of the
problem that we consider here is likely to greatly increase as additional physical
processes are included in the model.}  Indeed, this has been one of the major motivations 
for the development of the emulation techniques presented here 
(Oakley \& O'Hagan 2004). For a summary of 
state-of-the-art emulation techniques see the Managing Uncertainty for Complex Models 
website http://mucm.group.shef.ac.uk/index.html.

The emulator process identifies a small fraction of the total input space as generating
acceptable luminosity functions. The geometry
and extent of the region is, however, hard to comprehend. As we will see, some parameters 
(or parameter combinations) are poorly constrained: this can be viewed as telling us that 
these have little role in determining certain observable properties of galaxies. Conversely,
some parameter combinations are tightly constrained: these play a critical role, and
we can hope to use this to understand more about the interplay of the components in 
the GALFORM model, and thus to better understand the physics underlying the galaxy formation
process. 

As we have already stressed, this paper concentrates
on the Bow06 version of the GALFORM code. Future papers will explore the
much larger parameter space created by recent updates to the code, introducing 
new physical processes to the problem, such as better treatment of angular momentum,
a physical description of ram pressure stripping (cf., Font et al.\ 2008),
AGN heating of halo gas (Bower et al.\ 2008), and a variable 
stellar IMF (Baugh et al.\ 2005). We also
extend our new parameter search technique to use  a wider
range of calibration data from the outset (cf., Benson \& Bower 2009). 
 
The paper is laid out as follows. In \S2, we provide a brief overview of the GALFORM code,
and outline the physical meaning of the parameters that we vary in this project. In \S3,
we describe the model emulator technique on which our parameter exploration is based. 
{\rgb \S4 presents
the main results, \S4.1 focussing on our success in emulating the luminosity function
and its dependence on the model's parameters. Although it is not the primary
focus of the paper, it is obviously of interest to see whether additional data sets 
break the degeneracies evident in the luminosity function comparison. In \S4.2,
we briefly investigate the role of additional datasets.}
In \S5, we examine the physical implications of these results using a PCA to
identify important combinations of the input parameters. Finally, we present a discussion
of our work in \S6 and briefly summarise our main conclusions in \S7.
Throughout, we adopt a cosmology in which $\Omega_b = 0.045$, $\Omega_M = 0.25$, $\Lambda = 0.75$ and 
$\sigma_8=0.9$ at the present day. The model assumes $H_0 = 73\, {\hbox{km}\,\hbox{s}^{-1}\,\hbox{Mpc}^{-1}}$, although we quote 
luminosities and space densities in term of $h = H_0 / 100\, {\hbox{km}\,\hbox{s}^{-1}\,\hbox{Mpc}^{-1}}$

\section{Parameters of the GALFORM Code}

The GALFORM code contains many parameters. In Table~\ref{tab:params}, we list the parameters
used in the Bow06 version of the code, together with the range of plausible values
considered in our analysis.  
We have grouped the parameters by the physical processes that they are associated with.
Below, we briefly describe the parameters. For a full description, we
refer the reader 
to Cole et al.\ (2000), Baugh et al.\ (2005) and Bow06.

The first set of parameters are associated with star formation: $\epsilonStar$ determines
the normalisation of the star formation efficiency, while $\alphastar$ determines its
dependence on the disk circular speed:
\begin{equation}
SFR =\epsilonStar \left(V_{\rm c,disk}/200 \,{\rm km\,s^{-1}}\right)^{-\alphastar}
  \left(M_{\rm cold}/\tau_{\rm dyn, disk}\right) 
\end{equation}
where SFR is the star formation rate, $M_{\rm cold}$ the mass of cold gas in the 
galaxy, $V_{\rm c,disk}$ is the circular velocity of the disk and
$\tau_{\rm dyn, disk}$ its dynamical time. 
We calculate chemical enrichment using the instantaneous recycling
approximation, so the rate of ejection of newly synthesized metals
into the ISM is given by
\begin{equation}
\dot{M}_{\rm Z,ej} = \yield SFR
\end{equation}
where $\yield$ is the yield of metals, which depends on the IMF. For consistency with Bow06, we use a Kennicutt (1983) IMF throughout, but treat
$\yield$ as an adjustable parameter. In Font et al.\ (2008) and Bower et al.\ (2008), we showed 
that the match to the observed colours of galaxies was improved by adopting a 
higher yield (0.04) than the standard value (0.02). 

The second group of parameters are 
associated with the supernova-driven feedback: $\vhotdisk$ and $\vhotburst$ control
the normalization of feedback in quiescent star formation and bursts respectively; 
$\alphahot$ controls the dependence of the feedback on the circular velocity. For example,
the rate at which mass is returned from the cold phase to the halo during quiescent star formation
is given by
\begin{equation}
\dot{M}_{\rm outflow}= SFR \left(V_{\rm c,disk} \over \vhotdisk \right)^{-\alphahot}.
\end{equation}
Cold gas that is ejected from the disk becomes available to cool and form further
stars after a factor $\alphareheat^{-1}$ times the halo dynamical time. In low mass haloes
cooling is very rapid, and this parameter plays a key role in setting the disk
fueling rate.

AGN feedback is controlled by the parameters $\alphacool$, which effectively determines
the halo mass at which this form of feedback becomes effective, and 
$\epsilonSMBHEddington$\footnote{
Note that due to an error in Bow06, cooling luminosities were over estimated by a factor
$4\pi$. Thus, while the paper quotes the efficiency parameter $\epsilon_{\rm SMBH}$ 
as 0.5, this should have been $0.5/4\pi = 0.04$. With this correction the 
results of Bow06 are unchanged.}, 
which controls the maximum energy output possible for a central
supermassive black hole of given Eddington
luminosity $L_{\rm Edd}$.
                           Specifically, we only allow the AGN to regulate cooling if
\begin{equation}
t_{\rm cool}(r_{\rm cool})>\alphacool^{-1}\,t_{\rm ff}(r_{\rm cool})
\end{equation}
and
\begin{equation}
L_{\rm cool} < \epsilonSMBHEddington L_{\rm Edd},
\end{equation}
where $L_{\rm cool}$ is the radiative cooling luminosity of the halo gas.
Note that larger values of $\alphacool$ result in AGN feedback being effective in 
lower mass haloes.

Galaxy mergers are dependent on the rate of decay of satellite orbits due to 
dynamical friction and on the mass ratio of the merging objects. The normalisation of the
orbital decay rate is set by $\taumrg$ (see Cole et al.\ 2000), while
$\fellip$ and $\fburst$ are respectively  the mass
ratios needed to transform the morphology of the main galaxy and  to cause a burst
of star formation (see Baugh et al.\ 2005 and Malbon et al.\ 2007)\footnote{\rgb We note, 
however, that 
the parameter $f_{\rm gas, burst}$ is set to 0.1 in this study, and in Bow06,
so that almost all sufficiently high mass ratio mergers result in a burst of star formation.
In Malbon et al.\ this parameter was set to 0.75.}. 
The disk stability parameter, $\stabledisk$, sets the 
self-gravity threshold at which galaxy disks become unstable to bar modes (see Bow06).
This instability causes the cold disk gas to be consumed in a burst  of star formation. 
Smaller values of this parameter make disks more prone to bar instabilities.

Finally, the parameters $\VCUT$ and $\ZCUT$ encapsulate the effect of reionisation on cooling
in small haloes. For further discussion of this approximation, see
Benson et al. (2003b).
We will show that these parameters have little impact on the galaxy properties we consider here.

We list in Table~\ref{tab:params} the GALFORM parameters which we
allow to vary in our parameter space exploration, together with their
values in the Bow06 model and the ranges over which we allow them to
vary. 
{\rgb Ideally we would know in advance what range for each parameter
is physically meaningful or interesting, but this is only possible for
a subset of the parameters. For example, the parameters $\fellip$ and
$\fburst$ are constrained to lie in the range [0,1] by the way they
are defined, and numerical simulations of merging galaxies constrain
their values to an even narrower range. Similar arguements can be applied
to restrict the range of $\yield$, $\epsilonSMBHEddington$, $\taumrg$, $\stabledisk$,
$\VCUT$ and $\ZCUT$. On the other hand, theory does
not currently provide any useful guide as to the value of $\epsilonStar$, so the
value of this parameter is set purely by comparison with
observations and previous experience
with GALFORM. In these cases, we selected the range by posing the question 
``if an acceptable model was found outside this range, would it be interesting?''.
We answered no if the parameter value seemed inconsistent with the physical model
that component of the code was intended to describe. The range selected is intended
to be conservatively large, but is inevitably subjective.
In some cases the parameter value adopted in Bow06 is uncomfortably high 
(e.g. the $\vhotburst$ and $\alphahot$ parameters are in principle constrained by the amount of
energy available from supernova explosions) and we deliberately extended the search range
in order to bracket the value from Bow06.}

In the following analysis, it is often helpful to use scaled variables so that
each parameter covers the range $\pm1$. We denote scaled variables by $\tilde{\alpha}$ (etc)
where
\begin{equation}
\tilde{\alpha}= {\alpha - {1\over2}\left(\alpha_{\rm max}+\alpha_{\rm min}\right)\over
  {1\over2}\left(\alpha_{\rm max}-\alpha_{\rm min}\right)}.
\end{equation}

\begin{table}
\begin{center}
\begin{tabular}{|l|c|c|c|c|c}\hline
process& parameter&  Bow06&  min&  max&Active?\\
modelled&  name\\ 
\hline
Star Formation&  $\invepsilonStar$&   350&  10&1000&A \\ 
              &    $\alphastar$&     -1.5& -3.2&-0.3&A \\ 
              &       $\yield$&      0.02&  0.02&0.05&A \\ 
SNe feedback&   $\vhotdisk$&         485&   100&550&A \\ 
            &   $\vhotburst$&        485&   100&550&A \\ 
            &     $\alphahot$&       3.2&   2.0&3.7&A \\ 
            &  $\alphareheat$&       0.92&  0.2&1.2&A \\
AGN feedback&  $\alphacool$&         0.58&  0.2&1.2&A \\ 
    &$\epsilonSMBHEddington$&        0.04&  0.004&0.05 \\ 
Galaxy Mergers&   $\taumrg$&         1.5&   0.8&2.7&A \\
              &    $\fellip$&        0.3&   0.1&0.35 \\ 
              &   $\fburst$&         0.1&   0.01&0.15 \\
              &   $\FSMBH$&        0.005&   0.001&0.01 \\ 
Disk stability&   $\stabledisk$&     0.8&   0.65&0.95&A \\ 
Reionisation&      $\VCUT$&        50&    20&50 \\ 
              &       $\ZCUT$&       6&      6&9 \\ 
\hline
\end{tabular}
\end{center}
\caption{The parameters allowed to vary in our parameter space exploration. The first column 
provides an indication of the physical process associated with the parameter. The
second column gives the parameter name; column 3 gives the value of the parameter 
used in Bow06; and columns 4 and~5 the range of the parameter explored in this paper. Active variables 
 in the model emulator
(those that are important in capturing the behaviour of the $z=0$ luminosity function output
of the model, see \S3.5.2) are indicated in column~6.}
\label{tab:params}
\end{table}

The end result is that the model spans a 16 dimensional parameter space. However, it
is extremely important to stress that several of these parameters have little impact on
the \GALFORM\ output for the selected observables, 
and thus that it is initially possible for the emulator to
capture the behaviour of the model using many fewer parameters. Our
first step was to identify the
most important parameters whose values were key to matching the selected galaxy properties.
In terms of the match to the $b_J$ and $K$ luminosity functions, there
are 10 active parameters (at Wave~4, see \S\ref{sec:emulator}) 
that drive the majority of the variation in model outputs.
These are indicated by an A in column 4 of Table \ref{tab:params}.
As we will show, the parameter space of acceptable models is limited to a very small
fraction of this volume. Even though adequate fits can be obtained
for a wide range of parameter values, variations in parameters must be carefully 
traded off to keep the input parameter set on a narrow hypersurface.

Note, however, that a parameter that is inactive when the model is constrained
using the luminosity function data may play an important role in fitting other
data sets. For example, while the reionisation parameters $\VCUT$ and $\ZCUT$ have 
little effect on the global luminosity function (within the limits considered), these
parameters play a key role in determining the satellite galaxy population 
of the Milky Way (Benson et al.\ 2003b).

\section{The Model Emulator Technique}

\subsection{Bayesian Analysis of Computer Models}\label{ssec_BACM}

\begin{figure*}
\begin{center}
\begin{tabular}{cc}
\includegraphics[scale=0.4,angle=0]{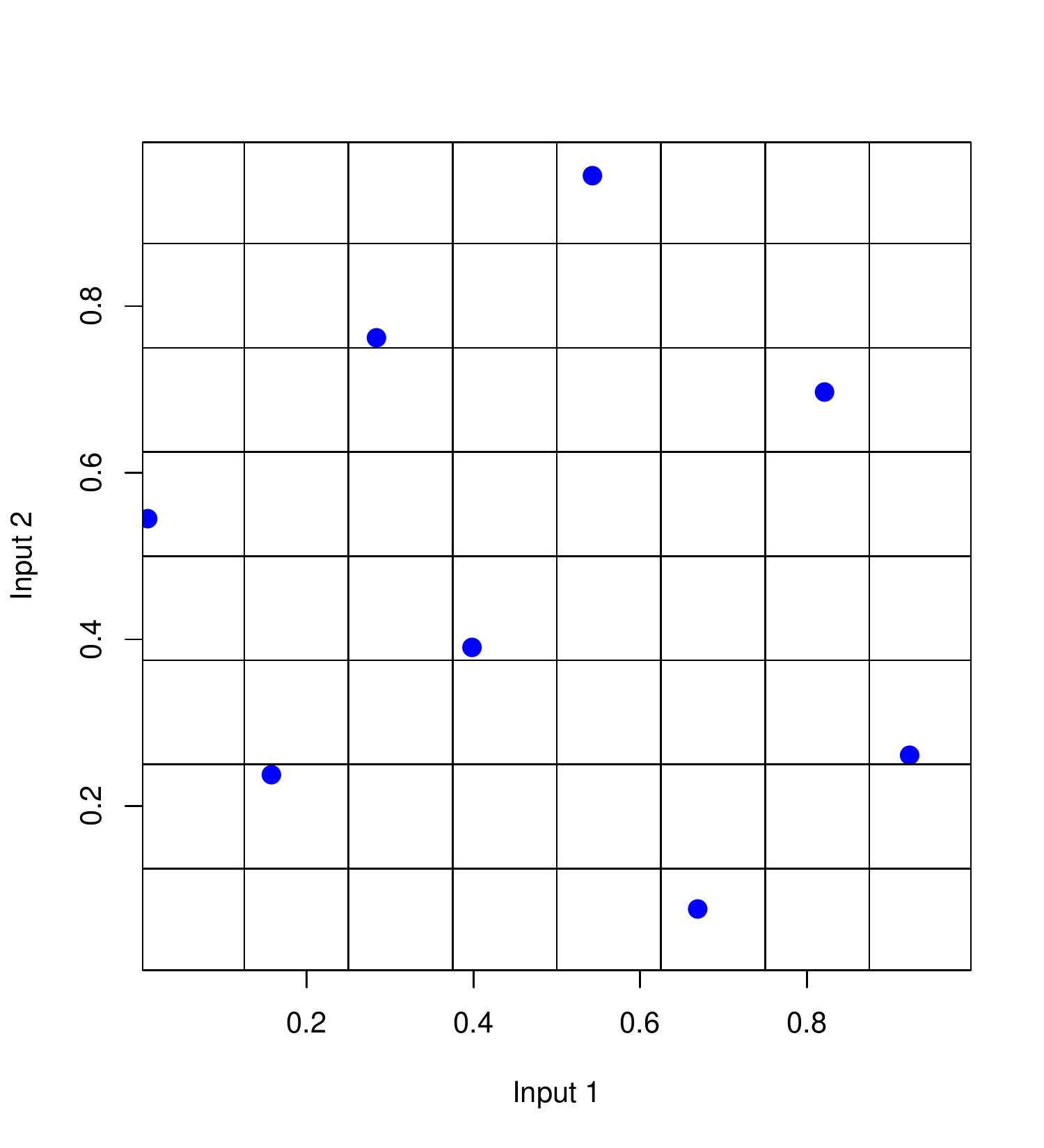} &
 \includegraphics[scale=0.4,angle=0]{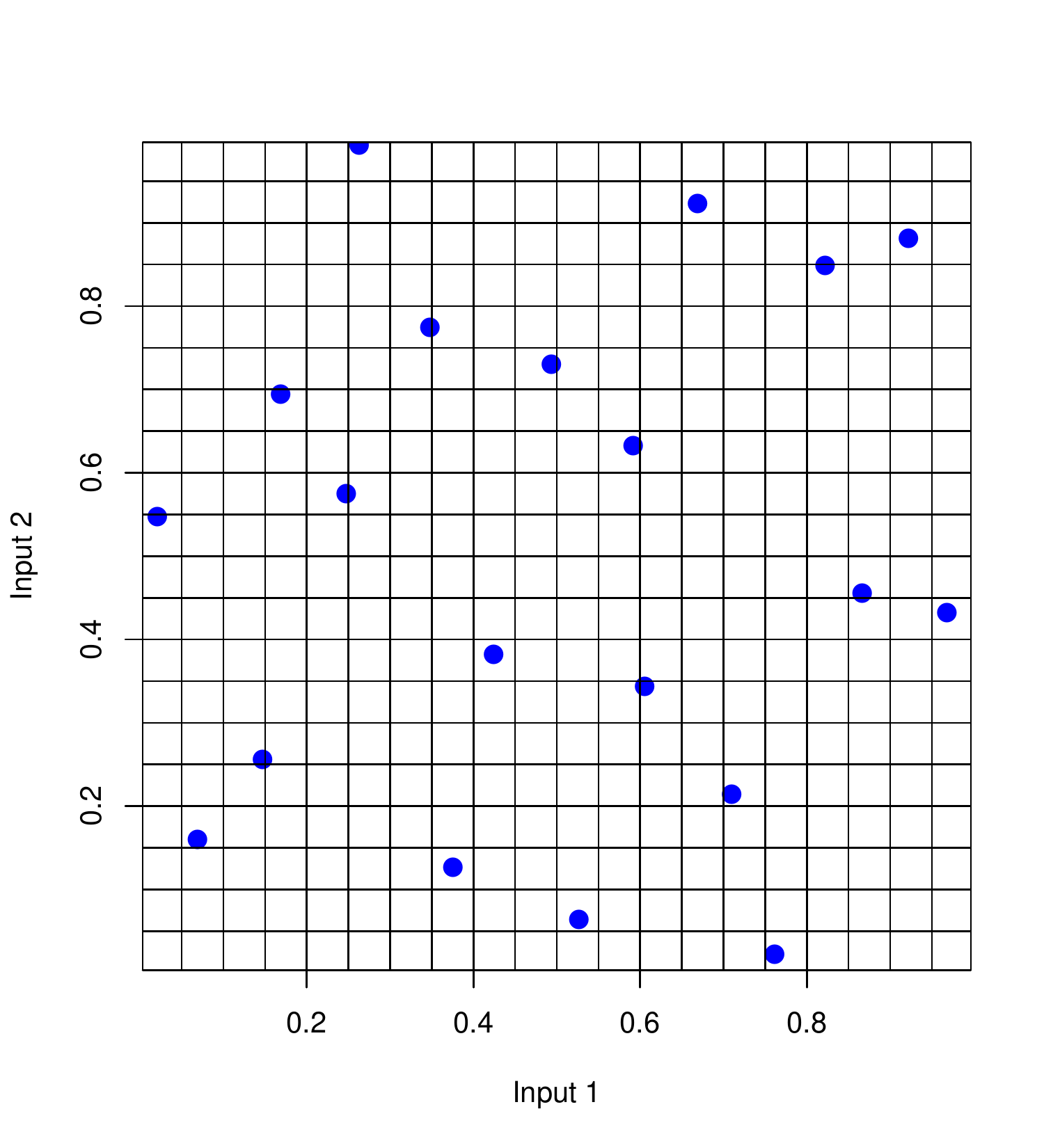}
\end{tabular}
\end{center}
\caption{Two examples of Latin Hypercube designs for a two dimensional
parameter space. The range of each input is divided into $n$ intervals where $n$ is the number of points. Note that only one point is placed in each of the $n$ intervals over Input 1 and Input 2, and that these points have been placed at random within each interval. 
The two panels show samplings with 8 and 20 points. The
Maximin strategy adopted in this paper ensures that the points are 
evenly spread throughout the region that we wish to sample.}
\label{fig_des}
\end{figure*}

There has been much interest in the statistics community in developing
techniques to help understand and analyse complex computer simulations
of real world processes, referred to generically as Computer Models (Currin et al.\ 1991,
Satner 2003, Craig et al.\ 1997, O'Haggan 2006). 
Such models, of which GALFORM is an example, generally take a significant time to run 
and require the specification of a large number of input parameters. They involve 
several distinct sources of uncertainty, all of which need to be assessed and 
combined in a unified analysis. These fall into 5 basic types: 

{\bf [1] Parameter uncertainty}:  We do not know the appropriate
values of the inputs to the simulator, and want to identify the class of inputs that give acceptable matches to the observed data. 

{\bf [2] Simulator uncertainty}: Due to the significant run time we cannot hope to cover the input space with a suitably  large number of model evaluations. Therefore, we will be uncertain as to the output of the model for regions of the input space where no evaluations have been performed. This uncertainty is handled through the use of an emulator as described in \S\ref{ssec_gestrat}. 

{\bf [3] Structural uncertainty}: This aspect, which is less familiar to
the astronomical community, refers to the fundamental
problem that, however carefully the model has been constructed, there
will always be a difference between the system (in this case the Universe) and the simulator. Simplifications in the physics, based on features that are too complicated for us to include, and simplifications and approximations in solving the equations
determining the system, lead to a discrepancy between the model and the system. We represent this through use of the ``model discrepancy'' term described 
in \S\ref{ssec_moddis}. 

{\bf [4] Observational error}: We do not know the properties of the
real Universe exactly, but instead have observational measurements with corresponding errors.

{\bf [5] Initial condition and forcing function uncertainty}: Most Computer Models require the specification of initial conditions and/or forcing functions, the form of which is most likely uncertain. 

In order to analyse the input space of the GALFORM model, and to determine which inputs are of interest, we need to address all of the above five sources of uncertainty in a unified manner. A Bayesian approach provides a natural framework for such an analysis. 
Powerful Bayesian techniques, centered around the idea of emulation, have been 
developed in the Statistics community for such problems, and have been successfully applied to models in several scientific disciplines (eg.\  Kennedy \& O'Hagan 2001, Oakley 2002, Higdon 2004, O'Hagan 2006, 
Schneider et al.\ 2008, Heitmann et al.\ 2009). However, 
employing a fully probabilistic Bayesian analysis (where every uncertain quantity is assigned a probability distribution) is often unnecessarily challenging and involves specifying prior distributions that are in some cases difficult to justify. Instead we employ the Bayes Linear approach (Goldstein \& Woof 2007) which is a more tractable version of Bayesian analysis that requires fewer assumptions, and that deals with only expectations and variances of all uncertain quantities (see \S\ref{ssec_BLA}). The Bayes Linear methods presented here have been successfully applied to several complex models 
including Oil Reservoir Models and Climate Models (Craig et al.\ 1997, Goldstein et al.\ 2009), and are well suited to the case of high-dimensional models.

\subsection{General Emulation Strategy}\label{ssec_gestrat}

 An emulator is a stochastic function that represents our
beliefs about the behaviour of a deterministic function at input
settings that have yet to be evaluated.  Representing a model such as
GALFORM as a function that maps a vector of inputs ${\vx}$ to a vector of
outputs ${\vf}({\vx})$, an emulator would give, for each input parameter
setting ${\vx}$, quantities such as an expectation and variance of the function: $\e{{\vf}({\vx})}$
and $\var{{\vf}({\vx})}$. In this way it represents the expected value of the
function at ${\vx}$ but also gives a measure of our uncertainty at this
point through $\var{\vf(\vx)}$. This uncertainty would be small at points
close to known model runs, and large at points far from known
runs. The expectation of the emulator will (in most cases, but not
always) interpolate the outputs of evaluated runs.

Emulators have many advantages, the most important being their speed:
in many cases an emulator will be many orders of magnitude faster
than the model it represents. Also, emulators are designed to cope with 
high numbers of input dimensions, far more than can be handled by more traditional 
methods such as Monte Carlo Markov Chains (eg.\ Heitmann et al.\ 2009).

Here we use emulation techniques to
identify the set of all inputs $\vx$ that will give rise to an
acceptable match (with respect to all relevant uncertainties) between
 the $b_J$ and K luminosity function outputs of GALFORM
and the corresponding observational data (Norberg et al.\ 2002 for the $b_J$
luminosity function and Cole et al.\ 2001 for the K band). 

The general strategy is as follows. Initially we design a suitable set
of 1000 runs of the GALFORM model chosen to be at parameter locations that
will cover the input space efficiently, and help the construction of an
acceptable emulator. Then we identify a subset of 11 outputs
(ie., the predicted values of the luminosity function at selected magnitudes) which
are representative of the $b_J$ and K 
luminosity functions, informative (about which regions of the input space
are unacceptable) and that are also straightforward to emulate. We
emulate each output by fitting a third order polynomial (defined over the
input space) to each of the 11 outputs, and then modeling the
residuals of this fit as a Gaussian process. Using the emulators and
assessments of all other relevant uncertainties (as described in 
\S\ref{ssec_BACM}) we then construct an
Implausibility Measure defined over the input space (see \S\ref{sec:ImpMeas}). Regions of the
input space that have a high Implausibility Measure are deemed highly
unlikely to give an acceptable match between the luminosity function
output of GALFORM and the observed data and are hence discarded from
further analysis. This defines a reduced region of input space that we
can explore further. We employ an iterative approach, and have reduced
the input space in four stages as is described below.

\subsection{Designing the First Set of Runs}

Determining a highly informative collection of points in input space to perform
evaluations of a Computer Model such as GALFORM is an important
task. The points must be space filling in addition to avoiding
repeated runs at similar values of one or more of the inputs (as
occurs regularly in a standard grid design). Maximin Latin Hypercube (Stein 1987)
designs fulfill both these properties and were used to generate
the initial set of runs. These designs are also approximately 
orthogonal: a desirable property when trying to fit polynomials to a
function, as is the case when building an emulator. To construct a
Latin Hypercube of $n$ points, the range of each of the inputs must be
divided into $n$ equal intervals; the points are then chosen randomly so that
no two points occupy the same interval for any of the inputs. Examples
of 2-dimensional 8 and 20 point Latin Hypercube designs are shown in
Fig.~\ref{fig_des}. 
A Maximin Latin Hypercube is constructed by creating a large number of Latin Hypercube 
designs, and then choosing the one that has the largest minimum distance between 
any pair of points within that design. {\rgb For each design, we generated 2000 hypercubes and
then selected the best one using the maximim criterion.} 
One thousand such runs of the GALFORM model were 
performed based on such a Maximin Latin Hypercube design, and these runs 
form the basis of Wave~1 of our analysis.

\subsection{Choosing Outputs}

\begin{figure}
\begin{center}
\includegraphics[scale=0.4]{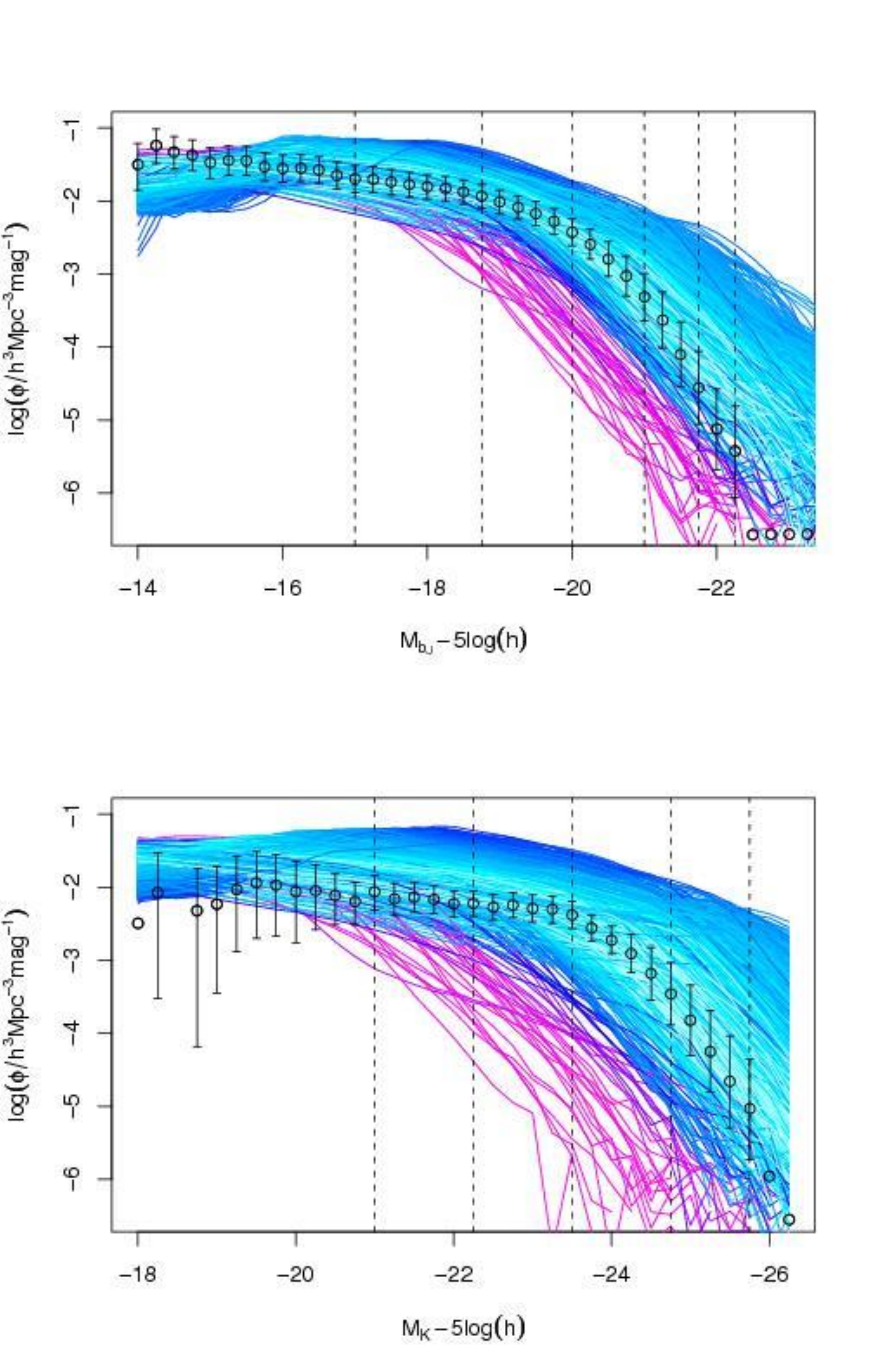}
\end{center}
\caption{The $b_J$ and K luminosity functions from the first 1000 runs of the model (Wave~1)
compared to observational data.  The data, from Norberg et al.\ (2002) and
Cole et al.\ (2001) respectively, are shown as black points with $2\sigma$ error bars which 
include all observational and model discrepancy uncertainties as described in \S\ref{ssec_moddis}.
Parameter values
were chosen using a Maximin Latin Hypercube design spanning the 8 most important
parameters (see Table~\ref{tab:bjAV}). The vertical dashed black lines show the 11 outputs chosen 
for emulation.  These provide a good characterisation of the luminosity function. Note
that below $\bJ=16$, some luminosity function calculations are affected by numerical resolution.
The colouring of lines indicates the quality of the match to the observed data,
{\rgb with blue colours indicating $I_M < 16$}.}\label{fig:11outs}
\end{figure}

Once the 1000 runs were completed, 11 outputs (ie., the values of the
luminosity function at selected magnitude points) were chosen for
emulation: 6 from the $b_J$ luminosity and 5 from the K luminosity
functions. These are shown as the vertical black dashed lines in
Fig.~\ref{fig:11outs}, along with the full outputs from the 1000 runs and
the observed data (the error bars contain all relevant uncertainties
as discussed below). These particular 11 magnitude outputs were chosen as they
represented the form of the luminosity functions well (and hence can
be used to reconstruct the luminosity function), they were easy to
emulate and, most importantly, were also sensitive to changes in
the input parameters implying that they are very informative with
regards to the input space. This last point implies that we can reliably cut
out regions of the input space using only these 11 outputs, and without
being forced into emulating the luminosity function in every
luminosity bin. {\rgb Analysis of the initial runs, showed that
adding additional outputs did not significantly improve the characterisation
of the luminosity function, but did risk weighting implausibility
measures too much towards the faint-end perfromance of a model.}
We also note that we did not attempt to emulate
the luminosity function for $b_J<16$~mag or $K<20$~mag because the
limited resolution of the Millennium simulation becomes important
for some parameter values in this region.

\subsection{Constructing the Emulator}

\subsubsection{A Simple Example}
\label{sec:simple_example}

Before we describe the construction of the emulator for the 
\GALFORM\  model, it is useful to briefly outline how the
method might be applied to a simple one-dimensional problem.

\begin{figure*}
\begin{center}
\begin{tabular}{cc}
\includegraphics[scale=0.4,angle=0]{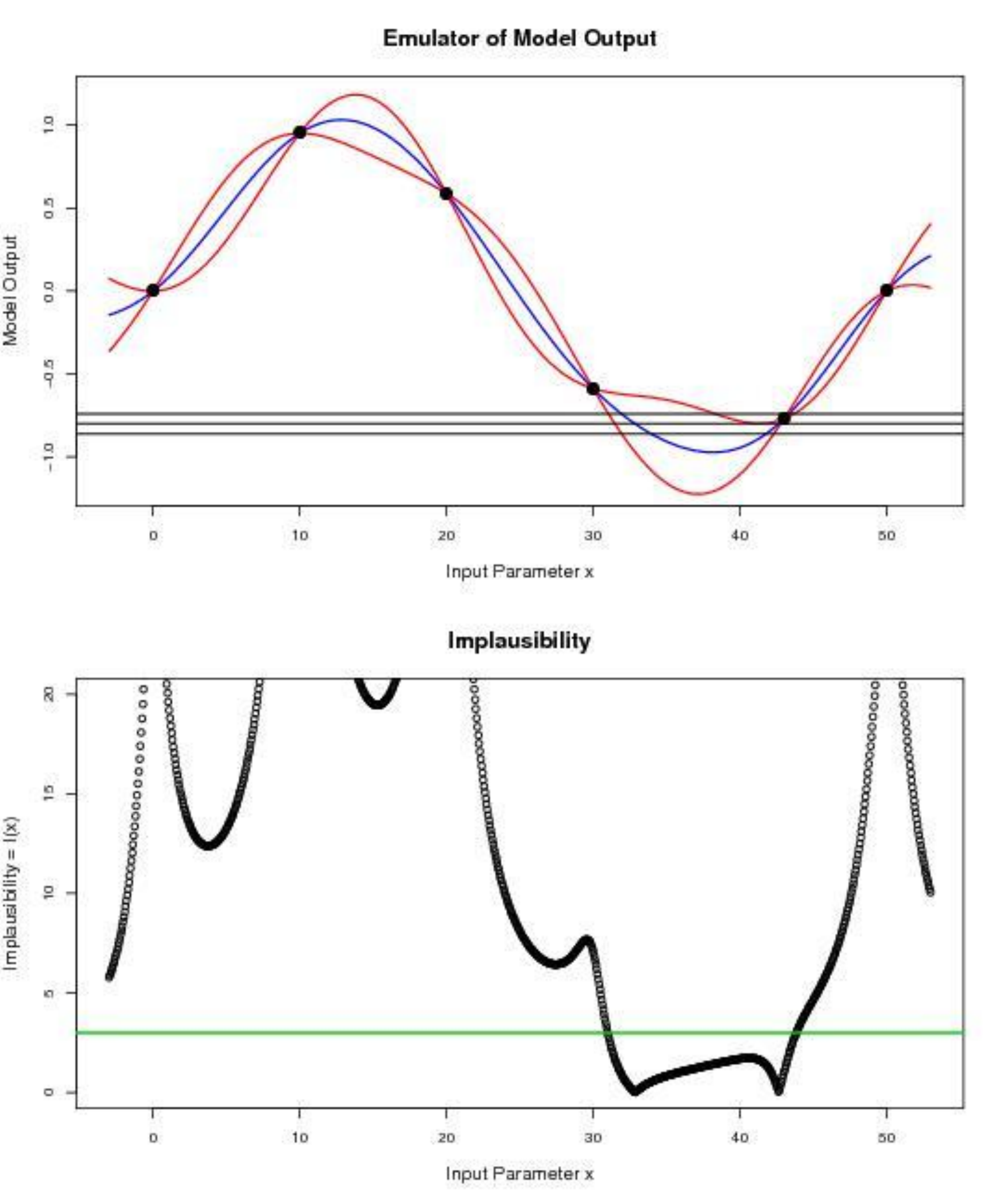}   &
\includegraphics[scale=0.4,angle=0]{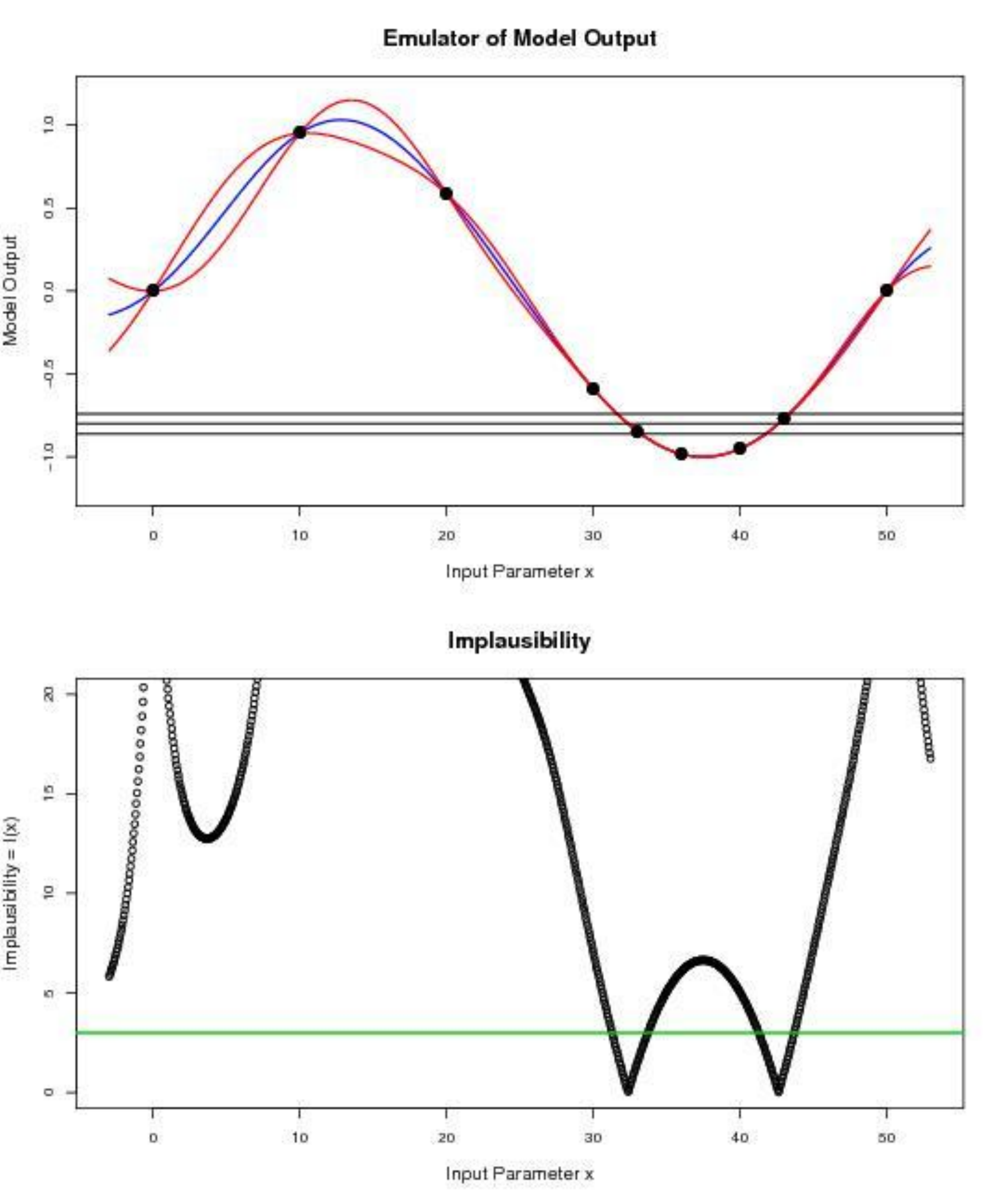} 
\end{tabular}
\end{center}
\caption{An example of the emulation of a 1-dimensional function (a
  sine wave). The top left panel shows the emulator $f(x)=u(x)$ after six
  evaluations of the function: the blue line is the $\e{f(x)}$, the
  two red lines define a credible interval of $\e{f(x)} \pm
  2\sqrt{\var{f(x)}}$ and the black dots are the 6 model outputs. Observational data $z$ is represented as the
  middle black line, with 2 sigma errors given by the top and bottom
  black line. The bottom left panel shows the implausibility function
  $I(x)$ in black, with the cutoff of 3 in green. Inputs along the
  $x$-axis are deemed implausible if $I(x)>3$. The top right and bottom
  right panel show the situation after three more runs have been
  performed in the non-implausible region. We re-emulate, and now, as
  the emulator is far more accurate, the implausibility naturally
  gives the two regions of the x-axis where the function matches the
  observed data (approximately around $x=33$ and $x=43$).
  }\label{fig:GP}
\end{figure*}

The first step is to construct an emulator of the simple
1-dimensional function shown in Fig.~\ref{fig:GP}. Imagine that the
function, $f(x)$, is a one parameter model for some measurable quantity.
In the left-hand panels,
the function (which is in fact a simple sine wave) has been evaluated at
$n=6$ input points denoted $x_i$. We use the function output at these 
points, $k_i = f(x_i)$, to construct an emulator based on
a random Gaussian process, $u(x)$, that is we say:
\begin{equation}\label{eq_fisu}
f(x) \;\;= \;\; u(x)
\end{equation}
where we assume the 
prior expectation and variance of the process 
$u(x)$ to be $\e{u(x)}=0$ and $\var{u(x)}=\sigma^2$, and that the prior covariance 
structure is defined to be of Gaussian form with correlation length $\theta$:
\begin{equation}
c(x,x') =  \cov{u(x)}{u(x')} = \sigma^2 \exp ( - (x - x')^2 / \theta^2 ).
\label{eq:cov}
\end{equation}
We can now update the emulator $u(x)$ using knowledge of the six evaluations of $f(x)$.
The updated emulator at a new point $x'$ now has expectation and variance given by:
\begin{equation}\label{eq_exp1}
{\rm E}[u(x')] = {\bf t}(x')^T A^{-1}{\bf k}
\end{equation}
\begin{equation}\label{eq_var1}
{\rm Var}[u(x')] = \sigma^2 - {\bf t}(x')^T A^{-1}{\bf t}(x')
\end{equation}
where ${\bf k}=(f(x_1),f(x_2),...f(x_n))^T$ (the vector of known function values),
${\bf t}(x') = ( c(x',x_1),c(x',x_2),...,c(x',x_n) )^T$ (the column vector of 
covariances between the new and known points) and $A$ is an $n\times n$ matrix 
with elements $A_{ij} = c(x_i,x_j)$ (the matrix of covariances between known points,
eg.\ Williams 2002). In a fully probabilistic analysis, equations (\ref{eq_exp1}) 
and (\ref{eq_var1}) would be derived from conditioning a Gaussian Process 
on the 6 known evaluations, and would give the mean and variance of the corresponding 
Normal distribution of $u(x')$ at the input point $x'$. However, here
we use a Bayes Linear analysis where equations (\ref{eq_exp1}) and (\ref{eq_var1}) 
are derived directly from the Bayes Linear update (described in 
\S\ref{ssec_BLA} and given by equations (\ref{BL1}) and (\ref{BL2})), and are 
considered as primitive quantities which are used directly to assess whether 
parts of the input space are acceptable.

The random process $u(x')$ quantifies the uncertainty in this fit: 
close to points at which the function has been
evaluated the uncertainty is small, while between points it is larger.
Making a suitable choice for $\sigma$ and $\theta$ is problem
specific. In this example, we set $\sigma$ to be 0.3
and chose $\theta$ to be $1/5$ of the range of the input variable $x$.

Now we suppose that we have some measurement for the quantity
being modelled. This is shown by the horizontal black lines (thin 
lines indicating the measurement uncertainty) in the figure. 
Using our emulator, we now try to identify the parameter values
at which an evaluation of the model might be compatible with
measured data.
While our emulator cannot guarantee that an evaluation
will successfully match the data, it identifies the regions 
at which a match is {\it implausible}.  This is quantified
through the use of an implausibility function, $I(x)$, which is discussed in more detail in \S \ref{sec:ImpMeas}. In this simple example $I(x)$ is defined as:
\begin{equation}
I^2(x)=|\e{f(x)} - z|^2/(\var{f(x)} + \var{\epsilon_{obs}}),
\end{equation}
where $z=-0.8$ is the observation (the middle horizontal black line), $\var{\epsilon_{obs}}$ is the 
variance of the observational errors which in this case were taken to be $0.03^2$, and $\e{f(x)}$ and $\var{f(x)}$ are given by equations~(\ref{eq_exp1}), 
(\ref{eq_var1}) and (\ref{eq_fisu}).

The value of $I(x)$ is shown in the lower left-hand panel of Fig.~\ref{fig:GP}: where $I(x)$
is large we reject the parameter values from further investigation. 
However, where $I(x)$ is below our cut-off value (in this case 3), we perform
a small number of additional evaluations of $f$. We then re-emulate using these 
additional evaluations, and this is shown in the 
right-hand panels. Now the uncertainty in the critical region is much reduced
and the only ``non-implausible'' regions are closely centered on
the regions where $f$ truely matches the measurement.  This iterative
approach is known as  ``history matching'' and is explained in more detail below,
but it can be see that the technique allows us to focus our evaluations
of $f$ on the regions where additional knowledge is critical.

Obviously, this one dimensional example is highly simplified, and
an emulator is not required to solve this problem.  In our \GALFORM\ 
application we are aiming to search a much higher dimensional 
parameter space and the ability to quantify our knowledge of the 
model is critical. Further, evaluation of the emulator is almost $10^7$
times faster than running an individual model, and thus this technique
can be combined with the Latin Hypercube scheme to explore
a seemingly vast parameter space.  
Note that for the simple example we have used an emulator consisting of purely a Gaussian Process, given by equation~(\ref{eq_fisu}). 
When we come to emulate the
GALFORM model we will use a more advanced emulator that contains polynomial regression terms, a Gaussian Process to model the residuals of 
the regression and a white noise term to model ineffective variables.
In the sections below, we describe
the construction of the GALFORM emulator in more detail. 

\subsubsection{The \GALFORM\ emulator}
\label{sec:emulator}

Here we describe the construction of the emulators for each of the 11
outputs identified in \S3.3. It must be stressed that although
we provide some detail, the construction of
the emulators and the subsequent analysis involve an extensive collection of 
statistical techniques, and we cannot hope to give a comprehensive
treatment here. A much more extensive description 
is given in Vernon et al.\ 2010. For examples of such techniques used in other
applications see Craig 1996, Goldstein \& Rougier 2006.

We view the GALFORM model as a function that maps the 16 inputs in
Table~\ref{tab:params} to the 11 identified outputs, and denote it by $\vf(\vx)$ where
$\vf$ is an 11 component vector of outputs and $\vx$ a 16 component vector
of inputs with $\vx = (x_1,x_2, x_3,...) = $ ($\vhotdisk$, $\alphareheat$,
$\alphacool$, ...).  (Note that we use the scaled variables directly,
without taking logs even when the variables cover a large range.) Great simplifications can be made in the
construction of emulators through the use of active variables. Often,
a subset of the inputs have strong effects on the outputs that are to
be emulated and we call these the Active Variables (see Table~\ref{tab:params}).  We
define $\vx_A$ to be a vector composed of active variables only and
model their effects on $\vf(\vx)$ in detail (Craig 1996). The remaining inputs
(the inactive variables) have only minor effects on the outputs so are
treated as contributing a noise process to the emulator. The form of
the emulator for component $i$ of $\vf(\vx)$ would then be:
\begin{equation} \label{eq:emul1}
    f_i(\vx) = \sum_j \beta_{ij}\, g_{ij}(\vx_A) + u_i(\vx_A) + w_i(\vx) .
\end{equation}
Here the $g_{ij}(\vx_A)$ are known functions chosen to be first, second or third order
polynomial terms in the active variables (for example, for output $i=1$ we might have 
terms of the form $g_{1j}(\vx_A) = x_1x_2^2$, $x_3^3$ or $x_1x_2x_3$, with different terms corresponding to 
different values of $j$); the $\beta_{ij} $ are coefficients
of the polynomial which will be fitted using regression
methods. $u_i(\vx_A)$ is a Gaussian Process\footnote{Technically we should refer to $u_i(\vx_A)$ as a weakly stationary random process in the Bayes Linear context, as we are making no assertions regarding the behaviour of higher order moments of $u_i(\vx_A)$.} which also
depends only on the active variables. The effects of the inactive
parameters are described by the $w_i(\vx)$ term, referred to as a
nugget, which is modelled as a random white noise process. The
regression term $\sum_j \beta_{ij}\, g_{ij}(\vx_A)$ on the right hand
side of equation (\ref{eq:emul1}) is included to capture the global
behaviour of the GALFORM function. The Gaussian process $u(\vx_A)$
represents localised deviations from this global behaviour, and a
simple specification is to suppose, for each $x$, that $u_i(x)$ has
zero mean, constant variance and $\cov{u_i(x)}{u_i(x')}$ which is a
function of $\|x-x'\|$, here chosen to be of Gaussian form
(see equation~(\ref{eq:cov}) below). 
As we perform evaluations of the model,
the expectation and variance of $f_i(\vx)$ at a given point is then
updated using the Bayes Linear analysis as 
described in \S3.5.4.

The above describes the general structure for all the emulators used
in this analysis. As we perform the reduction of input space
iteratively, and at each iteration (or ``wave'') we re-emulate
changing the specific form for the emulators (see the section on History
Matching below, \S3.8). At each wave the number of active variables increases
as the emulator becomes more accurate, and the random processes
$u_i(\vx_A)$ and $w_i(\vx)$ become less significant.

\subsubsection{The Wave 1 Emulator.}

Here we outline the construction of the Wave~1 emulators (full
details, {\rgb and extensive discussion,} 
are given in Vernon et al.\ 2010).  First, 8 of the 16
inputs were chosen as candidate active variables due to their clear
effect on the luminosity output in a set of initial test runs.  
In choosing the active variables, the aim is
to explain a large amount of the variance of $f_i(\vx)$ using as few
variables as possible.  {\rgb Initially, we ran GALFORM varying the 
primary parameters and holding the others fixed at their central 
values. The effect of the fixed variables is accounted for
through a contribution to the model discrepancy term (see \S3.6). 
(Note that in wave 4, as the region of parameter space becomes more 
restricted, we will allow the full set of parameters to vary.)} 
For each of the 11 outputs, the set of 8
parameters was initially reduced by backwards stepwise elimination,
starting with a model containing the 8 linear terms in $\vx$.
Then individual inputs were discarded in turn based
upon the significance of their main (i.e. linear) effect. Before an
input would be discarded, a full third order polynomial was fitted to
see the extent of variance explained with the current set of active
variables.  It was found that 5 active variables could explain
satisfactory amounts of the variance of $f_i(\vx)$ for each output $i$, based on the adjusted
$R^2$ of the polynomial fits ($R^2$ is the Coefficient of Determination of the fit and takes values between 0 and 1, with higher values implying that 
the fit explains more of the model output's behaviour). 
Note that the subset of 5 active variables is in general different for each output variable, as
shown in tables \ref{tab:bjAV} and \ref{tab:KAV}. Including more than 5 
variables would yield little extra benefit at this stage, while using fewer
than 5 leads to a significantly worse description of the main trends.

\begin{table}\label{tab:activeW1}
\begin{center}
\begin{tabular}{|l|c|c|c|c|c|c|}
\hline
  Output		  & $b_J^1$&$b_J^2$&$b_J^3$&$b_J^4$&$b_J^5$&$b_J^6$ \\
\hline
$\vhotdisk$       & x & x& x &  x& x&  x\\
$\alphareheat$      &  x& x& x&  x& x&  x\\
$\alphacool$    &   &  &  x&  x& x& x \\
$\vhotburst$     &  x&  &  x&  x& x&  x\\
$\epsilonStar$ &  x& x& x&  x&  &  \\
$\stabledisk$    &   &  &   &   &  x&  x\\
$\alphahot$      &  x& x&   &   &  &  \\
$\yield$             &   & x&   &   &  &  \\
\hline		
Adj.\ $R^2$  & 0.92 &  0.71 &  0.55 & 0.59  &  0.71 & 0.70  \\
\hline
\end{tabular}
\end{center}
\vspace{0.5cm}
\caption{The 8 candidate active variables for Wave~1 $b_J$ luminosity function
output (see Fig.~\ref{fig:11outs}), with the final 5 active variables for each bj output marked by an x. The adjusted $R^2$ value gives a measure of how effective the 3rd order polynomial is in capturing the global behaviour of the particular bj luminosity GALFORM output. Note the low adjusted $R^2$ of outputs 3 and 4: these were not used in the first wave analysis, but did feature in later waves.}\label{tab:bjAV}
\end{table}

\begin{table}
\begin{center}
\begin{tabular}{|l|c|c|c|c|c|c|}
\hline
  Output		  & $K^1$&$K^2$&$K^3$&$K^4$&$K^5$ \\
\hline
$\vhotdisk$       & x & x& x &  x& x\\
$\alphareheat$      &  x& x& x&  x& x\\
$\alphacool$    &    &   &   &  x& x\\
$\vhotburst$     &  x& x&  x&  x& x\\
$\epsilonStar$ &  x&   &    &    &   \\
$\stabledisk$    &   & x & x &  x&  x\\
$\alphahot$      & x& x & x &    &  \\
$\yield$             &   &    &   &     &  \\
\hline		
Adj.\ $R^2$  & 0.87 &  0.75 &  0.61 & 0.72  &  0.80  \\
\hline
\end{tabular}
\end{center}
\vspace{0.5cm}
\caption{The 8 candidate active variables for the Wave~1 K-band luminosity 
function output (see Fig.~\ref{fig:11outs}), with the final 5 active variables for each K-band output 
marked by an x. Note the low adjusted $R^2$ of output 3: this was not used in the first wave analysis, but did feature in later waves.}\label{tab:KAV}
\end{table}

Once the set of active variables has been determined, the full set of
regression terms (the $\beta_{ij}\, g_{ij}(\vx_A)$) can be chosen. This
was done by starting with the full 3rd order polynomial in the 5
active variables and using backwards stepwise elimination to remove
less significant terms from the model. Note that a large number
of model evaluations are required to enable the fitting of a third order
polynomial, although this greatly depends on the number of active
variables. 
Now that the
final regression terms have been chosen for each output $f_i(\vx)$,
estimates for the set of $\{ \beta_{ij} \} $ coefficients can be
obtained using Ordinary Least Squares, assuming uncorrelated
errors {\rgb (a reasonable assumption as we have a large number of runs 
sufficiently far apart in the input space that the residuals 
should not be strongly correlated).}
Note that it is important to check that the structure of the polynomials obtained from this process 
agree with physical intuition.

For the two contributions to the residual process $u_i(\vx_A)$ and
$w_i(\vx)$ we specify a correlation structure as follows. As the
$u_i(\vx_A)$ represent local deviations from the regression surface we
assume that there will be a large correlation between $u_i$ at
neighbouring values of the active inputs $\vx_A$, and specify the
following Gaussian covariance structure:
\begin{equation}
\cov{u_i(\vx_A)}{u_i(\vx_A')} \;\; = \;\; \sigma_{u_i}^2 \exp ( - || \vx_A - \vx_A' ||^2 / \theta_i^2 ),
\label{eq:cov}
\end{equation}
where $\sigma_{u_i}^2$ is the point variance at any given $\vx_A$,
$\theta_i$ is the correlation length parameter that controls the
strength of correlation between two separated points in the input
space (for points a distance $\theta$ apart, the correlation will be
exactly $\exp(-1)$), and $||\cdot||$ is the Euclidean norm.  As the
nugget process $w_i(\vx)$ represents all the remaining variation in
the inactive variables, it is often small and we treat it as
uncorrelated random noise with $\var{w_i(\vx)} =
\sigma_{w_i}^2 $. We consider the point variances of these two
processes to be proportions of the overall residual variance: $\sigma_i^2$ (which is obtained from the OLS regression fit), and write that
$\sigma_{u_i}^2 = (1-\delta_i)\sigma_i^2$ and $ \sigma_{w_i}^2=
\delta_i \sigma_i^2$ for some usually small $\delta_i$. 

Various techniques for
estimating the correlation length and nugget parameters $\theta_i$ and
$\delta_i$ from the data are available (eg.\ variograms, REML, maximum likelihood: Cressie 1991),
however an alternative is to specify them from previous
experience of computer models (Craig et al.\ 2001, Kennedy \& O'Hagan 2001)
which is the approach we adopt here. Specifically, in Wave~1 we
choose $\theta_i = 0.35$ and $\delta =0.2$, remembering that the inputs have all been scaled so that their range is $[-1,1]$. The choice for theta is motivated by 
the fact that we are fitting third order polynomials to the model output, and therefore the residuals from this fit, which are  
modelled by the stationary process, will behave like a fourth (or higher) order polynomial. This suggests a correlation length of 0.35 would be 
reasonable. The value for $\delta$ is assessed by examining the variance of the model output explained by the inactive variables by fitting various polynomials 
using only the inactive variables. It should be noted that the choices for $\theta$ and $\delta$ are more conservative than values obtained using alternative  estimation techniques, and that this was a deliberately cautious choice.  More details of
the motivation of these parameter choices, and description
of the diagnostic used 
to confirm the accuracy of this approach are given in Vernon et al.\ 2010.

The next step is to update the process $f_i(\vx)$ at a new point $\vx$,
with the information contained in the 1000 runs of the
model. We do this using the
Bayes Linear update formula discussed in the next section.

\subsubsection{Bayes Linear Approach.}\label{ssec_BLA}

For large scale problems involving computer models such as GALFORM, a
full Bayes analysis (involving probability distributions for all random quantities) 
is difficult for the following reasons. Firstly, it is
very difficult to give a meaningful full prior probability
specification over high-dimensional input spaces.  Secondly, the
computations for learning from both observed data and runs of the
model, and choosing informative runs, may be technically very
challenging.  Thirdly, in such computer model problems, often the
likelihood surface is extremely complicated, and therefore any full
Bayes calculation may be extremely non-robust.  However, the basic 
idea of capturing our expert prior judgments in
stochastic form and modifying them by appropriate rules given
observations, is conceptually appropriate.

The Bayes Linear approach is (relatively) simple in terms of belief
specification and analysis, as it is based only on the mean, variance
and covariance specification which, following de Finetti (1974), we take as
primitive. Therefore, a Bayes Linear approach proceeds by the specification and modification of 
mean and variance structures only.

We replace Bayes Theorem (which deals with full probability distributions)
by the Bayes Linear adjustment which is the appropriate updating rule
for expectations and variances.  The Bayes Linear adjustment of the
mean and the variance of a random quantity $B$ given data $D$ is:
\begin{eqnarray}
	{\rm E}_D[B] &=& \e{B}+\cov{B}{D}\var{D}^{-1}(D-\e{D}), \\ &&\label{BL1} \\  
	{\rm Var}_D[B] &=& \var{B}-\cov{B}{D}\var{D}^{-1}\cov{D}{B}.	\\ &&\label{BL2}\\
\end{eqnarray}
${\rm E}_D[B]$, ${\rm Var}_D[B]$ are the  expectation and variance for $B$
{\it adjusted by knowledge of} $D$\footnote{ ${\rm E}_D[B]$ and ${\rm Var}_D[B]$ are the corresponding Bayes Linear quantities to 
$\e{B|D}$ and $\var{B|D}$, the conditional expectation and variance of $B$ given data $D$ that would be extracted from a fully Bayesian analysis.}.
In equations~(\ref{BL1}) and (\ref{BL2}), $B$ and $D$ can represent scalars or vectors of uncertain quantities. In the later case (\ref{BL1}) and (\ref{BL2}) become matrix equations
where if $B$ is a vector of length $n_B$ and $D$ is a vector of length $n_D$, $\cov{B}{D}$ is a matrix of dimension $n_B \times n_D$ and $\var{D}$ a matrix of dimension $n_D \times n_D$.

The Bayes linear adjustment may be viewed as 
an approximation to a full Bayes analysis, or more fundamentally as 
the ``appropriate'' analysis given a partial prior specification based
on expectation. For more details see Goldstein \& Wooff (2007).

\subsubsection{Updating the Emulator.}

Equations~(\ref{BL1}) and (\ref{BL2}) give the rule for updating the emulator with knowledge of 
the 1000 model evaluations, where the random quantities $B$ and $D$ 
will represent an unknown output and the collection of 1000 known 
outputs respectively. We proceed with the update as follows. As we have a 
relatively large number of runs, we first assume that
the regression coefficients $\beta_{ij}$ in the emulator equation 
(\ref{eq:emul1}) are known and hence have zero variance. As $u_i(\vx_A)$ and $w_i(\vx)$ both 
have zero expectation, equation (\ref{eq:emul1}) gives the expectation 
of model output $i$ at input $\vx$ to be:
\begin{equation}\label{eq:expf}
\e{f_i(\vx)} = \sum_j \beta_{ij}\, g_{ij}(\vx_A)
\end{equation}
and the variance of $f_i(\vx)$ to be:
\begin{equation}\label{eq:Vf}
 \var{f_i(\vx)} = \var{u_i(\vx_A)} + \var{w_i(\vx)} = \sigma_{u_i}^2 + \sigma_{w_i}^2 = \sigma_i^2.
\end{equation}
As the $u_i(\vx_A)$ and the $w_i(\vx)$ terms are uncorrelated, the covariance between output $i$ at two different 
inputs $\vx'$ and $\vx$ can now be written (using equations~(\ref{eq:emul1}) and (\ref{eq:cov})):
\begin{eqnarray}
c(\vx',\vx) \!\! &=& \!\!\cov{f_i(\vx')}{f_i(\vx)}   \nonumber \\ 
\!\! &=& \!\! \cov{u_i(\vx'_A)}{ u_i(\vx_A)} +\cov{w_i(\vx')}{ w_i(\vx)} \nonumber  \\
\!\! &=& \!\! \sigma_{u_i}^2 \exp ( - || \vx'_A - \vx_A ||^2 / \theta_i^2 ) + \sigma^2_{w_i}\delta_{\vx'\vx}   \label{eq:cov1}
\end{eqnarray}
where $\delta_{\vx'\vx}$ is a Kronecker delta, equal to 1 when $\vx'=\vx$ and zero otherwise. 
The second term in equation (\ref{eq:cov1}) comes from the nugget $w_i(\vx)$ which gives a zero contribution 
except when $\vx'=\vx$.

We can now define the following quantities corresponding to the $n=1000$ model 
evaluations. We write the locations of the $n$ runs in input space as $\vx_j$ with 
$j=1,..,n$ where each $\vx_j$ represents the vector
of inputs for the $j$th run. Similarly $\vx_{A,j}$ is defined to be the vector of Active 
Variable inputs for the $j$th run. We define 
$D_i = (f_i(\vx_1),f_i(\vx_2),..., f_i(\vx_n))^T$, that is the column vector of the n evaluation 
outputs for output $i$, the prior expectation of which ($\e{D_i}$) 
can be found using equation~(\ref{eq:expf}). 

Replacing the random quantity $B$ in the Bayes Linear update equation~(\ref{BL1}) with the unknown output $f_i(\vx)$ at input $\vx$ gives
the adjusted expectation ${\rm E}_{D_i}[f_i(\vx)]$ to be: 
\begin{equation}
{\rm E}_{D_i}[f_i(\vx)] = \e{f_i(\vx)}+\cov{f_i(\vx)}{D_i}\var{D_i}^{-1}(D_i-\e{D_i}), \nonumber
\end{equation}
which becomes (using equation~(\ref{eq:expf})),
\begin{equation}\label{eq:adjE}
{\rm E}_{D_i}[f_i(\vx)] =  \sum_j \beta_{ij}\, g_{ij}(\vx_A) + {\bf t}(\vx)^T A^{-1} (D_i-\e{D_i}), 
\end{equation}
where now ${\bf t}(\vx)  = ( c(\vx,\vx_1),c(\vx,\vx_2),...,c(\vx,\vx_n) )^T = \cov{f_i(\vx)}{D_i}$ is the column vector of covariances between the
new and known points, and $A$ is the matrix of covariances between known points: an $n\times n$ matrix with elements $A_{jk} = c(\vx_j,\vx_k)$.
The Adjusted Variance ${\rm Var}_{D_i}[B]$ can similarly be found from equation~(\ref{BL2}) and (\ref{eq:Vf}) giving:
\begin{eqnarray}\label{eq:adjV}
&&{\rm Var}_{D_i}[f_i(\vx)] =  \nonumber \\
&&= \var{f_i(\vx)}-\cov{f_i(\vx)}{D_i}\var{D_i}^{-1}\cov{D_i}{f_i(\vx)}, \nonumber \\
&&= \sigma_i^2- {\bf t}(\vx)^T A^{-1}{\bf t}(\vx) .
\end{eqnarray}

The Adjusted Expectations and Variances, ${\rm E}_{D_i}[f_i(x)]$ and ${\rm Var}_{D_i}[B]$, 
given by equations~(\ref{eq:adjE}) and (\ref{eq:adjV}) form 
the basic ingredients in the construction of the Implausibility measure used to reduce 
the input space to a much smaller ``non-implausible'' volume. Note that if we had 
chosen a simple emulator of the form $f_i(\vx) = u_i(\vx)$, such as was used in the 
1D example in \S\ref{sec:simple_example}, then equations~(\ref{eq:adjE}) and (\ref{eq:adjV}) would reproduce 
exactly equations~(\ref{eq_exp1}) and (\ref{eq_var1}).

\subsection{Linking the Model to the System: Structural Uncertainty and Model Discrepancy }\label{ssec_moddis}

In order to declare regions of the input space as ``implausible'', and to
then exclude them from the analysis, we need to formally link the
GALFORM model $\vf(\vx)$ to the observed luminosity function data which we
represent as the 11 component vector $\vz$. We do this by linking both $\vf(\vx)$ and $\vz$ to the
actual system (in this case the real Universe) represented by $\vy$, taking into consideration the
Structural Uncertainty. In a rigorous Bayesian
approach, this step is key in order to justify any further
uncertainty statements; it is, however, a relatively unfamiliar
process to many scientists.
We employ a description that is widely used in
computer modeling studies (eg., Craig 1996; Kennedy \& O'Hagan 2001; Goldstein \& Rougier 2009). This 
involves the notion that when we evaluate GALFORM at the actual system properties, $\vx^*$ say, 
then we aim to reproduce the actual system behaviour $\vy$. This does not mean that we would 
expect perfect agreement between $\vf (\vx^*)$ and $\vy$. Although GALFORM is a highly sophisticated 
simulator, it still offers a necessarily simplifed account of the evolution of galaxies, and involves various 
numerical approximations. The simplest way to view 
the difference between $\vf ^*= \vf (\vx^* )$ and $\vy$ is to express this as: 
\[ \label{md}
\vy = \vf^* + \vepsilon_{md},
\]
where we consider the 11-vector $\vepsilon_{md}$ as a random variable uncorrelated with $\vf^*$. 
The ``Model Discrepancy'' term $\vepsilon_{md}$ represents the Structural Uncertainty. It comes 
from our judgments regarding the accuracy of the model and determines how close a fit
between model output, $\vf^*$, and an observation of $\vy$ we require for an
acceptable level of consistency between theory and observation.

As GALFORM is an approximation
to the physical processes that occur during galaxy formation in the
real Universe, we must acknowledge and attempt to quantify the level
of this approximation, represented by $\vepsilon_{md}$, in order for further analysis to be
meaningful. While this is a difficult task, ignoring the model
discrepancy will lead to all future statements being conditional on the current version of 
GALFORM being a perfect model of the Universe. Since we know that
this is not the case, it is essential that we build in some degree
of fuzziness into the comparison between the model and data. Failure
to do so may result in us prematurely rejecting regions of parameter 
space in which a solution of interest resides. As the level of approximation
in the model is reduced (by considering an improved version of the model say), 
$\vepsilon_{md}$ will become smaller. 
In principle, this process of model improvement can be built into our
statistical emulation so that we use our knowledge
of previous versions of the \GALFORM\ code both to speed up parameter
exploration in newer versions, and to obtain more realistic representations 
of the Structural Uncertainty (Goldstein \& Rougier 2009); however, we have not explored this
possibility here.

As we are employing a Bayes Linear approach we only need to specify expectations and variances for $\vepsilon_{md}$: 
we give a summary of this process here, the full details of which can be found in (Vernon et al.\ 2010). 

We decompose $\vepsilon_{md}$ into three uncorrelated contributions, each of which are 11-vectors that are assumed to have zero expectation:
\[ \label{mdcomp}
\vepsilon_{md} \;=\; \vPhi_{IA}  \; + \; \vPhi_{DM} \; + \; \vPhi_{E}.  
\]
Here $ \vPhi_{IA} $ represents the discrepancy due to the eight inactive parameters that we did not model in detail 
in the initial waves of the analysis (that is, the parameters that do not feature in Table~\ref{tab:activeW1}). 
We assessed $\var{\vPhi_{IA}}$ from a small set of runs over 
the 8-dimensional inactive parameter space and found for $i=1,..,11$ that  $0.015^2<  \var{\Phi_{IA,i}}<0.32^2$. In later waves, we performed runs
across all 16 inputs and hence the $ \vPhi_{IA} $ term was then set to zero as this model discrepancy was now absorbed directly into the emulator.

$\vPhi_{DM}$ is the discrepancy due to the finite number (40) of sub-volumes
used for the model runs: $\var{\vPhi_{DM}}$ was found by analyzing the sample variance of the 1000 Wave~1 runs across sub-volumes, and 
it was estimated for $i = 1,...,11$ that $0.014^2< \var{\Phi_{DM,i}}<0.022^2$.
{\rgb For a sub-set of 100 runs, we confirmed this estimate by drawing a random set of 40 sub-volumes.}

$\vPhi_{E}$ summarises the structural deficiencies of the full GALFORM model itself, and is the component derived from subjective judgments regarding 
model accuracy. 
{\rgb We proceeded by generating a random test set of luminosity functions by perturbing
the observational data and smoothly interpolating. These were then compared
to the observational data using an interactive tool which asked the expert user to judge
whether to reject the corresponding region of parameter space on the basis of the comparison and 
previous experience of improvements to the GALFORM code and changes in cosmological models.
Summarising these tests, we concluded that a credible interval around the observational data, within 
which runs would be deemed acceptable would be approximately 
a factor of 2 wide in terms of galaxy counts, and hence $\pm{1\over 2}\log_{10}(2)$ on the log scale used 
throughout this paper (see for example Fig.~\ref{fig:11outs}). 
Relating this to a $\pm 2 \sigma$ interval (a conservative choice) this leads to 
the variance of each of the 11 components of $\Phi_{E}$ being assigned values: $\var{\Phi_{E,i}} = (\log_{10}(2)/4)^2=0.0753^2$ for $i=1,..,11$. Reassuringly, this results
in the Bow06 model being close to the boundary of acceptable solutions. Examples of the 
range of fits that are deemed acceptable are illustrated in Fig.~\ref{fig:LFGood113}.} 
The expectation was again set to $\e{\Phi_{E,i}} =0$, as it was thought that there were no significant asymmetries concerning this component of the Model Discrepancy.
It is important to realise that while this assessment for $\vPhi_{E}$ is necessarily subjective, it was also chosen to be deliberately conservative. Once the volume of acceptable inputs has been identified corresponding to all uncertainties discussed in this section, it is then possible to explore the effect of reducing the size of $\var{\vPhi_{E}}$.

Finally, we must make allowance for the uncertainty in observational 
measurements. Since, we
cannot observe the system $\vy$ (i.e. the actual Universe) without measurement
error, we link it to the observations $\vz$ by:
\[ 
\vz = \vy + \vepsilon_{obs}, 
\]
where $\vepsilon_{obs}$ is again a random quantity that represents the
observational errors. It has expectation zero, and variance composed
of contributions from the luminosity calibration uncertainty, the
normalisation uncertainty, $k+e$ errors and Poisson errors (see Norberg et al.\ 2002
for a discussion of how these terms are estimated).  Fig.~\ref{fig:11outs} shows the 2-sigma 
error bars formed from the combination of all components of $\var{\vepsilon_{md}}$ and $\var{ \vepsilon_{obs}}$. It should be noted 
that in most cases the Model Discrepancy terms dominate over the observational errors.
(Fig.~\ref{fig:testlf} shows the same error bars minus the $\vPhi_{IA}$ component which 
is no longer relevant, as by Wave~4 we have modelled the 
effect of the remaining inactive variables within the emulator directly.)

With this structure linking $\vf(\vx)$, $\vy$ and $\vz$ in place we can now proceed to learn about acceptable values of $\vx$.

\subsection{Implausibility Measures}
\label{sec:ImpMeas}

We want to learn about which values of the input parameters $x$ are
likely to give an acceptable match between model output and
observational data. 
We do this through use of an Implausibility
Measure $I(\vx)$ defined over the input space. The Implausibility
Measure describes the magnitude of the difference between the expected
value of the GALFORM outputs and the observational data, standardized
with respect to all relevant uncertainties. The basic idea is that for
a particular value of $\vx$, if $I(\vx)$ is large then we can discard this
value of $\vx$ as it is highly unlikely to yield a good match between
model output and the observational data.

Using the emulator, the model discrepancy and the measurement errors
we define the Univariate Implausibility Measure, at any input
parameter point $\vx$, for each component $i$ of the computer model
$\vf(\vx)$ as:
\begin{equation} \label{uimp}
I^2_{(i)}(\vx)=|\ed{D_i}{f_i(\vx)} - z_i|^2/\vard{D_i}{\ed{D_i}{f_i(\vx)}-z_i}
\end{equation}
where $ \ed{D_i}{f_i(\vx)} $ and $ \vard{D_i}{f_i(\vx)} $ are the emulator expectation
and variance adjusted by $D_i$ and $z_i$ is the observed data for component $i$. Introducing the
model discrepancy and observational error terms, this
can be re-written as:
\begin{equation}\label{eq:imp2nd}
I^2_{(i)}(\vx)=\frac{|\ed{D_i}{f_i(\vx)} - z_i|^2}{(\vard{D_i}{f_i(\vx)} + \var{\epsilon_{md, i}} + \var{\epsilon_{obs, i}} )}
\end{equation}
where $\var{\epsilon_{md,i} }$
and $\var{\epsilon_{obs,i}}$ are the (univariate) Model Discrepancy variance and
Observational Error variance.

When $I_{(i)}(\vx)$ is large this implies that, even given all the
uncertainties present in the problem, we would be unlikely to obtain a
good match between model output and observed data were we to run the
model at input $\vx$. This means that we can cut down the input space by
imposing suitable cutoffs on the implausibility function (a process
referred to as History Matching). Regarding the size of $I_{(i)}(\vx)$,
if we assume that for fixed $\vx$ the appropriate distribution of
$(f_i(\vx^*)-z_i)$ is both unimodal and continuous, then we can use the $3\sigma$ rule which
implies that if $\vx=\vx^*$, then $I_{(i)}(\vx) < 3$ with a probability of
approximately 0.95. This is a powerful result that applies to any distribution
that is unimodal and continuous, even if it is asymmetric.
It suggests that values higher than 3 would imply that the point $\vx$ should be discarded. This is
still a very conservative bound: we would expect the distribution of $(f_i(\vx^*)-z_i)$ to be somewhat 
better behaved and hence choose slightly tighter bounds, as discussed in \S~\ref{ssec_HM}.

It should be noted that since the implausibility relies purely on
means and variances (and therefore can be evaluated using Bayes Linear
methodology), it is both tractable to calculate and simple to use to reduce the input space.

One way to combine these univariate implausibilities is by maximizing over outputs:
\[ \label{eq:maximp}
I_M(\vx)=\max_i I_{(i)}(\vx)
\]
We can similarly define $I_{2M}(\vx)$ and $I_{3M}(\vx)$ to be the second and third highest of the 11 univariate implausibility measures 
at the point $\vx$. {\rgb These are clearly more conservative measures since a 
model will not be deemed implausible on the basis of a single bin.}

If we construct both a multivariate emulator and multivariate model discrepancy (as is described in detail in Vernon et al.\ 2010), then we can define 
the corresponding multivariate Implausibility measure:
\[ 
I^2_{MV}(\vx) = (\ed{D}{\vf(\vx)} - \vz)^T \vard{D}{\ed{D}{\vf(\vx)}-\vz}^{-1} (\ed{D}{\vf(\vx)} - \vz), 
\]  
which becomes:
\begin{eqnarray}
&& \!\!\!\!\!\!\!\!\!\!\!\! I^2_{MV}(\vx) = (\ed{D}{\vf(\vx)} - \vz)^T  \times    \\
&&  \!\!\!\!\!\!\!\!\!\!\!\! (\vard{D}{\vf(\vx)} + \var{\vepsilon_{md}} + \var{\vepsilon_{obs}})^{-1}  (\ed{D}{\vf(\vx)} - \vz) .  
\end{eqnarray}
where $\vf(\vx)$ is the full 11-vector model output and $\vard{D}{\vf(\vx)}$, $\var{\vepsilon_{md}}$ and $\var{\vepsilon_{obs}}$ are all 
$11 \times 11$ covariance matrices.
Again, large values of $I_{MV}(\vx)$ imply that we would be unlikely to
obtain a good match between model output and observed data were we
run the model at input $\vx$. Choosing a cutoff for $I_{MV}(\vx)$ is more
complicated. As a simple heuristic, we might choose to compare $I_{MV}(\vx)$
with the upper critical value of a $\chi^2$ distribution with degrees
of freedom equal to the number of outputs. For further discussion of
implausibility measures, see Vernon et al.\ 2010.

\subsection{History Matching via Implausibility}\label{ssec_HM}

History Matching is the process of identifying the set $\mathcal{X}^*$
of all possible values of $\vx^*$, that is the set of points that would
give acceptable matches between model output and observational
data. Identifying $\mathcal{X}^*$ is a difficult task as often
$\mathcal{X}^*$ represents a complicated object in a high-dimensional
space. $\mathcal{X}^*$ could also comprise disconnected volumes,
which could even possess non-trivial topology. In many applications
$\mathcal{X}^*$ occupies an extremely small fraction of the original
input space, with large volumes of input space leading to very poor
matches to the observed data. 

We employ an iterative technique where the
Implausibility Measures are used to perform the History Matching
process. The basic strategy is based around discarding values of $\vx$
that are highly unlikely to yield acceptable matches between model
output and observational data. This is done by applying a cutoff on
the Implausibility Measures defined in \S~\ref{sec:ImpMeas}. As
the Implausibility Measures are constructed using the emulator, they
are fast to evaluate and therefore we can efficiently identify values
of $\vx$ that will be discarded, for example, in Wave~1 we discard all values of 
$\vx$ that do not satisfy both:
\begin{equation}\label{eq:cut1}
I_{2M}(\vx) \;\; < \;\; I_{2cut}  \;\; {\rm and} \;\; I_{3M}(\vx) \;\; < \;\; I_{3cut}
\end{equation} 
where $I_{2M}(\vx)$ and $I_{3M}(\vx)$ are the second and third highest univariate implausibility measures defined
in \S~\ref{sec:ImpMeas} and $I_{2cut}$ and $I_{3cut}$ are the corresponding
implausibility cutoffs. Table~\ref{tab:waves} shows all the implausibility measures used in each of the waves
along with the corresponding cutoffs. Note that in early waves we make the conservative choice of using only 
$I_{2M}(\vx)$ and $I_{3M}(\vx)$ (and not $I_M(\vx)$), so that the cutoff we impose is not sensitive to the 
possible failings of an individual emulator point on the luminosity function. 
This allows slightly tighter cuts to be chosen for $I_{2cut}$ and $I_{3cut}$ as
is shown in table~\ref{tab:waves}.

Equation~(\ref{eq:cut1}) defines a volume of input space that we refer to
as non-implausible and denote $\mx_1$. This non-implausible volume
should hopefully contain the set $\mx^*$, that is $\mx^* \subset
\mx_1$. In the first wave of the analysis which we are describing
here, $\mx_1$ will be substantially larger than $\mx^*$. This is
because it will contain many values of $\vx$ that only satisfy the
implausibility cutoff given by equation~(\ref{eq:cut1}) because of a
substantial emulator variance $\var{\vf(\vx)}$.  If the emulator had a high degree of
accuracy over the whole of the input space so that $\var{\vf(\vx)}$ was
small compared to the Model Discrepancy and the Observational Error
variances, then the non-implausible volume defined by $\mx_1$ would be
comparable to $\mx^*$ and the History Match would be complete. However, to construct such an accurate emulator for any
realistic computer model (and especially for GALFORM) would require an
infeasible number of runs of the model. Even if such a large number of
runs was possible it would be an extremely inefficient method: we do
not need the emulator to be highly accurate in regions of the input
space where the outputs of the model are clearly very different from
the observed data.

This is the main motivation for our iterative approach. In each wave
we design a set of runs over the current non-implausible volume denoted $\mx_i$,
emulate using these runs, calculate the implausibility measure and
impose a cutoff to define a new (smaller) non-implausible volume denoted $\mx_{i+1}$
which should satisfy $\mx^* \subset  \mx_{i+1}  \subset  \mx_i $. 
As
we progress through each iteration the emulator at each wave will
become more and more accurate, but will only be defined over the
previous non-implausible volume defined by the previous wave's
implausibility.

As we proceed through waves of emulation the volume being emulated
decreases, and the GALFORM function should become smoother over the
restricted range of interest. As a result, it becomes easier to capture
more of the behaviour using the regression terms in the emulator (ie.,
by fitting a cubic polynomial to the GALFORM output). Furthermore, 
the density of runs that inform us about the models
behaviour increases and the Gaussian Process part of the
emulator becomes more accurate. As the output of the runs have been restricted and 
the effects of certain dominant variables limited, it becomes easier to identify additional active variables which are 
then used in both the regression and the Gaussian Process terms, further
increasing the accuracy of the emulation.

This iterative process is continued until the emulator variance is
smaller than the model discrepancy variance and observational error
variance. We have completed 4 iterations or Waves in this analysis,
and the subsequent results in this paper are from Wave~4.

\subsection{Projection Pursuit}

The end result of the emulator analysis is to identify a region of parameter space in which models 
produce an acceptable fit to the $b_J$ and $K$-band luminosity functions. However, comprehending 
the resulting space is rather challenging. As we will show, while the parameter space of acceptable 
model occupies only a small fraction of the overall parameter space,
acceptable fits can be found over a wide range of input parameters. This situation arises
because the acceptable space takes the form of a thin curved hyper-surface.

The aim of projection pursuit is to select a suitable co-ordinate system that 
allows the geometry  of the acceptable region to be better understood. We achieve
this using principal component analysis (PCA, eg.\ Jolliffe 2002, Zito et al 2009). 
However, in contrast to many applications
of PCA, we are primarily concerned with the components with smallest variance.
These components define an optimal set of projections for displaying the data,
and the relation between the PCA vectors and the input parameters. The latter
connection has the potential to inform us about the physics of galaxy formation.

\subsection{Exploring Constraints from Additional Datasets}

We have adopted a strategy in which
the primary calibration of our model comes from the local $b_J$ and $K$ band 
luminosity functions.
Nevertheless, we wish briefly to explore whether adding additional data
sets would impose further constraints on the range of acceptable model parameters.
In this paper, we do not aim to make an exhaustive exploration of the possible
data sets and limit our attention to just a small fraction of the possible local 
data.  We use a simple $\chi^2$ statistic to assess the relative performance
of models in these additional tests and ask about the region of parameter space 
which matches the additional data at a similar level of performance to Bow06
(as well as adequately matching the observed luminosity functions). As we show in 
\S5.2, the model experiences contradictory pressures from 
the observed disk sizes and the normalisation of the Tully-Fisher relation,
possibly indicating that a revised treatment of angular momentum is
required in the Bow06 version of the GALFORM code. This clearly illustrates the 
need to carefully define the model discrepancy terms for these additional
data sets before they are used to exclude regions of parameter space.
We present further exploration of additional data sets in a future paper 
(Benson \& Bower, 2009).

\section{Results}

\begin{table}
\begin{tabular}{|c|c|c|cccc|c|}
\hline
Wave & Runs & \#Act. &  $I_{\rm cut}$ & $I_{2\rm cut}$ & $I_{3\rm cut}$ & $I_{MV\rm cut}$ &  \% Space\\
\hline
 1 & 1000 & 5 & - & 2.7 & 2.3 & -&   14.9 \% \\
  2 & 1400 & 8 & - & 2.7 & 2.3 & - & 5.9 \% \\
 3 & 1600 & 8 & - & 2.7 & 2.3 & 26.75 &    1.6 \% \\
 4 & 2000 & 10 & 3.2 & 2.7 & 2.3 & 26.75 &    0.26 \% \\
 5 & 2000 & - & 2.5 & - & - & - & (0.014\%) \\
\hline  
\end{tabular}
\caption{The fraction of parameter space considered acceptable
in each wave of emulation. Column 1: the wave; Column 2, the number of
model runs used to construct the emulator; Column 3, the number of active
variables; Column 4-7, the implausibility cut-off threshold;
Column 8, the fraction of the parameter space estimated to be 
acceptable. Note that in Wave~5 we do not construct an emulator, but we impose a 
cutoff of $I_{M} < 2.5$ on the Wave~5 runs to generate the 
113 acceptable runs used in \S\ref{sec_projpur}.
}
\label{tab:waves}
\end{table}

The results described below were obtained with 4 waves of emulation. The
implausibility cut-off threshold for each wave is shown in Table~\ref{tab:waves}, 
together with the fraction of the parameter space considered acceptable after the 
emulator has been constructed.  The table also gives the number of runs
used during each wave.

After 4 waves of emulation, the uncertainties in the emulator are small and the
implausibility of each run is becoming dominated by the intrinsic model discrepancy $\vepsilon_{md}$. 
Uncertainties in the observational measurement of the luminosity function 
make almost negligible contribution, and the dominant contribution to the
model uncertainty comes from the 
model discrepancy term, $\vPhi_E$ (see discussion in \S\ref{ssec_moddis}).

At this point, the emulator suggests that only 0.26\% of the initial parameter 
volume is ``not implausible''. While this volume is small, 
we will show below that an ``acceptable'' fit can be obtained with a wide 
range of values for some parameters. Note that we are being careful in our use of language 
here.  We do not know for certain what the outcome of running the model
will be at a particular set of parameter values, except close to values at
which we have already performed a model run. We do, however, have a prediction
for the expectation and variance. Fig.~\ref{fig:testcompare}
shows a comparison of the expectation of the emulator and its uncertainty
with the results of actual model runs and we discuss this comparison further in
\S\ref{sec:testcompare}. However,
as we should expect, only a fraction of the runs within the ``not implausible''
region actually result in sufficiently good fits to the luminosity function
to be considered ``acceptable''. There are two factors involved here. Firstly, we are tightening
the required implausibility from 3.2 to 2.5. In 16 dimensions, this results in a large reduction in 
the surface of the interesting region. Secondly, for many runs the expectation 
of the emulator is that the model implausibility lies above 2.5, but the residual emulator variance cannot
rule out the region as unacceptable without direct evaluation.
{\rgb This uncertainty arises from emulator variance, and not from observational
error or model discrepancy terms.}

\subsection{Emulating the Luminosity function}

\begin{figure*}
\includegraphics[scale=0.6]{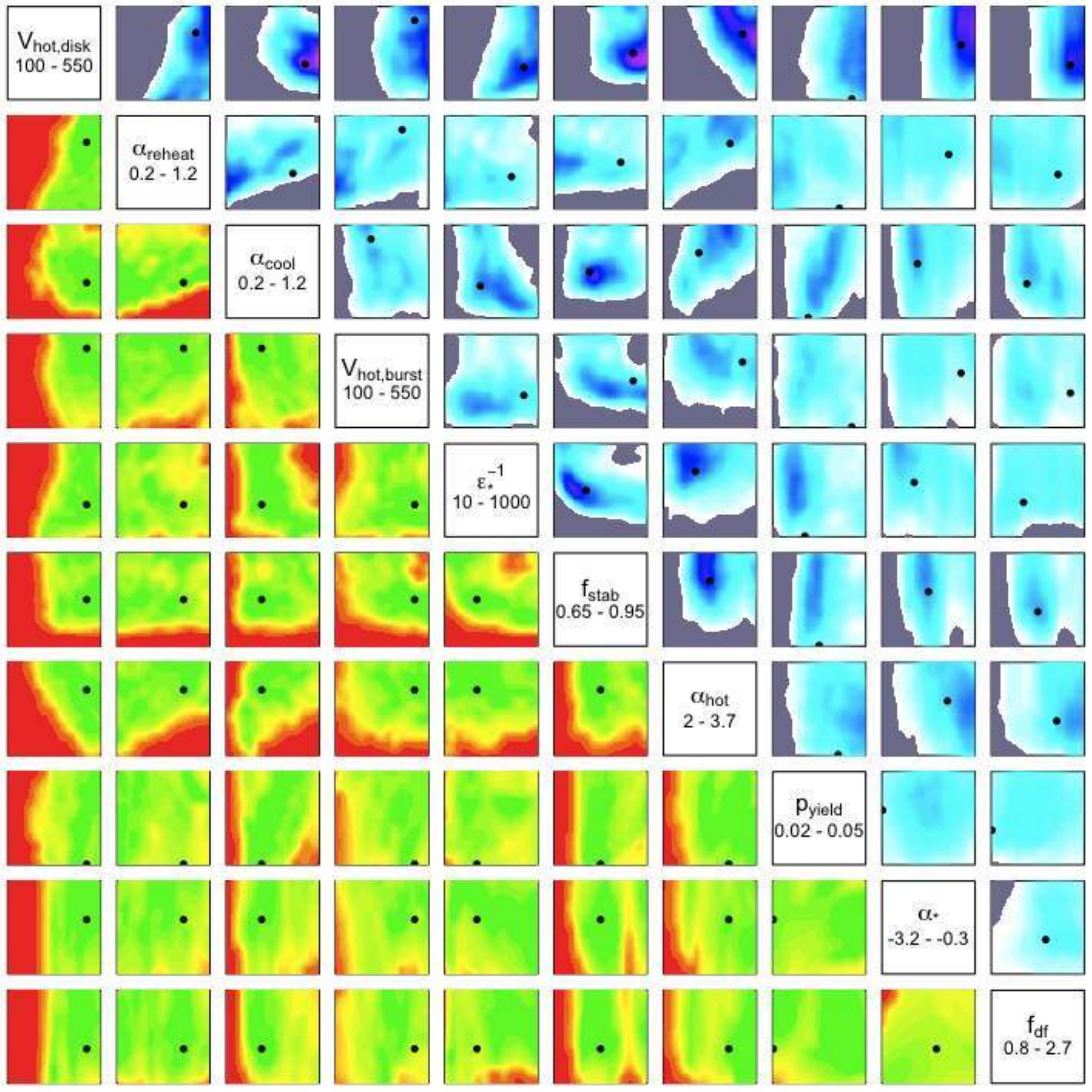}
\parbox[b]{2cm}{
\vspace*{-5cm}
\vbox{\includegraphics[scale=0.3]{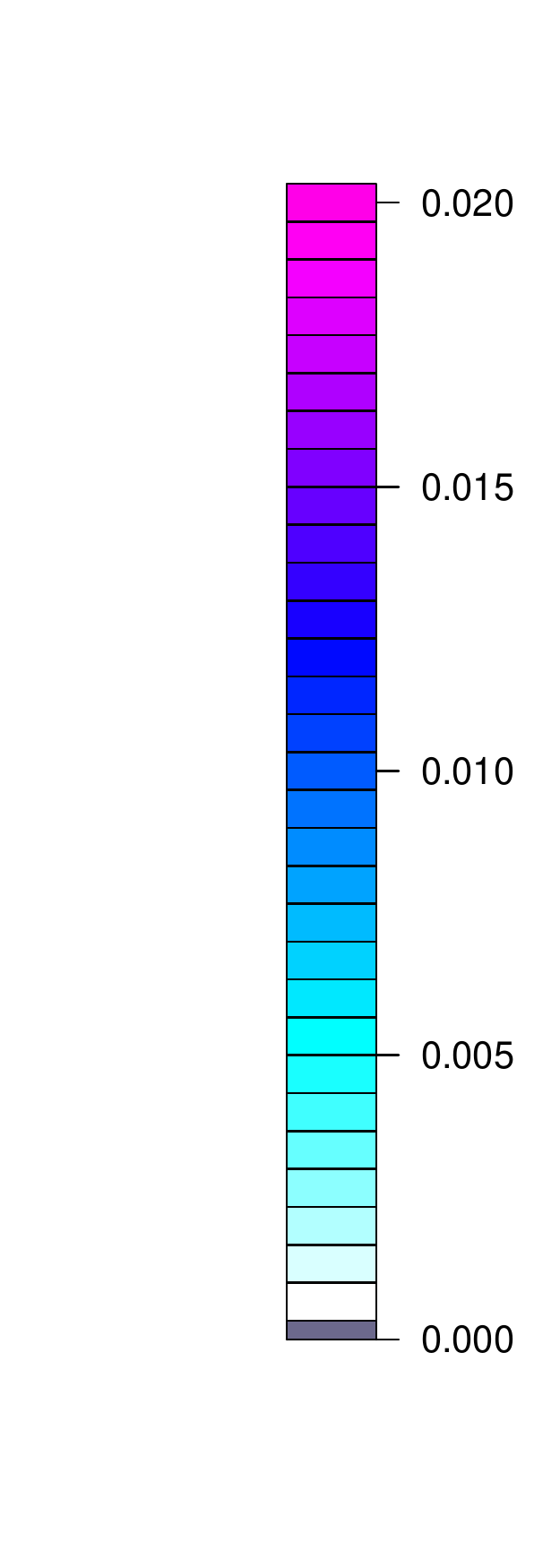}}
\vbox{\includegraphics[scale=0.3]{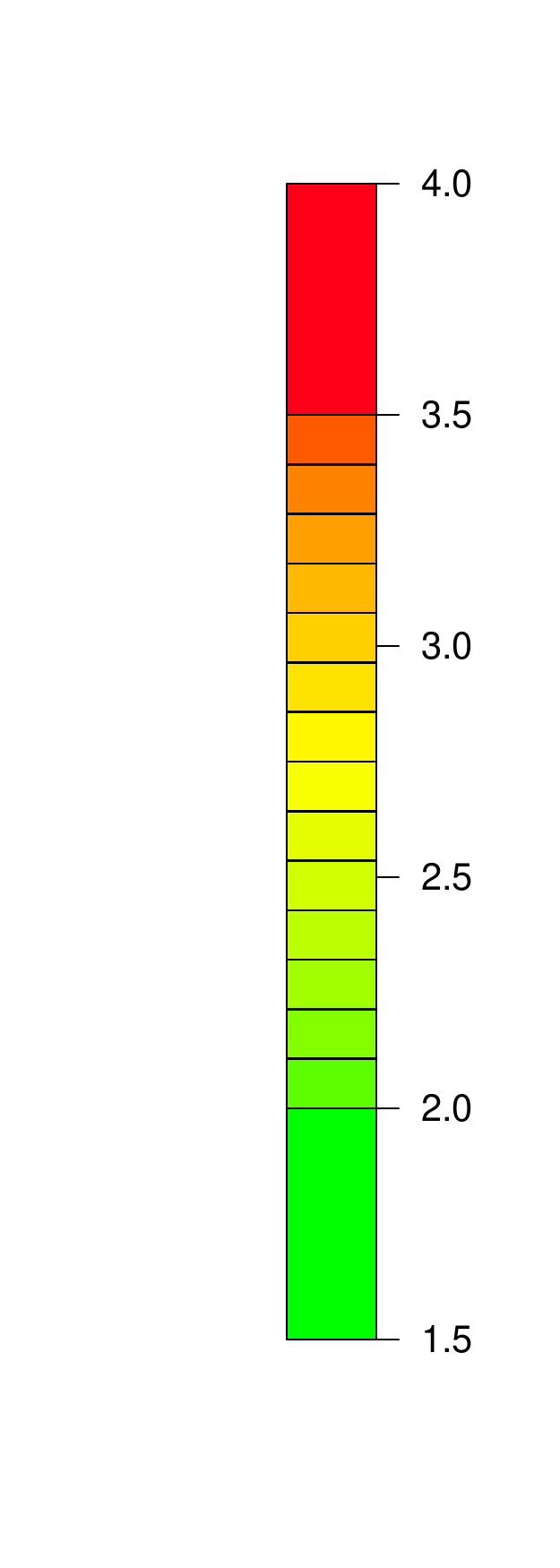}}
     }
\caption{2-D Projections of the implausibility landscape of the Wave~4 emulator
for some of the major parameters.
Plots below the diagonal show the projected minimum implausibility predicted by the
Wave~4 emulator. 
We sample data points in the hidden dimensions using a latin hypercube design, accumulating 
the minimum implausibility and the fraction of points lying below the implausibility
cut-off (see Table.~\ref{tab:waves}). The emulator suggests that green regions 
are likely to give  acceptable fits to the luminosity function data, for some choice of the hidden parameters (given our model 
discrepancy). In the red region the emulator confidently suggests that acceptable
matches are implausible, regardless of the values of the hidden parameters. For comparison, the Bow06 model is shown as a black point.
Plots above the diagonal give an impression of the line-of-sight depth of the acceptable
region with blue/purple regions indicating that a high fraction of the ``not implausible''
points are aligned at this projected position.
Note that plots above the diagonal have x and y
axes transposed to make comparison with the maximum implausibility plots more apparent. 
Although it seems that only a small
region of the projected space is ruled out as implausible, the projected space frequently 
has a very thin, but extended, geometry.
}
\label{fig:impproj}
\end{figure*}

\begin{table}
\begin{center}
\begin{tabular}{|l|c|c|c|}\hline
 &Min&Max \\ 
 \hline
$\invepsilonStar$&46&1000 \\ 
$\alphastar$&-3.2&-0.3 \\ 
$\yield$&0.02&0.05 \\ 
$\vhotdisk$&300&550 \\ 
$\vhotburst$&190&550 \\ 
$\alphahot$&2.3&3.7 \\ 
$\alphareheat$&0.2&1.2 \\ 
$\alphacool$&0.38&1.2 \\ 
$\taumrg$&0.8&2.7 \\ 
$\stabledisk$&0.73&0.95 \\ 
\hline
\end{tabular}
\end{center}
\caption{This table shows the parameter ranges of models which gave
acceptable luminosity function matches. The inactive variables range
over their full range given in Table~\ref{tab:params}.}
\label{tab:notimprange}
\end{table}

A serious problem for such high-dimensional parameter sets is to find a way of 
representing the implausibility map. Projecting the full 10 dimensional map 
(of active variables) down to 2 dimensions so that it can be printed loses
considerable information. We can try to compensate for this by showing the 
full set of projections as a matrix (this is commonly referred to as a ``pairs plot'').
Fig.~\ref{fig:impproj} shows the implausibility space projected onto pairs of parameters
in this way. Note that only active variables are shown so that there
are 45 plots of 10 variable pairs. The parameters have been scaled to range over 
$\pm1$ using the initial range given in Table~\ref{tab:params}.   

Plots below the diagonal show the projected minimum implausibility surface. The colour code is
set so that green indicates that the region is ``not implausible''.
 The implausible region is shown in red.  The minimum
implausibility is determined by evaluating the emulator over a grid of values for the
two ``visible'' parameters and a Latin hypercube of parameters in the unseen variables.
Because the hyper-volume of acceptable solutions is very thin in some projections, 
a large number of evaluations are required in order to obtain a reliable projection
of the minimum value. Even though each evaluation is almost $10^7$ times quicker than 
performing a \GALFORM\ model evaluation, this means {\rgb that such plots cannot be made
interactively}.

It is immediately apparent that many of the projections contain a substantial fraction
of green covering a large fraction of the parameter space. For example, the $\alphacool$
parameter was allowed to vary in the range 0.1 to 1.2, and it is not implausible to find 
acceptable fits through-out this region. {\rgb This is the result of projecting
over a large number of hidden parameters, however,} and it is apparent that varying 
the visible parameters had physical effects that can be compensated for by variations in the
other parameters. 
In order to better appreciate the underlying geometry, therefore, it is helpful to plot the
``optical depth'' of the projected hyper-volume. Therefore above the diagonal,
we show the fraction of the evaluations (for each pair of fixed visible parameters)
that resulted in low implausibility values.  This allows the viewer to 
distinguish regions that have a low minimum implausibility but require very precise
coordination of the unseen parameters, from regions in which a low implausibility is
obtained for a wide range of the unseen parameters. One has the intuitive sense
that the ``best'' solution lies in a region that has a large depth, but this assertion
does not necessarily hold. Thus the appearance of regions in this optical depth
plot needs careful interpretation.
For reference, the Bow06 model is shown by a black point. It is often centered in 
a ``deep'' region of
low implausibility, but in some projections it is offset from the depth weighted centre
of the region. It is apparent from this that the Bow06 is not a ``typical'' model that matches
the zero redshift luminosity function: illustrating the limitations of searching parameter
space to find a single acceptable model. 

Higher dimensional projections can also be used to reveal more of the underlying structure. 
The typical hyper-surface geometry is that of a thin, slightly curved
plane. 
For a few variables, it is nevertheless informative to look at the range of 
the ``not implausible'' region in a 1 dimensional sense, and this is given in Table~\ref{tab:notimprange}.
While this region covers a large fraction of the initial range for many parameters, 
a few of the parameters are significantly constrained. For example, the emulator shows 
that it is implausible that runs with low values of $\vhotdisk$ (below $300\kms$) will result in 
acceptable fits to the luminosity functions. Similarly, the emulator suggests that
the disk stability parameter cannot be reduced below 0.73, showing that disk instabilities
are a key component of the model and that the role of instabilities cannot be replaced
by altering the sensitivity of galaxies to high mass ratio mergers (Parry et al.\ 2009).

The parameters that are shown as inactive in Table~\ref{tab:params} have no
clear effect on the luminosity function and are treated as an additional source of
uncertainty in the emulator. Runs close to the implausibility cut-off may result
in acceptable fits to the luminosity functions if these parameters are carefully
chosen. However, our procedure is conservative and this is taken into account 
in deciding whether a region is implausible or not. 
However, while the emulator can identify regions for which an acceptable fit is 
implausible, it does not guarantee that a run in the remaining region will actually
result in a good fit to the luminosity function.  This is a consequence of our 
conservative approach: the ``not implausible'' region has a cut-off threshold of 
$I_{M}<3.2$, while we only deem a model to be ``acceptable'' if $I_{M}<2.5$. 
Therefore, in order to demonstrate that an
acceptable match to the luminosity function can be obtained, we must perform 
a model run at the point in question.

\subsection{Comparing the Emulator with Model Runs}
\label{sec:testcompare}
\begin{figure}
\includegraphics[scale=0.4]{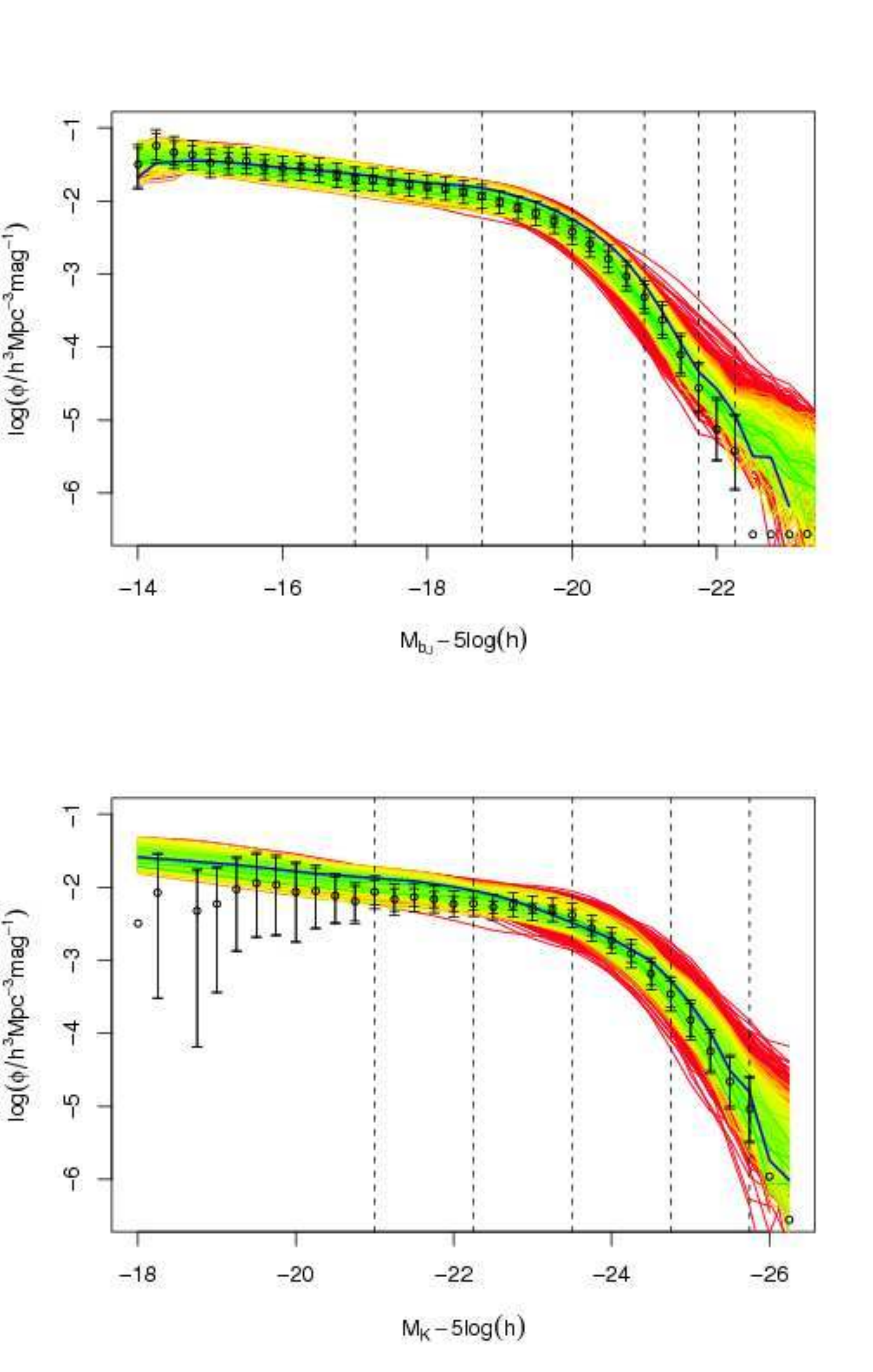}
\caption{The $b_J$ and K luminosity functions of the model runs evaluated in Wave~5.
Runs are colour coded by the model run implausibility. 
For comparison, the Bow06 model is shown as a blue line. Note that 2$\sigma$ error bars are shown.
{\rgb The inner error bars include the contributions of observational errors and repeatability,
but exclude the subjective $\vPhi_{E}$ term. Outer error bars also include the full model discrepancy term.} 
Green curves are considered acceptable fits and are tested against other observational constraints
in \S\ref{sec:addconstr}. Vertical lines show the magnitudes at which the model is compared to the 
observational data in order to determine its implausibility.}
\label{fig:testlf}
\end{figure}

\begin{figure*}
\includegraphics[scale=0.6]{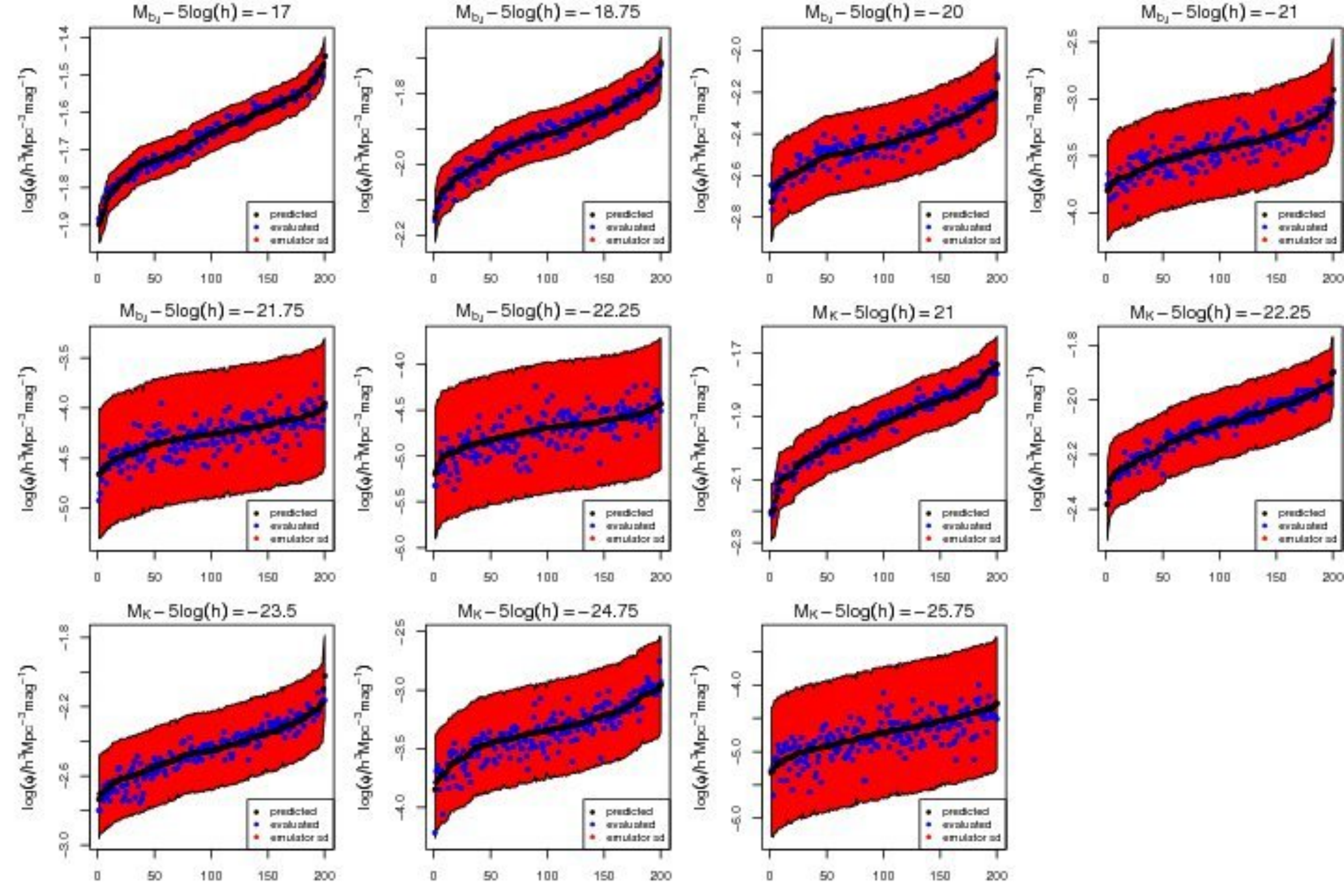}
\caption{Values of the $b_J$ and K-band luminosity functions for 
  direct model evaluations
compared with the expectation and variance predicted by the
emulator. The figure shows 200 runs selected at random from Wave~5. 
The panels are labeled with the
relevant $b_J$ or $K$ magnitude at which the luminosity function is sampled:
these correspond to the vertical lines in Fig.~\ref{fig:testlf}.  The x-axis
shows the run number ordered by the value of the luminosity function
predicted by the emulator
The y-axis shows the value of the 
luminosity function.  The solid black line shows the
expectation value of the emulator: because the runs have been ordered, the black line traces a 
smooth curve (but adjacent points need not be close in parameter space). The red shaded 
regions indicates the 3.2$\sigma$ range of uncertainty in the emulator output. The open circles show 
the values of model runs with each parameter set. Note that some regions of the luminosity
function are easily emulated, while other regions, particularly the bright end of the
$b_J$ luminosity function, show significantly greater variance.} 
\label{fig:testcompare}
\end{figure*}

\begin{figure*}
\includegraphics[scale=0.48]{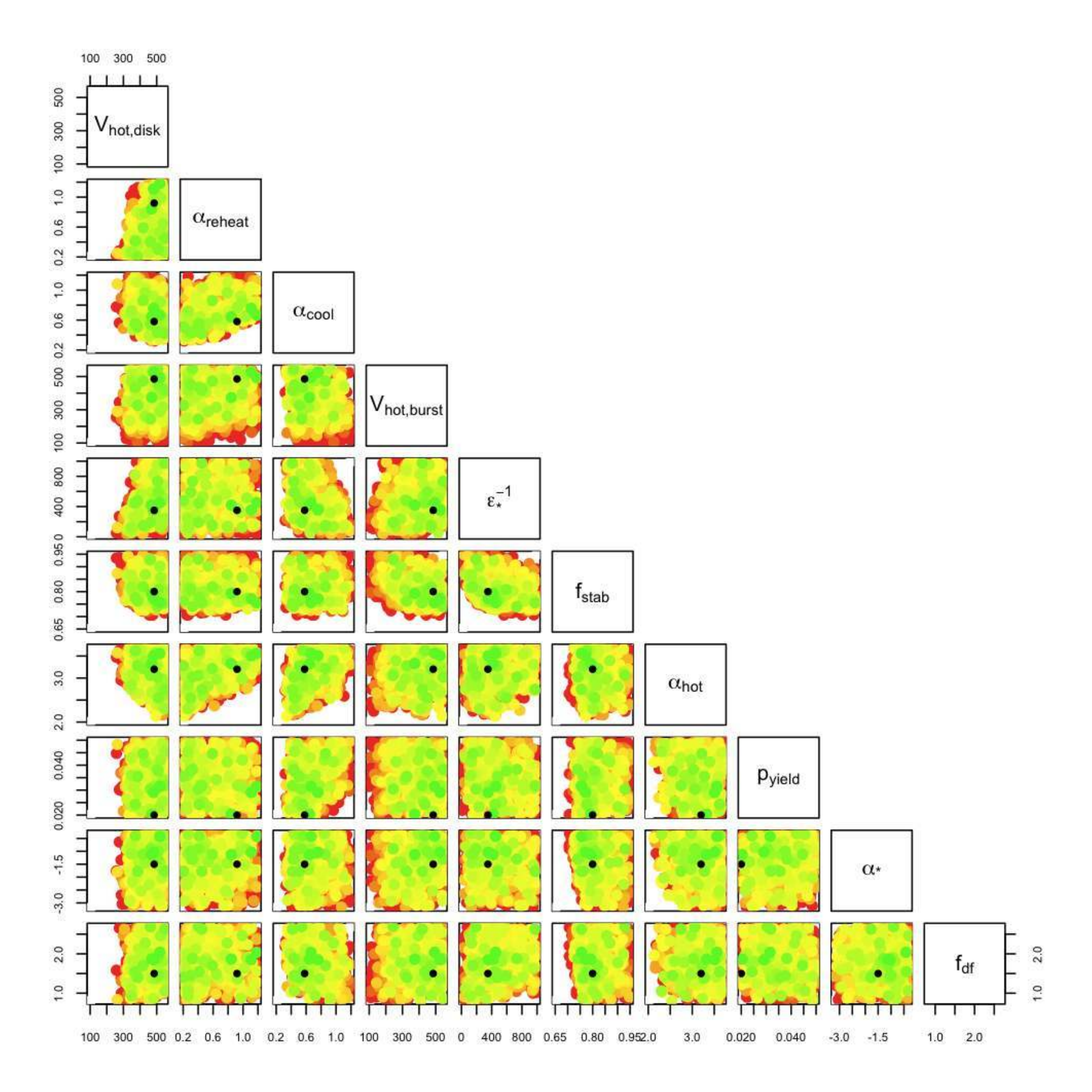}
\caption{The Wave~5 model runs plotted on a set of 2-d projections.
The points are colour coded by the model run implausibility, rather than
the emulator's predicted implausibility. Note that the model has only been
evaluated in the ``not implausible'' region defined by the Wave~4 emulator.
Green points resulted in acceptable luminosity
functions given the model variance and repeatability. 
The Bow06 model is shown as a black point.}
\label{fig:testproj}
\end{figure*}

In the previous section, we illustrated the shape of the low implausibility region
in parameter space. It is important to stress that this is a region in which
good fits {\it might} be obtained, but that an acceptable match is not guaranteed.
To proceed, we investigate the success of the emulator technique by randomly
generating 2000 model runs (which we refer to as ``Wave~5'') with various parameter values
within the {\it not implausible} region of Wave~4. Our aim is twofold. 
Firstly, we wish to show that 
the model runs do indeed return matches to the luminosity function within the
expected uncertainties, and that good descriptions of the luminosity
function are indeed found throughout the range of parameter space illustrated in 
Fig.~\ref{fig:impproj}.  Secondly, we wish to identify a set of model runs that 
give acceptable matches to the observed luminosity functions, and that
can be used as the basis for our exploration of the constraints imposed by
additional datasets.

Since we are now comparing genuine model evaluations with the 
data, the emulation stage is no longer required and the emulator variance term, 
$\vard{D_i}{f_i(\vx)}$ in eq.~\ref{eq:imp2nd}, is replaced by
a small factor based on the stochastic variation in model evaluations. We 
denote this revised implausibility measure $I_M'$.
The denominator in $I_M'$ thus includes only contributions from the 
model discrepancy, observational errors and the repeatability of model runs.
Because we are now referring to actual model evaluations, we can refer to a model 
as ``acceptable'' rather than ``not implausible''. We identify ``acceptable'' models 
as those for which $I_M' < 2.5$. It is important to note that we distinguish acceptable
models by a lower implausibility cut than the threshold used to reject implausible regions
(see Table~\ref{tab:waves}).

The luminosity functions of the Wave-5 model runs are shown in Fig.~\ref{fig:testlf}.
The lines have been colour code to reflect the implausibility derived for the run
using the colour scale of Fig.~\ref{fig:impproj}. Note that
the $2\sigma$ error bars shown on the
observational points include the effect of the model discrepancy and repeatability. Around the
knee of the luminosity function the statistical errors on the luminosity function are
small and are dominated by the model discrepancy term. 
Vertical lines show the points at which we have emulated the luminosity function. 
An acceptable luminosity
function must pass close to the error bars at the lines. As can be seen, this {\rgb generally} 
provides a good description of the shape of the luminosity function and justifies our
assertion that the shape is well described by the 11 outputs chosen for emulation.
{\rgb In common with previous version of the code, acceptable models tend to lie
above the lowest luminosity measurements.}
The distinction between green
(``acceptable'') and yellow (``not quite acceptable'') models is that the later tend
to miss one or more data points. It is clear, however, that
the points at the bright end of the luminosity function, particularly in the $b_J$-band, present 
the greatest challenge for the galaxy formation model.

We can first use the model evaluations to confirm the success of the emulator in 
describing the luminosity function behaviour. In Fig.~\ref{fig:testcompare}, we compare
the values of the luminosity function in different absolute magnitude
bins from direct evaluations of the model with predictions from the Wave~4 emulator.
 The panels are 
labeled by the absolute magnitude in either the $b_J$ or K-band at which the luminosity
function is sampled.
The runs are shown in order of their emulator expectation, which is drawn
as a think black line.  Note that adjacent points are not adjacent in parameter space. The 
result of evaluating the model for each parameter set is shown as a blue circle. If the 
model evaluation agreed exactly with the expectation value, the blue points would lie
perfectly on the line. However, we expect the points to scatter about the line
as a result of the uncertainties in the emulation and the Monte-Carlo nature of the model. 
The extreme of the predicted range of these uncertainties are
shown as the red shaded region. This is formally a 3.2$\sigma$ deviation (although we
do not necessarily expect the tails of the distribution to be Gaussian) that
is used to define the {\it not implausible} region.  Note that some regions of the luminosity
function are easily emulated, and the points lie close to the expected value.
In other regions, particularly at the bright end of the
$b_J$ luminosity function, there is significantly greater variance. {\rgb This is
telling us that it is hard to precisely emulate the behaviour of GALFORM in
these regions.} The uncertainty is, however, well described
by the emulator's predicted uncertainty. The ability of the emulator to capture this
variance is key since it has allowed us to efficiently cut down the full parameter space
of models. Note that a series of similar diagnostics were performed for each emulator at each wave of 
the analysis, and in each case the emulator was found to provide the expected levels of accuracy 
(see Vernon et al.\ 2010 for details).

Since we are confident that the emulator has successfully directed the selection
of Wave-5 parameter sets, we continue to compare the parameter space of acceptable
models with that suggested by the emulator.
Fig.\ \ref{fig:testproj} shows a pairs plot of the Wave-5 evaluations. 
Each dot shows the implausibility of the model evaluation, colour coded to match the 
implausibility colour scale of Fig.~\ref{fig:impproj}.
The points have been superposed so as to illustrate the most acceptable in each region.
Acceptable luminosity function fits (green points) are found throughout the 
parameter space, and the figure bears striking similarity to the analogous
Fig.~\ref{fig:impproj}, which was based on emulator predictions rather than model
evaluations. This again confirms that the emulator successfully captures the
behaviour of the \GALFORM\ model.
 
Of the 2000 model evaluations, 113 resulted in acceptable models according to
the criterion $I_M'<2.5$. 
The reduction in volume is not surprising since
we have tightened the implausibility threshold and eliminated the emulator variance.
Furthermore, many of the evaluations were performed for 
marginal models for which the emulator gave only a small chance of an acceptable match
(recall that the surface  of the acceptable parameter space is huge in 10 dimensions).  
Despite this apparent inefficiency, if we wish
to make a systematic investigation of parameter space it is important that marginal models
are evaluated. Of course, with only a relatively small number of evaluations,
some parameter space projections suffer from considerable shot noise, particularly the low ``optical depth''
regions seen above the diagonal in  Fig~\ref{fig:impproj}. Further evaluations of the model
could be used to fill in these regions if required. It is already apparent, however,
that acceptable fits to the luminosity functions can be found throughout the wide
range of parameter values suggested by the emulator analysis.

We conclude that the emulator method provides an accurate scheme for identify model 
parameter sets that are very likely to yield acceptable fits to the luminosity functions. 
Before we examine the physical
links between the parameters in the acceptable regions, we use the 
113 acceptable runs to examine how well models which
make acceptable fits to the luminosity functions perform in matching other low redshift data sets.

\subsection{Further Constraints from Additional Data}
\label{sec:addconstr}

\begin{figure}
\includegraphics[scale=0.4]{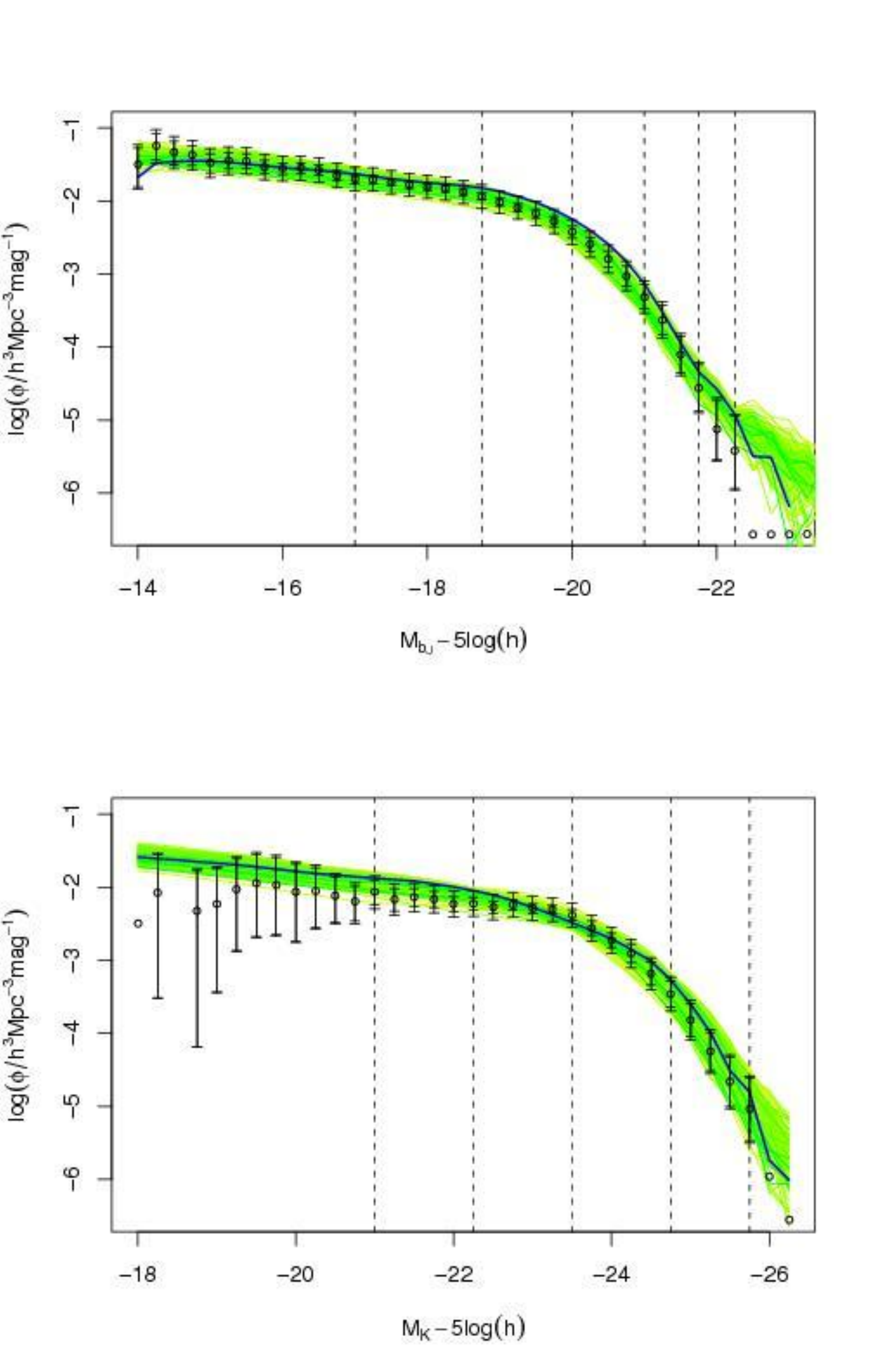}
\caption{The $b_J$ and K luminosity functions of the 113 acceptable model evaluations found in 
Wave~5. The axes, colour scale, error bars and lines are the same as Fig.~\ref{fig:testlf}.}
\label{fig:LFGood113}
\end{figure}

\begin{table}
\begin{center}
\begin{tabular}{lcccc} 
  data set & $N_{\rm bin}$& $\chi^2_{\rm Bow06}$& $\chi^2_{\rm min}$ & $\chi^2_{\rm cut}$\\
\hline
  disk size   &     40 &  7160&  443&  10000  \\ 
  TF relation &     20 &   124&   44&    150  \\ 
  gas metalicity&   4&    21&     1&    27   \\
  gas mass to $L_B$  &     9&  3090&    57&  3500   \\
  SDSS colours  & 2400& 16600&  5150& 17000\\
  BH mass       &     8&    5&      3&  60 \\
\end{tabular}
\end{center}
\caption{Additional data sets used to constrain the model fits. Column 1: the
property considered; column 2: the data bins used in the comparison. 
The $\chi^2$ of the Bow06 model is given in column 3.
Column 4 gives the $\chi^2$ value for the best fitting model for each separate
comparison. In order to selected an interesting set of models for further
comparison between data sets, we used the $\chi^2$
cut off value listed in column 5. Note that the cut-off
values used for the disk size and Gas mass to $L_B$ constraints are far in excess of those expected
for a statistically acceptable model.}
\label{tab:addconstr} 
\end{table}

\begin{figure*}
\vspace*{-2cm}
\includegraphics[scale=0.7]{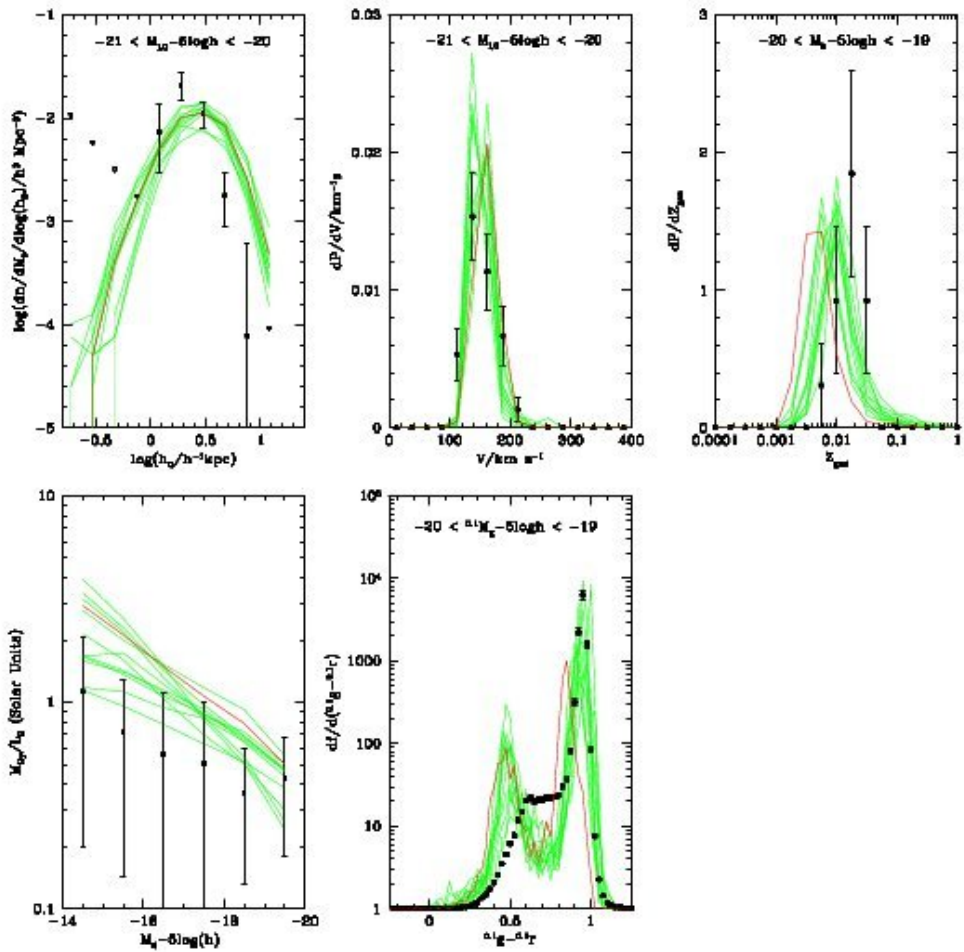}
\vspace*{-4cm}
\caption{This figure illustrates the comparision of models with additional datasets.
Models which give an acceptable match to the luminosity function data, and produce 
fits with $\chi^2 < \chi^2_{\rm cut}$ (see Table~\ref{tab:addconstr})
for all of the additional datasets considered are shown as green lines. The same models are 
highlighted as green points in Fig.~\ref{fig:chi2add}. The Bow06 model is shown as
a red line for comparison, and the observational data are shown as black points
with error bars or upper limits (denoted by triangles). The source of the
observational data is described in the text. With the exception of the gas mass to
luminosity comparison, the panels show the comparison for a single magnitude slice. 
The total $\chi^2$ values shown are derived by summing the contibutions from several 
magnitude slices.}
\label{fig:other_props}
\end{figure*}

The main focus of this paper is to investigate the constraints derived from the $b_J$ and
$K$ luminosity functions. This follows the methodology of Bow06. However, it is interesting
to briefly examine how the model may be constrained by including additional data in the
comparison. In order to make an initial investigation, we apply these constraints as 
a second phase, so that we only consider models which
survive the primary criterion of generating an acceptable match to the luminosity functions,
and we base the further exploration on the 113 fully acceptable models
that were identified in the previous section. 
The luminosity functions derived from these models are shown in 
Fig.~\ref{fig:LFGood113}. 

We now outline the additional data-sets that we consider. Note that the first four physical
properties listed below were already used by Cole et al. (2000) and Baugh et
al. (2005) in choosing the parameter values in their respective
versions of GALFORM, though we have here updated some of the
observational data used. {\rgb Further details of the data sets and our approach
to the comparison are given in Benson \& Bower 2009.}

\begin{itemize}
\item{\bf disk size}: We compare to disk size data from De Jong \& Lacey (2000).
We compute the $\chi^2$ values in 
a series of bins in both magnitude and size. The ideal model would therefore
match not only the sizes of local galaxies, but also the spread in size. The predicted
disk size distribution depends strongly on the angular momentum of accreted gas
which has a complex dependence on halo growth  
\item{\bf TF relation} Galaxy formation models have traditionally struggled to match 
the Tully-Fisher relation. The normalisation and slope of the relation depend both
on the relationship between stellar mass and halo mass, and on the baryonic contraction
of the halo. We compare with $i$-band  data from Pizagno et al.\ (2007). 

\item{\bf gas metallicity}: This is an important constraint 
on the effectiveness of supernova-driven feedback.
We compare with data from Tremonti et al.\ (2004) on the oxygen
abundance of gas in late-type galaxies in the SDSS.
\item{\bf Gas mass to $L_B$}: The cold gas reservoir is sensitive to the rate at which gas
is accreted, the rate at which it is converted into stars and the effectiveness of 
supernova-driven feedback. We compare with a compilation of HI data from Huchtmeier \& 
Richter (1988) by computing the mean and standard deviation of the
ratio of the HI
gas mass to B-band luminosity as a function of 
B-band magnitude. We only consider model galaxies with a bulge to total ratio of less
than 0.4 and gas mass fraction greater than 3\%.
\item{\bf SDSS colours}: We compare with the overall distribution of galaxy $g-r$ colours from 
Weinmann et al.\ (2006). The
vast amount of data available for this test results in a large number of bins.  As discussed
in Font et al 2008, this match can be substantially improved by adjusting the stellar
yield of the model.
\item{\bf BH mass}: We compare to data from Haering \& Rix (2004).  This is a weak test for our model
since although the parameter $\FSMBH$ has little effect on the luminosity
function output, its value can be adjusted to fine tune the 
black hole mass normalisation. Since the parameter is inactive, we could have made this adjustment
without having discernible effect on the luminosity function. We show the comparison 
for completeness only, and we have not undertaken this fine tuning step.  
\end{itemize}

For each additional data set, we reduce the comparison to a single $\chi^2$ value. Except where
indicated, this is achieved by comparing
the binned distributions of model and observed galaxies. The $\chi^2$ values 
for the Bow06 model and an indication
of the number of bins used in each test is given in
Table~\ref{tab:addconstr}. Some examples of the fits to these
additional datasets are shown in Fig.~\ref{fig:other_props}.
Further examples are shown in Benson \& Bower (2009).   As has been extensively 
discussed, the $\chi^2$ value gives only a coarse indication of the fit of the model to the data, and
makes no allowance for model discrepancy terms that estimate the level at which
we expect the model to perform in each test. For this reason, we 
focus on the performance of each model relative to Bow06. It should be noted however that Bow06 provides a
poor match to the disk size data, the gas mass to luminosity data and SDSS colour data.
In the case of the
disk size data (cf. Gonzalez et al.\ 2009), although the best fitting models significantly improve
the match to average sizes, they still fail to describe the variation with luminosity well,
suggesting that the treatment of angular momentum  in the
code may need improvement.  Bow06 also fails to reproduce
the normalisation of the gas mass to luminosity relation and the colours. However, the better fitting
models result in a substantial improvement to these comparisons. The colours are significantly improved
by increasing the yield $\yield$, although the required value is
larger than the best estimates based on stellar evolution models 
(c.f. Cole et al. 2000) but within the range of their plausible 
uncertainty (cf., Font et al., 2008, Benson \& Bower 2009).
We see below that the gas mass to luminosity
ratio is sensitive to the assumed star formation efficiency (as
was earlier found by Cole et al.\ 2000).

In a future application of the emulator method, we will apply the full emulator 
technique to encapsulate and hence emulate all of these statistical outputs. This needs to 
be combined with a careful analysis of the underlying statistical assumptions and
the realistic model discrepancy terms.  For the moment
we aim only to make an indicative comparison, and we will only distinguish models which perform comparably
well to Bow06 (the cut-off $\chi^2$ values are given in the table). {\rgb It is 
important to note that $\chi^2$ cut-off values are not intended to denote a 
statistically acceptable fit to the data.} As we will see, 
combining all these data sets in this way already restricts the parameter space substantially.  

\begin{figure*}
\includegraphics[scale=0.6]{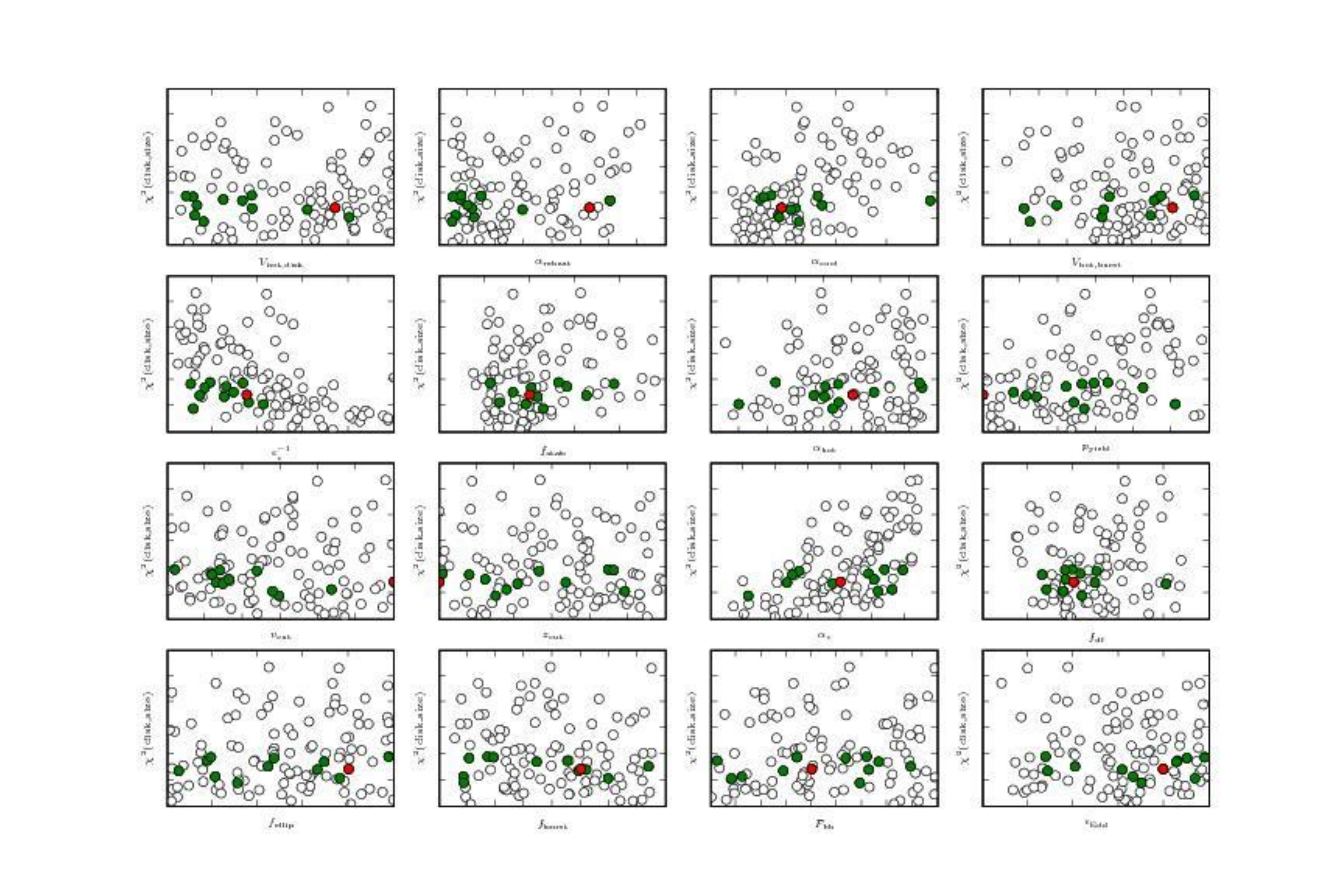}
\caption{This plot shows the $\chi^2$ values of the models compared to the
observed sizes of galaxy disks as a function of some of the more important parameters.
The y-axis ranges from 0 to 30000 in each case, while the model parameters cover the range given in 
Table~\ref{tab:params}.
The Bow06 model is shown as a red filled circle. All the open points shown provide an
acceptable match to the  $b_J$ and K luminosity functions. Green points highlight
models which have 
$\chi^2$ measures similar to or better than Bow06 when compared to all
of the observational constraints listed in Table~\ref{tab:addconstr}.
All the points shown (filled and open) 
produce acceptable fits to the $b_{J}$ and K-band luminosity functions.
}
\label{fig:chi2add}
\end{figure*}

\begin{figure*}
\begin{center}
\includegraphics[scale=0.6]{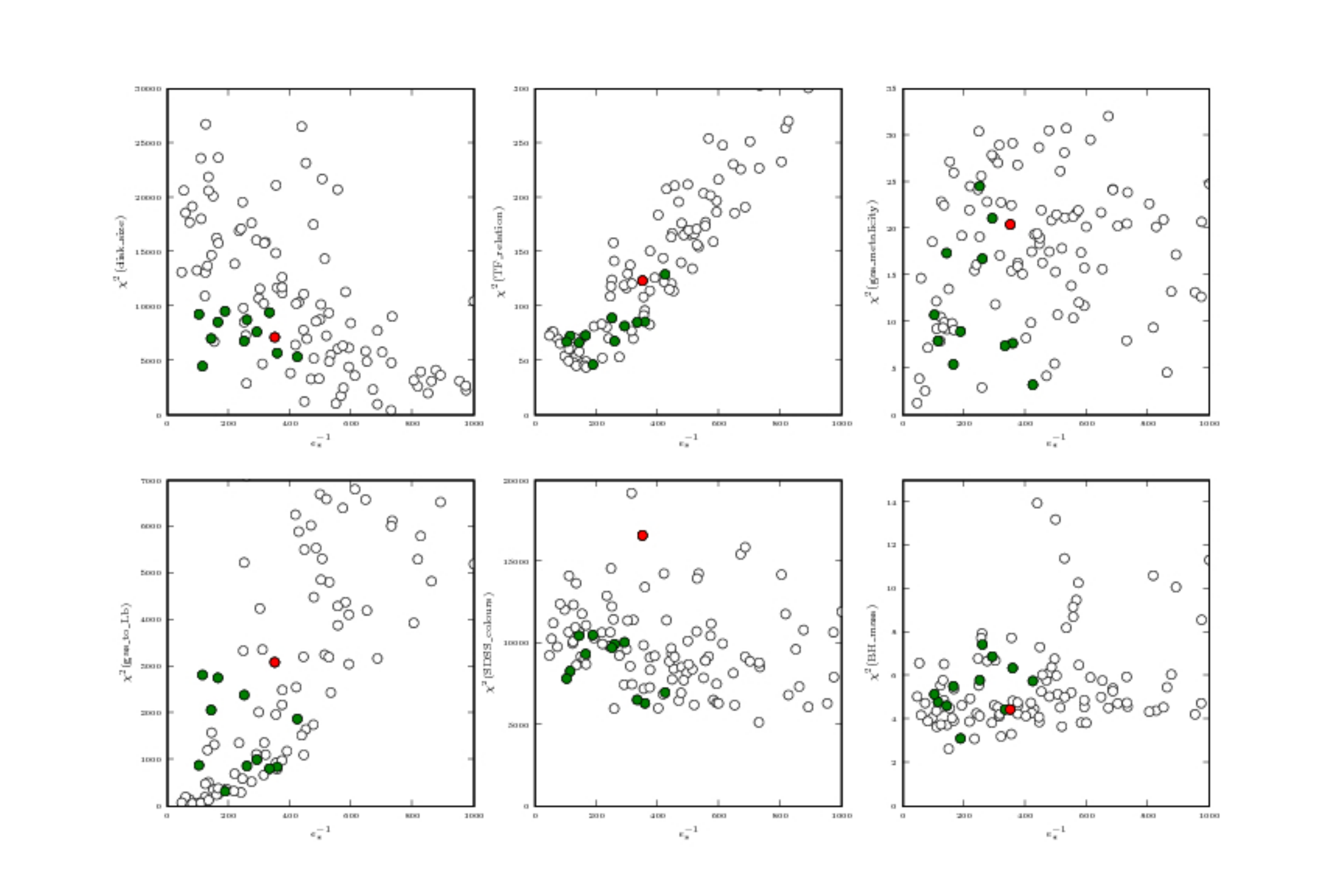}
\caption{The variation in $\chi^2$ for various data constraints as a function of
the $\invepsilonStar$ parameter. From top left to bottom right, the y-axes show
the $\chi^2$ derived from comparison of disk sizes, the Tully-Fisher relation,
gas metalicity, gas mass to luminosity ratio, SDSS colours and black hole mass. The colours 
of data points are the same as Fig.~\ref{fig:chi2add}.
}
\label{fig:chi2add_onepar}
\end{center}
\end{figure*}

The dependence of $\chi^2$ for one of the data sets (disk sizes) is illustrated in 
Fig~\ref{fig:chi2add}. Each panel shows the $\chi^2$ value as a function
of one of the 16 input parameters. The Bow06 model is highlighted in red, while 
green points highlight models which fall below the cut-off $\chi^2$ value in all of the 
data sets tested. While a significant trend
of $\chi^2(\hbox{disk size})$ with $\invepsilonStar$
is apparent, it is hard to discern a pattern in most
of the panels. The lack of dependence may arise because the model output is not 
strongly dependent on a parameter, or because the dependence is masked by dependence 
on the other parameters. We describe our approach to systematic identification of
interesting parameter combinations in \S\ref{sec_projpur}.

We have plotted the parameter
dependence of $\chi^2$ for each of the other data sets with similar  results,
and so do not reproduce each of the figures here. However, many
properties exhibit a strong  dependence on the parameter $\invepsilonStar$,
 and we illustrate this in 
Fig.~\ref{fig:chi2add_onepar}.
The contradictory pressures on $\invepsilonStar$ are
apparent. {\rgb Looking at the open and filled points together}, it is clear 
that a better match to disk sizes is obtained by increasing this
parameter; however, higher values of $\invepsilonStar$ tend 
to worsen the match to the Tully-Fisher relation
and the gas mass to luminosity ratios of the galaxies. The models highlighted in green
are models which pass the $\chi^2$ cut-off in all of the data sets as well
as provinding an acceptable match to the luminosity functions. Thus it is
nevertheless possible to balance 
these opposing pressures, picking lower values of $\invepsilonStar$ and 
exploiting other parameter dependencies to obtain adequate fits to the 
disk size.  We discuss the implication of these results for future directions
of the model in \S\ref{sec:disc_pp}.  It is notable that the green points all
produce better matches to the SDSS colour data than Bow06 even though we
did not enforce this in their selection {\rgb (of course, we have enforced some
colour information by requiring that models match both the $b_J$ and $K$-band
luminosity functions).}

A final point to note from Fig.~\ref{fig:chi2add} is the hint of 
bimodality in the 
distribution of green points in the $\alphareheat$ and $\vhotdisk$ parameters. The two 
families of acceptable models correspond to models with very strong supernova
feedback, but a cycling time for the galactic fountain close to the halo
dynamical time, and models with weaker feedback but with cycling times substantially
longer than the halo dynamical time. Bow06 belongs to the first family of models, while
the second family is more typical of the earlier Durham models (Cole
et al. 2000; Baugh et al 2005) and models from the Munich group (eg.,
Croton et al.\ 2006; De Lucia \& Blaizot 2007). {\rgb Further model runs, and fuller
treatment of the additional datasets are required to explore this point further.}

\section{Projection Pursuit}\label{sec_projpur}

\subsection{Luminosity function constraints}

With such a high-dimensional data-set, plotting the dependence of model runs
as a function of one or two input parameters conveys only a small fraction
of the complexity of the underlying parameter space. This bi-variate approach assigns a 
special significance to the input parameters and thus, although it is not
easily possible to present higher dimensional projections, we can make a more
informed choice of the projection vectors. In particular we can 
optimise the projection to show (1) the reduced dimensionality of the parameter space enforced
by the constraints we have applied, and (2) maximise the additional leverage of the
other data sets that we have considered in the previous section. This optimisation
is often referred to as ``projection pursuit''.

One approach is to project the parameter space of ``acceptable'' models
along its principal components. This effectively corresponds to rotating the 
region of acceptable models to align with
the directions of greatest to least variation of parameters.
In many situations, principal component analysis (PCA) is used to find the 
directions with greatest variance;
in our case, we are more interested in the directions with least variance.
These are the parameter combinations that are most tightly constrained by the 
observational data. Using PCA we can align the hyperplanes revealed by our parameter
space exploration so that their cross-section is viewed. It should be stressed, however,
that PCA is inherently linear, and that the projected cross-section may 
hide a much thinner, but warped, relationship between the variables.

\begin{table*}
\begin{center}
\begin{tabular}{|c|cccccccccc|} \hline
 &Var 1&Var 2&Var 3&Var 4&Var 5&Var 6&Var 7&Var 8&Var 9&Var 10 \\ 
 \hline
$\tvhotdisk$&0.198&0.112&-0.231&-0.283&0.334&0.311&0.0637&-0.114&-0.256&{\bf 0.724} \\ 
$\talphacool$&-0.386&0.361&0.136&-0.00109&-0.00943&-0.0733&0.0977&-0.265&{\bf 0.713}&0.33 \\ 
$\tstabledisk$&-0.149&-0.147&0.247&-0.312&-0.0218&0.331&-0.104&{\bf 0.781}&0.236&0.103 \\ 
$\tvhotburst$&-0.115&-0.167&0.414&-0.234&0.102&0.173&{\bf 0.783}&-0.181&-0.135&-0.168 \\ 
$\ttaumrg$&-0.203&0.0895&0.0975&{\bf 0.729}&0.0545&{\bf 0.627}&-0.0033&0.00626&-0.116&0.0238 \\ 
$\tyield$&{\bf -0.581}&0.043&-0.263&0.0517&{\bf 0.632}&-0.3&0.0585&0.223&-0.183&-0.109 \\ 
$\talphastar$&0.214&0.0194&{\bf 0.718}&0.0465&0.49&-0.152&-0.392&-0.127&-0.0453&0.0239 \\ 
$\talphareheat$&0.0702&{\bf 0.780}&-0.031&-0.313&0.051&0.273&-0.0775&0.0104&-0.104&-0.44 \\ 
$\tinvepsilonStar$&{\bf 0.589}&0.111&-0.166&0.297&0.348&-0.102&0.383&0.302&0.387&-0.082 \\ 
$\talphahot$&-0.013&0.421&0.27&0.213&-0.333&-0.41&0.229&0.341&-0.379&0.337 \\ 
\hline
Range Ratio&0.598&0.554&0.496&0.469&0.487&0.357&0.343&0.21&0.167&0.0945 \\ 
\hline
Mean&0.00341&0.00371&-0.265&0.0368&-0.244&0.00854&-0.132&0.153&-0.279&0.697 \\ 
Rel.\ Std.\ Dev.&0.679&0.65&0.564&0.487&0.418&0.393&0.323&0.224&0.145&0.0854 \\ 
\end{tabular}
\end{center}
\caption{Principal components for the acceptable space of luminosity
functions. The columns give the PCA variables ordered by decreasing relative 
standard deviation, where the relative standard deviation is the standard deviation 
of the component when the initial range of variables has been scaled to the range $\pm1$.
Small relative standard deviations correspond to components that
are tightly constrained by the requirement of producing a good luminosity
function. Dominant input variables in each of the vectors are
highlighted in bold font. The variables have been ordered so that the
most constrained components appear last.}
\label{tab:pca_lf}
\end{table*}

\begin{figure*}
\includegraphics[scale=0.6]{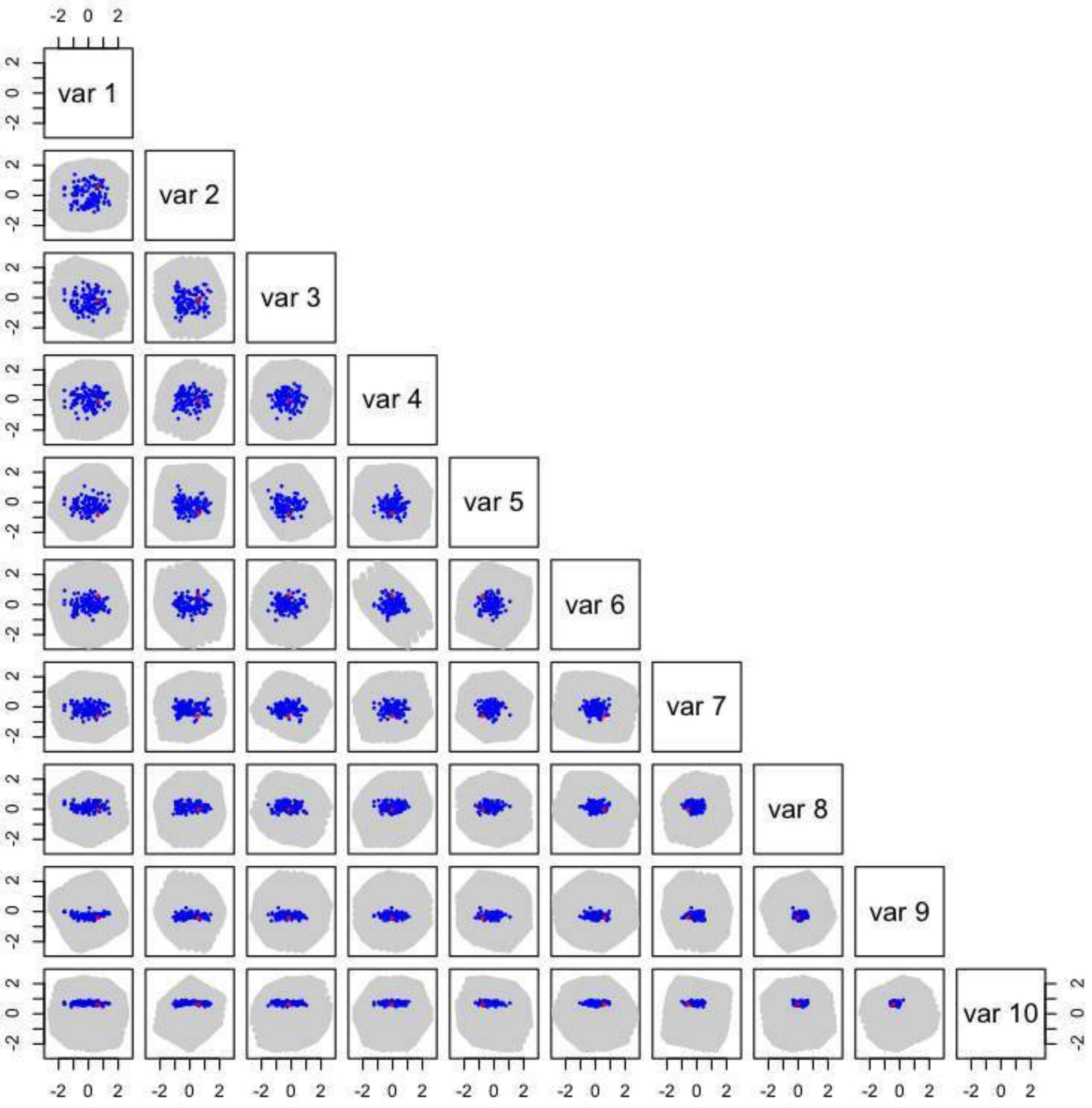}
\caption{
The space of acceptable models that reproduce the local $b_J$ and K
luminosity functions projected along pairs of principal components.
Blue points show models which produce luminosity functions with
implausibility less than 2.5. Grey points show the geometry of the
initial allowed parameter space in each of these projections. 
The principal components are chosen to minimise the variance of the 
models with low values of $\chi^2$. The highest components (eg., Var~10) 
are the most constrained. The Bow06 model is shown as a red filled circle.
This figure illustrates the reduction in the dimensionality
of parameter space resulting from requiring a good match
to a particular LF.}
\label{fig:pca_lf}
\end{figure*}

\begin{figure}
\begin{center}
\includegraphics[scale=0.5]{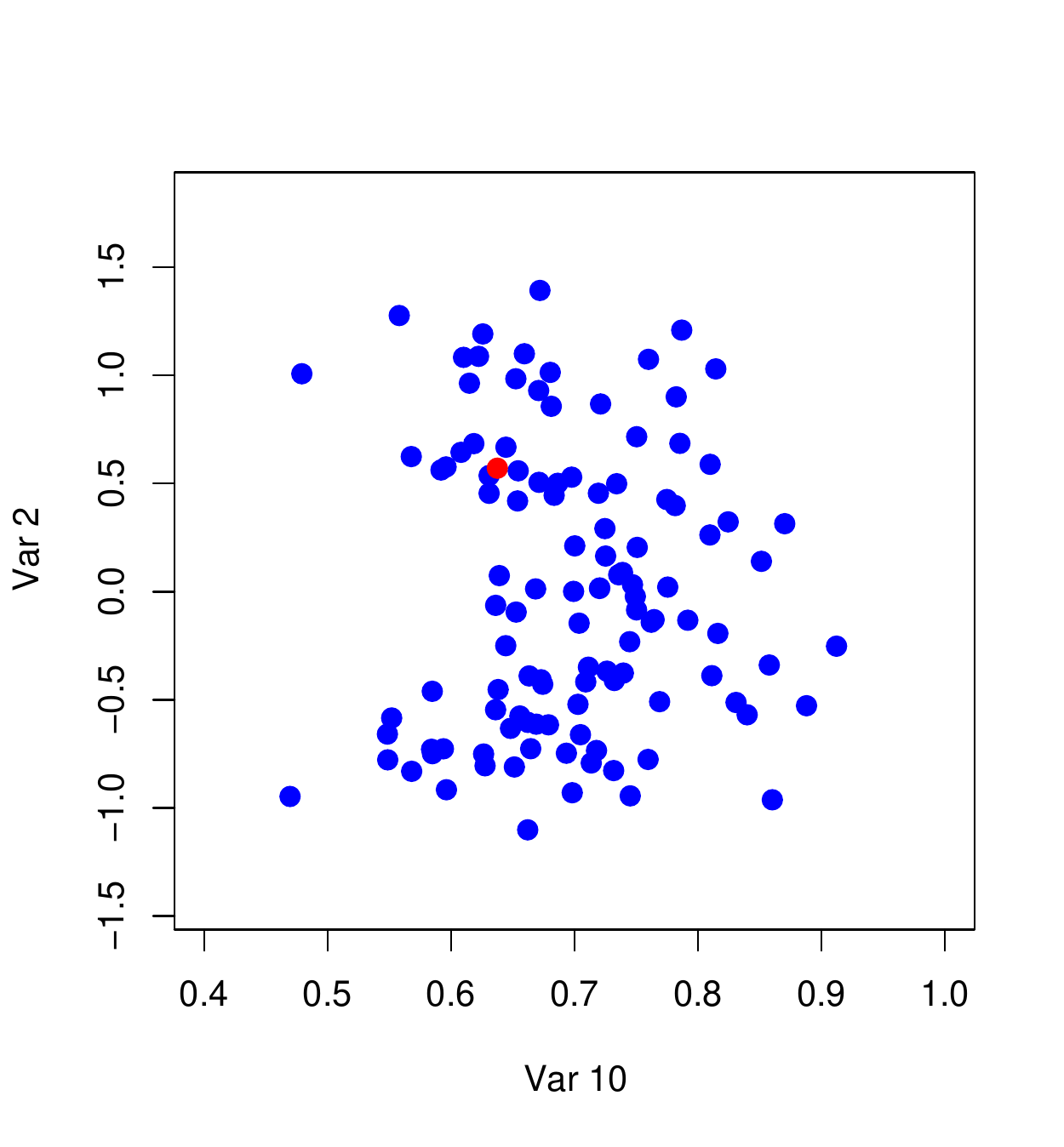}
\end{center}
\caption{
Close up of one PCA projection, showing that the acceptable space is not
linear, and cannot be accurately described by the principal components}
\label{fig:pca_example}
\end{figure}

One potential problem of PCA is that the
input variables must be scaled so that the variance along different 
axes can be compared. This is somewhat arbitrary. Our approach is
to scale the variables by the initial search range,
rather than restricting it to the range over which fits were found
to be possible. By restricting our range in this way, we are aiming 
to determine whether the high-dimensionality of the acceptable space 
can be reduced by suitable combination of parameters. 

A principal component with low variance
implies that this particular combination of the parameters is tightly
constrained if the model is likely to produce an acceptable luminosity 
function fit.  Of course, even if this constraint is satisfied, a good
model is not guaranteed; rather we can be confident that if it not
satisfied the fit is unlikely to be good.  When analysing the acceptable
region in this way, we also need to bear in mind that the PCA
assumes that the relationships are linear, whereas we have seen that
the actual acceptable space is curved.  This will prevent any of the
suggested projections being arbitrarily thin and limit the accuracy
of constraints.

We begin by focusing on the the luminosity function data alone.
Fig.~\ref{fig:pca_lf} shows the result of applying the PCA 
analysis to the 113 model runs which resulted in acceptable matches
to the luminosity function. As with previous plots derived from
luminosity function data alone,
we consider only the 10 ``active'' variables identified by the 
emulator analysis. The relationship between the PCA variables
and the  original model parameters is given in Table~\ref{tab:pca_lf}
(see below).

We have previously seen that, by trading between 
parameters, acceptable solutions can be found over almost the 
full range of the input parameters. 
The PCA reveals a different story, however.
In the plot, the gray region illustrates the original parameter
space projected onto each pair of PCA components. The region
is not square because the new coordinate system is not aligned
with the axes of the original hypercube. The blue
regions show the projection of models for which the luminosity
function was acceptable.
In this space, it is seen that many parameters are strongly constrained
compared to their initial range. In the projections of the three most
constrained parameters, the good fits shrink towards a point.
For comparison, we show the Bow06 model as a red point. In many 
projections, it is evident that it lies well to the side of the
main parameter space.

\begin{figure}
\begin{center}
\includegraphics[scale=0.4]{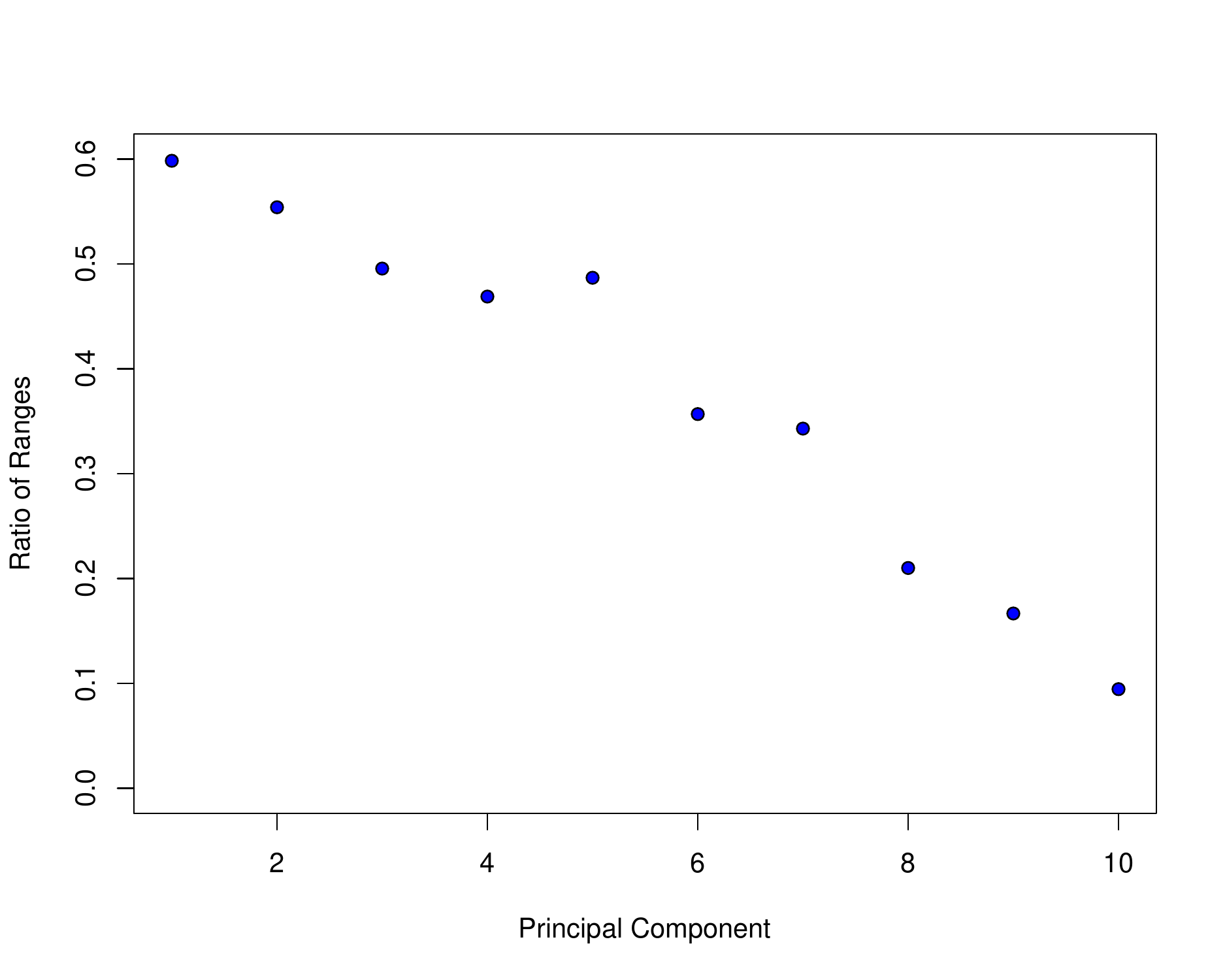}
\caption{
The reduction in the range of the PCA components from requiring that
the model produce a low implausibility luminosity function. The
PCA components are ordered by their variance, while the vertical axis
shows the ratio of the range of acceptable models to the range of 
the PCA components in the initial parameter space.}
\label{fig:pca_ratio}
\end{center}
\end{figure}

We can quantify the degree to which parameters have been constrained
by comparing the range of the PCA components in the initial parameter space to 
the range covered by the acceptable models. Note that although all
the input variable have been scaled to $\pm 1$, the range of the 
PCA components may be greater because the component may be aligned with a 
diagonal of the hypercube.  The comparison of the ranges
is shown in Fig.~\ref{fig:pca_ratio}.
The figure quantifies the degree to which the components are constrained.
We consider the reduction in range, since the ratio of the standard
deviation is sensitive to the orientation of the original hypercube.
However, similar results are obtained when we consider the ratio of standard 
deviation of the PCA components.
Three components stand out in particular, with ranges that are less than 
1/4 of their initial range.
In order to have a adequate match to the luminosity function these
components must have quite precisely determined values. However, simply
matching these values does not guarantee a fit to the data, since
the other components also need to be considered. Indeed, all 
of the PCA components cover a reduced range compared to the initial parameter
space. 

It is important to emphasise again that PCA cannot extract the 
full interdependence of the parameters since the analysis is
intrinsically linear. Fig.~\ref{fig:pca_example} illustrates the 
limitations. The crescent shape arises because the observational
constraints generate a nonlinear dependence 
between the components of Var.~2 and Var.~10. The PCA
has selected these directions to minimise the linear 
variance. On the other hand, although the emulator analysis includes terms 
up to 3rd order, it is unclear how the physical dependencies 
can be extracted. Thus, despite its limitations, PCA provides 
a good means of gaining insight into
the physical processes driving the match to the luminosity function.

The relationship between the PCA variables
and the  original model parameters is given in Table~\ref{tab:pca_lf}.
This expresses each PCA component as a vector direction in the scaled input
variables. The components have all been normalised to unit length in the 
10-d space so that a large coefficient indicates that the PCA component direction is
closely aligned with the input variable. Without first applying the scaling
given in Table~\ref{tab:params} the relative importance would be obscured.
We consider the weight of the input variables in the most constrained components below:
although none of the components are completely aligned with the 
input parameters, most have a dominant component that can be
identified with one particular input variable. These are highlighted in the 
table. Below we consider the three most constrained PCs, focusing on components with 
contribution greater than 0.3.
 
\begin{itemize}

\item[Var.~10] This component is dominated by a 70\% contribution from
$\vhotdisk$. This parameter controls the amount of mass ejected from the 
galaxy disk to the halo for each solar mass of stars formed.
However, an adequate match to the luminosity function may
be maintained if an increase in $\vhotdisk$ is offset by an increase
in $\alphareheat$ (decreasing the timescale for
the material to become available for reaccretion), decreasing 
$\alphacool$ (increasing the mass scale at which AGN feedback 
becomes important) and/or decreasing  $\alphahot$ (reducing 
velocity scaling of the mass ejected from the disk). 

\item[Var.~9] is dominated by a 70\% contribution from $\alphacool$. 
Increases in $\alphacool$ can be offset by reducing the star formation
efficiency (through $\epsilonStar$) or increasing the mass dependence
of the disk feedback (through $\alphahot$). This PCA component also has important contributions
from $\vhotdisk$ (with the opposite sign to the cross term in Var.~10) and 
the disk stability criterion.
The physical significance of the Var.~9 becomes clearer if we consider the
effect of adding a small
contribution from Var.~10 to create a new component that is independent
of $\vhotdisk$. This strengthens the dependence on $\alphacool$, while
slightly weakening the dependence on $\alphahot$. We explore the idea
of optimally rotating principle components in \S\ref{sec:disc_pp}. 

\item[Var.~8] is dominated by the disk stability criterion, $\dstab$,
with significant contributions from the star formation rate efficiency, $\epsilonStar$,
and the mass dependence of feedback efficiency, $\alphahot$. 

\end{itemize}

Looking at the less constrained components, we see that these often
have dominant input variables too. Var.~6 and Var.~4 are dominated by $\taumrg$
(the dynamical friction timescale); Var.~5 by the yield (increasing the 
metal abundance normalisation); Var.~3 by $\alphastar$ (the star formation law exponent);
Var.~2 by $\alphareheat$ (the timescale for re-incorporating ejected disk gas). 
Var.~1, the least constrained component, has equally 
strong contributions from the yield and the star formation rate efficiency.

We have ordered the variables in Table~\ref{tab:pca_lf} so that the most
constrained dominant parameters appears first. This emphasises that the most 
important parameters are $\vhotdisk$, $\alphacool$ and $\stabledisk$. It is also
notable that $\alphahot$ (the mass dependence of the feedback) is not
dominant in any particular component, but makes an important
contribution to many of them.

\subsection{Important directions from other data}

\begin{table*}
\begin{center}
\begin{tabular}{c|c|cccccccccccccccc} 
        &
    Relative std.\ dev.&    
    $\tvhotdisk$ &  
    $\talphacool$ &
    $\tinvepsilonStar$ &
    $\tstabledisk$ &
    $\tyield$ &
	 $\talphastar$ &
	 $\tepsilonSMBHEddington$\\
\hline
disk size   &  0.49&  0.301& -0.519&  0.468& -0.119& -0.449& -0.269&  -0.021\\
Gas mass to $L_B$&  0.40& 0.16&   -0.251&  {\bf 0.764}& -0.282& -0.197& -0.334&  -0.205\\
TF relation &  0.29& 0.264&  -0.33&  {\bf 0.775}&  -0.207&  -0.202&   -0.24&  -0.196\\
\end{tabular}
\end{center}
\caption{
The most additionally constrained directions using
various complimentary data sets in addition to the luminosity function
constraints. For each data set, the table shows the most
constrained component vector together with the
factor by which the standard deviation is reduced. Directions
are quoted relative to the normalised parameter range: the scaling
is given in Table~\ref{tab:params}.
These components provide physical insight into the effect of introducing the 
additional data sets. Only data sets which resulted in a significant additional constraint
are shown: the other data sets considered in \S4.3 do not result in strong additional
linear constraints on parameter combinations. Parameters with no contribution greater
than 0.2 have been suppressed.  Note that the best constrained directions
are similar for the Tully-Fisher and gas mass to luminosity data sets.}
\label{tab:pca_add}
\end{table*}

\begin{figure*}
\includegraphics[scale=0.4]{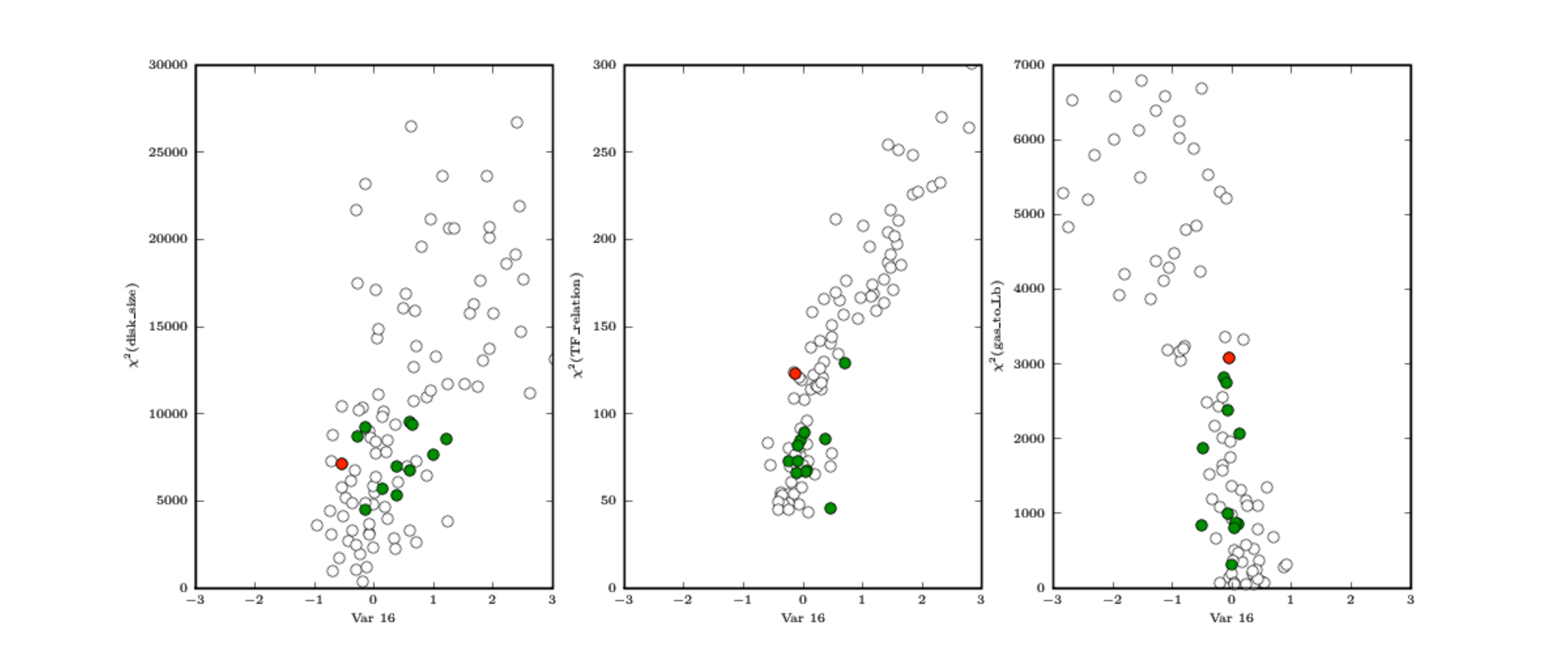}
\caption{We show the $\chi^2$ values of the models as functions of some 
of the most additionally constrained principal components identified in 
Table~\ref{tab:pca_add}. In each panel, we show the most constrained component 
(x-axis) resulting from requiring the model to match the data sets shown 
on the y-axis. Note that the vector direction represented by the axis
is different in each plot (see Table~\ref{tab:pca_add}).
The Bow06 model is shown in red. Green solid points highlight models which gave fits comparable
to, or better than, Bow06 in all the tests given in Table~\ref{tab:addconstr}.
Note how the chosen projection results results in a bunching of the low $\chi^2$
models campared to the full set.
All the points plotted produce an acceptable match to the luminosity function data.}
\label{fig:pca_add}
\end{figure*}

Having shown that projection pursuit using PCA provides
a useful way of capturing the geometry of the region producing
acceptable luminosity functions, we proceed to apply a second
level of analysis to examine how the introduction of additional
constraints further restricts the allowed region of parameter
space.

In order to perform this analysis, we first express 
each model as a function of the principal components identified
by the luminosity function constraints, as described in \S5.1. 
We then renormalise these components so that their variance 
is equal (and recentre the distribution on the mean).  
This has the effect of mapping the distribution of acceptable
parameter region (according to the luminosity function criterion) into
a roughly spherical distribution centered on the origin.
We then further restrict the
data set using the additional data, keeping only those runs
with low $\chi^2$, and characterise this reduced space
in terms of its new principle components.  
We then  map these vectors back into the space defined by the scaled 
input parameters, providing insight into the physical differences between 
models that satisfy both the
luminosity function constraints and the additional data and
those that do not.  The coefficients quoted in Table~\ref{tab:pca_add} are the
coefficients of the scaled input variables in the additionally constrained
direction. The scalings of the original input
parameters are given in Table~\ref{tab:params}.

We find that the analysis reveals significant constraints that
arise from adding data on the disk sizes, the $M_{HI}$ to $L_B$ ratio,
and the Tully-Fisher relation. 
The dependence of $\chi^2$ on the most constrained component
is shown for each of the data sets in Fig.~\ref{fig:pca_add}. 
The importance of these data sets was already apparent because of the 
clear dependence on $\invepsilonStar$ in Fig.~\ref{fig:chi2add_onepar}; however, the PCA is
capable of revealing constraints that are based on combinations
of parameters that would not have been evident in a simpler analysis. 
For example, contrast the evident bunching of the low $\chi^2$ points in the 
first panel (which shows $\chi^2(\hbox{disk size}$) with absense of clear trends 
in Fig.~\ref{fig:chi2add}.

The reduction in the standard deviation due to the PCA is shown in Table~\ref{tab:pca_add},
together with the model's additionally constrained directions.
It is immediately apparent that the directions due to the gas mass to $L_B$ ratio
and the Tully-Fisher relation constraints are similar.   Not only
do both have dominant contributions from $\invepsilonStar$, but the
weighting of most other variables are also similar. 

While the disk size constraint also depends strongly on $\invepsilonStar$,
the overall direction of the constraint is different from that implied
by the Gas to $L_B$ ratio, and the Tully-Fisher relation,  and the constrained
direction also depends equally strongly on the yield and $\alphacool$ parameters.
Thus, models which match all the data
sets tend to have low values of $\invepsilonStar$, selecting particular
values for the other parameters to compensate for this. This is
evident in Fig.~\ref{fig:chi2add_onepar}, and explains the trend for the
accepted models to be squeezed into the lower left corner of the
first panel. This tension has also been apparent in earlier versions of the 
\GALFORM\ code (eg.\ Cole et al.\ 2000); however, the analysis scheme
presented here provides an objective means of identifying the interplay
between observational constraints and model parameters.

\section{Discussion}

In this paper, we have set out to explore the parameter space of 
the \GALFORM\ semi-analytic model.  The model implements the key
physical processes that define the formation of galaxies, including
the hierarchical growth of dark matter haloes, gas accretion
and cooling, and the feedback effects of supernovae and AGN.
The model we use contains 16 parameters that describe these
processes. Some parameters have strongly constrained, physically
plausible values; others represent poorly understood physical
processes and are only weekly constrained by the observational data
used here.
Bow06 identified a largely successful point in this parameter
space by matching the local $b_J$ and $K$-band luminosity functions. 
In this paper we set out to find out how unique this
point is and to explore whether other parameter combinations
could perform equally well (or better) at reproducing local
galaxy properties.

\subsection{The Model Emulator Technique}

We have adopted the {\it Model Emulator} technique  to efficiently
but conservatively search the 16-d parameter space. The essence of the 
approach is to build a statistical predictor for model results on the 
basis of a limited set of model runs. This is a Bayesian approach, where
we are aiming to quantify the information we derive from the runs.
We use the statistical model to identify regions where we are confident 
that an acceptable fit will not be found. Such regions are denoted as 
`{\it implausible}' and excluded from further consideration. A second wave
of emulation is then performed to better characterise the surviving
portion of parameter space. After 4 waves of emulation, we are left with
an accurate emulation of the model that defines a 'not implausible' region
containing 0.26\% of the original volume.  A run within this region is 
not guaranteed to give a statistical match to the observational data,
and subsequent runs within this region are needed to identify acceptable
model realisations.
 
In this paper, we have focused on matching to the observed $b_J$ and $K$-band luminosity
functions of galaxies in order to parallel the approach used in Bow06. An important issue has been
to quantify the observational and model uncertainties. A key
concept has been that of {\it model discrepancy}. This term accounts for the 
expected accuracy of the model itself: since the semi-analytic model is inherently
an approximation to the true physical process, we are willing to accept models
which come close to the observational data but do not match it exactly. 
{\rgb This is a key concept in our Bayesian approach, and it differs from 
most previous work in which the philosophy is to adapt the model
parameters to find fits consistent within the observational uncertainties alone.
We argue that it is important to explicitly account for the approximate nature of the
model, and that ignoring the model discrepancy term will lead to over-zealous exclusion
of some regions of parameter space. In this sense, the model discrepancy term makes
our approach conservative. Naturally, it is potentially of interest to investigate
the effect of reducing the model discrepancy term in order to guide improvements to the
model. However, we feel that improvements in the model are currently better driven
by comparison to additional data sets, where the tensions in the model can be
more easily exposed (as illustrated in \S4.3). Formal discussion of the process of ``model 
reification'' can be found in Goldstein \& Rougier 2008.}
 
We have accepted models at the 2.5$\sigma$ level, which equates to models that come 
within a factor of approximately 1.54 of the observed luminosity 
function (ie., a factor of $10^{2.5\times0.0753}$, see \S3.6). 
Since the dynamic range of the luminosity function covers almost 5 
orders of magnitude in space density this term appears relatively small
in any visual comparison, as can be seen from Fig.~\ref{fig:LFGood113}. Acceptable luminosity
functions are generally very good matches to the luminosity function around
the knee of the luminosity function. It is evident that many models struggle to match
the very bright part of the $b_J$ luminosity function, even though they provide a good description 
of the break in the K-band luminosity function. Physically, these models tend
to allow a small amount of star formation to take place in the most massive
systems.

Nevertheless, the model emulator suggests that plausible matches to the 
luminosity functions may be found over a large fraction of the parameter
ranges. However, this results from a fine-tuned interplay
between parameters and the acceptable models are confined to thin hyperplanes.
Out of the 16 input variables, we find that 10 need to be taken into account in order
to make the emulator predictions. These are labeled 'A' in Table~\ref{tab:params}.
The remaining parameters ($\epsilonSMBHEddington$,  $\fellip$, $\fburst$, $\FSMBH$,
$\VCUT$ and $\ZCUT$) have a weak
effect on the luminosity functions (although it may have significant
impact on other aspects of the model). Although we have achieved a large
reduction in parameter space, further model runs need to be performed to determine
whether each point actually returns a statistically acceptable 
luminosity function. Out of a sample of 2000 runs surviving the 
Wave~4 implausibility cut-off, we find 113 runs that provide acceptable fits
to the luminosity function, implying that only 0.014\% of the initial input parameter
space is compatible with the combined observational constraints and model discrepancy
terms. 

Although the model emulator technique is gaining favour in other areas of
scientific modeling, we could have applied the Markov Chain Monte Carlo 
method to the problem. This approach has been pursued by Henriques et al.\ 2009
and Kampakoglou et al.\ 2008. However, they consider only
a small subset of the possible parameters, analysing a 6d and 7d space respectively,
while we consider 16d space. {\rgb We note that while ``only'' 10 active
variables are included in our fitting functions, the full parameter space is considered
by the emulator. The active variables do not need to be guessed in advance of
the minimsation process, and are all varied in all our GALFORM runs. Moreover, the
approach we adopt is readily adapted as introducing new physical processes in the GALFORM
model generates even higher dimensionality (eg., Baugh et al.\ 2005; Benson \& Bower, 2009).}

Our galaxy formation model is most comparable to that used by Henriques et al.\ 2009 
(which is based on Croton et al.\ 2006). Comparison
of the results is, however, complicated because several of their
chosen parameters do not have direct equivalents in our code and their analysis
does not allow for any model discrepancy term.  Nevertheless, their results
for luminosity function constraints are qualitatively similar to ours, given the
lower dimensionality of their investigation. A significant difference is that they
find that including constraints on the black hole mass -- bulge mass correlation
strongly constrains the parameterisation. In contrast, we find that 
this relation adds little constraint.  This arises from the very different treatment
of black hole feedback in the two models: with the larger number of parameters
that we consider, we find that the black hole mass may
be adjusted largely independently of the galaxy luminosity function.
This illustrates the potential danger of including too few dimensions in the 
analysis, since this artificially restricts the dimensionality of the problem. 
As we have stressed, parameters should be allowed freedom to vary even if their
prior distributions are significantly constrained. If the emulator finds that
the prior range has little impact on the model output, it will reject the parameter
from further consideration. This is not a qualitative decision that should
be made at the outset. 
In future models, we will also allow the cosmological parameters to vary within
their prior distributions. This is difficult if the model is driven by 
N-body simulations, but is possible if it is driven by the improved Monte Carlo
merger tree generators such Parkinson et al. (2008).
In principle, the MCMC method (eg.\ Trotta 2008)
could be applied to higher dimensionality problems, however, it then becomes hard
to drive convergence of the MCMC chains.
In the statistical model fitting literature, these 
scaling problems have meant that MCMC is falling out of favour, being 
replaced by the emulator techniques that we have applied here, which were
specifically designed to deal with high-dimensional models 
(eg.\ Oakley \& O'Hagan 2004, Heitmann et al.\ 2009). As the dimensionality
of the \GALFORM\ model increases further (eg., Benson \& Bower, 2009) the
advantages of the emulator technique become even more relevant.

Our experience in 16-d space suggests that the techniques could be extended
to even higher dimensions, with a relatively modest increase in computing
effort. To arrive at the emulation presented here, we performed 5500
model runs. This was by no means the minimum number of evaluations,
and we have tended to be very conservative in the strategy adopted. A number of
techniques could be used to speed up the calculations, for example performing
smaller, low accuracy runs for the initial waves, increasing the accuracy
as the {\it not implausible} region shrinks. 
An important aspect of the emulator technique is that variables 
are only explicitly included in the 
emulator, when the statistical improvement in doing so is justified.  Thus
adding irrelevant (and even degenerate) variables does not unduly handicap
the method.

\subsection{Projection Pursuit}
\label{sec:disc_pp}

The emulator technique provides a reliable statistical description
of the GALFORM model.  This allows rapid estimation of the likely
success of different regions of parameter space. However, the statistical
model does not easily provide insight into the physical interactions
between parameters. PCA of the successful runs provides 
a complimentary analysis, allowing us to orient the parameter space
so as to reveal the most tightly constrained model parameters.
We find that 3 directions are strongly constrained, corresponding
to linear combinations of the $\vhotdisk$, $\alphacool$, $\invepsilonStar$, 
$\alphareheat$ and $\alphahot$. This makes good physical sense: increasing
the strength of disk feedback needs to be balanced by reducing the 
timescale for reheated gas to fall back to the disk, for example.
Setting these parameters to an appropriate combination is necessary to 
arriving at good models but it is not sufficient, and
we stress that all 10 {\it active} parameters play a significant
role.

As well as providing physical insight,
the PCA components provide a means of orienting future
parameter space exploration.
For example, if we were to start from a single acceptable model, we
can attempt to generate another by moving along the PCA hyperplanes. If we 
select a value for $\tvhotdisk$, Var.~10 (in Table~\ref{tab:pca_diag}) suggests 
we adopt a particular combination
of $\talphacool$, $\talphareheat$ and $\talphahot$ (plus smaller contributions from the 
other parameters). Since $\vhotdisk$ is fixed, Var.~9 and Var.~8 provide two additional 
linear equations linking (primarily) $\talphacool$, $\tinvepsilonStar$ and $\talphahot$. 
By choosing one of the 4 variables, we can invert the system of 
equations and arrive at a good ``guess'' for an acceptable model. For example,
we have already suggested that angular momentum transport in the model
needs to be improved. This would result in denser gas disks and thus a change
in the disk dynamical time. However, from outside the galaxy this change
would be broadly like a shift in the effective value of the $\invepsilonStar$
parameter.

\begin{table}
\begin{center}
\begin{tabular}{|c|ccc|} \hline
                 &Var$''$ 8&Var$''$ 9&Var$''$ 10 \\ 
 \hline
$\tvhotdisk$&           &    &       1\\
$\talphacool$&          &   1\\
$\tstabledisk$&        1&    \\
$\tvhotburst$&       {\bf 0.30}&  -0.14&    0.13\\
$\ttaumrg$&          0.03&         -0.11&   -0.08\\
$\tyield$&            0.17&  {\bf 0.32}&   -0.03\\
$\talphastar$&       0.16&          0.0&    -0.05\\
$\talphareheat$&    -0.14& {\bf 0.26}&   {\bf -0.47}\\
$\tinvepsilonStar$&{\bf 0.44}& {\bf -0.29}&  {\bf -0.30}\\
$\talphahot$&        {\bf 0.37}& {\bf 0.43}&   {\bf 0.62}\\
\hline
Mean&             0.29&  -0.13&   0.99\\
\end{tabular}
\end{center}
\caption{The 3 most constrained PCA components have been combined
to diagonalise the constraints on $\vhotdisk$, $\alphacool$ 
and $\stabledisk$. We use the notation Var$''$ 8 (etc) to emphasise
that these variables are distinct from those in Table~\ref{tab:pca_lf}.
This rotation makes the physical dependence of
these strongly constrained variables evident (see text for details).
Coefficients larger than 0.2 have been highlighted in bold.
Note that these directions are not normalised.}
\label{tab:pca_diag}
\end{table}

\begin{table*}
\begin{center}
\begin{tabular}{ccl}
 $\vhotdisk$ & $=$ & $550 +  110\,\talphareheat + 70\,\tinvepsilonStar -140\,\talphahot \ \ (\kms)$\\  \\
 $\alphacool$ & $=$ & $0.64 + 0.16\,\tyield + 0.13\,\talphareheat - 0.14\,\tinvepsilonStar +0.22\,\talphahot$ \\ \\
 $\stabledisk$ & $=$ & $0.84 - 0.04\,\tvhotburst - 0.07\,\tinvepsilonStar - 0.06\,\talphahot$\\
\end{tabular}
\end{center}
\caption{Equations describing the broad brush interactions
between the most constrained parameters. These relations have been derived from
Table~\ref{tab:pca_diag}. Note that these relations are highly approximate and only
capture a small fraction of the behaviour described by the model emulator.}
\label{tab:approx_physical_eq}
\end{table*}

We can investigate this further by noting that the variance of Var.~8, 9 and 10
is similar. Together, they define the directions of a 7-dimensional 
hyperplane within which the model is significantly less well constrained,
while it is strongly constrained in the 3 perpendicular directions. However, 
we can select a new linear combination of these
components and define the 7-plane equally well\footnote{If the variances had
been equal, the PCA components would be degenerate and all linear combinations
would be equivalent. We could visualise a 3-disk lying within the 7-plane.
Since the variances are not exactly equal, the 3-disk is distorted slightly into an
ellipsoid.}. In particular, we can choose
to ``diagonalise'' the contributions from $\tvhotdisk$, $\talphacool$
and $\tstabledisk$, the variables with the largest contributions to the 
PCA components. The result of applying the diagonalisation procedure is
to define the new component vectors given in Table~\ref{tab:pca_diag}. 
If we now drop terms with coefficients 
less than 0.2, we can arrive at a simple (but approximate) description 
of how $\tvhotdisk$, $\talphacool$ and $\tstabledisk$ should vary as a function
of the other active variables in order to keep within the 7-plane.  
These equations are given in Table~\ref{tab:approx_physical_eq},
where we have translated the scaled variables back to their original units
in order to make the physical dependence more apparent. Because we ellimated
terms with small coefficients, $\ttaumrg$ and $\talphastar$ do not appear
in these expressions. Examining the equations gives useful insight into the 
interplay of the parameters. 
For example, it shows that most successful models require very high values 
for the feedback parameter $\vhotdisk$, close to the maximum value
allowed in our analysis. In order to match the luminosity
function with a lower value, low values must be chosen for $\talphareheat$ 
(the timescale on which reheated gas fall back into the galaxy must be increased)
and $\tinvepsilonStar$ (the disk star formation rate must be reduced) 
and high values for $\talphahot$ (the relative strength of feedback in low mass
galaxies is increased). Similarly, we see that high values for $\talphahot$
imply that $\talphacool$ must be increased (increasing the mass at which the 
radio-mode feedback operates); however, decreases in both  $\talphareheat$
and $\tinvepsilonStar$ have a compensating effect on $\alphacool$.
The appropriate choice of $\alphacool$ is also dependent on $\tyield$, as would be
expected from the metalicity dependence of the cooling function.

These trends can be confirmed by looking at Fig.~\ref{fig:impproj}, however
it is evident that the description of the trends in terms of a few 
components loses a great deal of the true complexity of the underlying parameter
space. Thus, while consideration of the PCA components provides a useful guide to how 
we should remap the parameter space in order to capture the best fitting parameter
region; it must be remembered that PCA is fundamentally linear and its description
is very approximate. The restriction to linear dependencies does not apply to the
emulator technique.

\subsection{Additional datasets}

In this paper, we have followed the approach of Bow06, primarily requiring that
the model be able to reproduce the $b_J$ and $K$ luminosity functions at an
acceptable level of accuracy, only examining the other data sets for the
models which passed this test. Other published models have often
included additional observational constraints from the start (e.g. Cole et al
2000). Clearly an alternative strategy to that followed here would be to 
emulate the full range of data at the outset, perhaps including both high and
low redshift constraints. We have been initially cautious of this approach
since the model implausibility must be combined across different data sets in
a carefully thought-out manner and
weighting the data from prior knowledge would require much consideration. Moreover, if the 
two data sets are in contradiction, this can lead to over-confident exclusion of 
parameter space regions, rather than pointing to a particular area in which
the model needs improvement. The tension between the disk size and Tully-Fisher
relation data sets is a clear case of this. Applying a strict requirement that
the model match both data sets greatly restricts the acceptable parameter 
region. The physical cause of this tension is the baryonic contraction
of the halo, something that is perennial problem in models of disk galaxies
(cf.\ Cole et al.\ 2000). However, the degree of contraction is strongly
dependent on the angular momentum distribution of the accreted material
and the treatment of contraction in disk instabilities (see Benson \& Bower 2009).
It is therefore questionable if we should reject
models on the basis of this tension at the outset, and an approach of
considering data sets separately may be preferable.

We find that the disk size, Tully-Fisher and gas mass to luminosity ratio data sets 
provide the strongest additional constraints (ie., in addition to the luminosity
function constraints). Qualitatively similar conclusions were reached previously by
Cole et al. (2000), using an earlier version of the GALFORM model, and
following the traditional approach of perturbing parameters away from their
``best fit'' values and visually inspecting the results. In contrast,
the methods we develop here are objective and do not rely on visual assessment.
Although, data on colours and metalicity provide additional 
restrictions, at the level that is considered here, they provide
little additional leverage over comparison of the luminosity
functions. 
Future model changes that improve treatment of angular momentum and baryonic contraction
should allow these data sets (and others that we have not considered 
here, such as the X-ray luminosities of groups and clusters, 
Bower et al.\ 2008) to play a more consistant role. Inclusion of high-z 
data sets is also possible; however, there is clearly a balance to be struck
between using all the available data to calibrate the model and holding some 
data sets in reserve to act as a test of the model's  predictive (or rather ``post-dictive'') 
power.

\section{Conclusions}

The GALFORM model is a semi-analytic model of the galaxy formation process 
that has been extensively used to understand the formation of
galaxies. Different versions of GALFORM have used varying
prescriptions for some of the key physical processes, and been tuned to match
different, though overlapping, observational datasets (Cole et
al. 2000, Baugh et al. 2005; Bow06). In particular,
the Bow06 version of the model, incorporating AGN feedback, 
has been successful in providing a good description of both the
$b_J$ and $K$-band luminosity functions of present-day galaxies and
the evolution of the $K$-band luminosity function
with redshift. However, the Bow06 model represents one selection of parameters from
a vast 16-dimensional parameter space (which is in turn a subspace of
the full GALFORM parameter space), and it is natural to ask how unique the 
model is. In particular, we would like to identify the region of parameter space
that is constrained in this way, allowing us to understand the degeneracies between the
input parameters and their relative importance.
Unfortunately, direct evaluation of the model with a uniform and sufficiently dense covering 
of the parameter space is not feasible because of the high-dimensionality of the 
parameter space and the relatively long run-time of the model. This is a common 
problem in many computer-modeling disciplines, such as climate change modeling, and there
is great interest in developing efficient mathematical techniques that optimise the
use of the available computing resource.

In this paper, we have used the {\it model emulator} technique (Craig et al.\ 1996; Craig et al.\ 1997; Kennedy \& O'Hagan 2001) to explore the 
full 16 dimensional parameter space of the Bow06 galaxy formation model. Rather than
trying to evaluate the model directly at all points in parameter space, we run the 
model at a sparse sampling of points and then construct an emulator of the model 
that allows us to interpolate between these points.  A key aspect of this construction
is that we are not only able to establish an expectation value for the model's performance
between runs, but we are also able to encapsulate the degree of uncertainty 
in this estimate.  Thus we can thus use the emulator to identify regions in which the
run outcome is uncertain and target additional evaluations there. By proceeding
in waves of iterations, we focus the model evaluation down on a smaller and smaller 
region in which run outcomes are likely to accurately reproduce the observational data.

{\rgb Another important aspect of our approach is that we introduce the concept of ``model discrepancy''.
This is a small additional variance term that is included to explicitly account for
the level of approximation inherent in the GALFORM model. This term means that model
luminosity functions which lie within a factor of 1.54 of the observed luminosity functions
are deemed acceptable, even if the observational errors are much smaller than this. This term 
ensures that we do not attempt to over fit the model and rule out regions of parameter
space when this is not justified (given the approximate nature of the model).}

The method is shown to be highly effective. We find that
  0.014\%  of the input parameter
space produces model luminosity functions that are acceptable matches to the
observed $b_J$ and $K$-band luminosity functions from Norberg et
al. (2002) and Cole et al. (2001) {\rgb (although, in common with previous versions of
the code, acceptable models tend to lie slighly above the faintest $K$-band measurements)}. However, 
we find that although the region of parameter-space is small, acceptable fits 
can be obtained as parameters are adjusted over a large fraction of their input range.
We show that the choice of parameters in the Bow06 model is not unique in reproducing the observational data and that
other choices of parameters perform at least equally well: changes in one parameter 
may be compensated by adjusting several other parameters to leave the predicted
luminosity functions almost unchanged. However, while some parameters play a vital role in 
adjusting the luminosity function to match the observational data, others 
have little effect on this (when adjusted within the range explored). Interestingly these inactive 
parameters include the re-ionisation parameters ($\VCUT$, $\ZCUT$), suggesting that these affect only
galaxies below the faintest luminosities included in our luminosity function, and the
merger parameters (these are masked by the dominance of disk instabilities
in the model and, to some extent, the lack of morphological constraints).

In order to explore the parameter dependencies further, we have investigated using principal
component analysis to identify optimal projections of the data.  These allow us to
identify the hyperplane onto which the data is constrained by the 
luminosity function data.  This analysis reveals physically interesting interactions
between the data, although each plane generally includes contributions from a 
large number of parameters. We have briefly explored how these directions can be 
rotated to reveal simple relations between the parameters that must be obeyed in 
order to obtain a well fitting model. For example, we are able to quantify the 
relationship between the feedback parameters ($\vhotdisk$, $\vhotburst$, 
$\alphahot$), the return timescale for reheated gas ($\alphareheat$) 
and the star formation parameter ($\invepsilonStar$). However, although these relations 
can provide a useful
check on physical intuition and understanding, they do not reproduce the full non-linear 
complexity of the model that is captured by the model emulator.

We have also briefly explored the impact of adding additional data sets to 
constrain the model by evaluating a simple $\chi^2$ statistic against observational
data sets describing disk sizes, the Tully-Fisher relation, gas metallicity, the 
gas mass to light ratio, the galaxy colour distribution and the 
black hole mass -- bulge mass correlation.  We find that the disk sizes, 
the gas-to-light ratio and the Tully-Fisher relation place the strongest additional
constraints on the model. Similar conclusions were reached by Cole et
al. (2000) using an earlier version of GALFORM, who followed the
traditional approach of manually varying parameters and then visually
inspecting the results. The great advantage of the new approach
presented here is that it is automated, objective, and reveals the
couplings between different parameters in their effects on observable quantities.
However, we find that the direction of constraint from
the disk size data is in contradiction to the constraints imposed by the
gas mass to luminosity ratio and Tully-Fisher relation
data.  The disk size data presents a particular challenge for the Bow06 model,
and the issue has been explored in detail in Gonzalez et al. 2009.

The model considered here was deliberately restricted to the version
of GALFORM used in Bow06. In future papers we will apply the same
general methods to a broader class of GALFORM models, including
processes such as feedback from  galaxy super-winds (Benson et al.\
2003a), variations in the IMF (Baugh et al.\ 2005), ram-pressure
stripping  (Font et al.\ 2008), and X-ray emission from hydrostatic haloes
(Bower et al.\ 2008), greatly increasing
the model parameter space. This leads to a new challenge --- but it is one that we
now have the statistical techniques to meet.

\section*{Acknowledgments}

RGB acknowledges the support of a Durham-University Christopherson-Knott
Fellowship. IRV and MG acknowledge the support of the Basic Technology initiative as part of the 
Managing Uncertainty for Complex Models project. IRV acknowledges the support of an EPSRC mobility fellowship.
AJB acknowledges the support of the Gordon \& Betty Moore
Foundation. CGL acknowledges support from the STFC rolling grant to
the ICC.

\end{document}